\documentclass[11pt]{article}
\usepackage{hyperref}
\usepackage[T1]{fontenc}
\usepackage{graphicx}
\usepackage{amsmath,amssymb}
\usepackage[sort&compress,numbers]{natbib}
\usepackage{authblk}
\usepackage{graphicx}
\begin{document}

\title{Dangerous Questions in Astronomy Education}
\author[1]{Michael Fitzgerald\thanks{Corresponding author: psyfitz@gmail.com}}
\author[2]{Rachel Freed}
\author[3]{Dan Reichart}
\author[4]{Kate Meredith}
\author[5]{Kalée Tock}
\author[6]{Daryl Janzen}
\author[7]{Saeed Salimpour}

\author[8]{Jennifer Lynn Bartlett}
\author[9]{Matthew Beaky}
\author[10]{Art Borja}
\author[11]{Ken Brandt}
\author[12]{Jim Buchholz}
\author[13]{Patricia Craig}
\author[14]{Anthony Crider}
\author[15]{Richard Datwyler}
\author[16]{Marta Dark-McNeese}
\author[17]{Anna DeJong}
\author[18]{Donovan Domingue}
\author[19]{Debbie French}
\author[20]{Oliver Fraser}
\author[21]{Amy L.\ Glazier}
\author[22]{Enrique Gomez}
\author[23]{Erika Grundstrom}
\author[24]{Nicole Gugliucci}
\author[25]{Kevin Healy}
\author[26]{Ardis Herrold}
\author[27]{Ian Hewitt}
\author[28]{Jack Howard}
\author[29]{Katherine Hunt}
\author[30]{Yashashree Jadhav}
\author[31]{Jonathan Keohane}
\author[32]{Noah Kearns}
\author[33]{Lancelot Kao}
\author[34]{Brian Kloppenborg}
\author[35]{Kelly Kosmo O’Neil}
\author[36]{Kevin Lee}
\author[37]{Ulrike Lahaise}
\author[38]{Sandy Liss}
\author[39]{David McKinnon}
\author[40]{Adam McKay}
\author[41]{Stephen McNeil}
\author[42]{Mariel Meier}
\author[43]{Jackie Milingo}
\author[44]{Qurat-ul-Ann Mirza}
\author[45]{Abbas Mokhtarzadeh}
\author[46]{Raúl Morales-Juberías}
\author[47]{Sean Moroney}
\author[48]{Rhone O’Hara}
\author[49]{Angela Osterman Meyer}
\author[50]{Imad Pasha}
\author[51]{Bradley W.\ Peterson}
\author[52]{Luisa Rebull}
\author[53]{Digesh Raut}
\author[54]{Christine Russell}
\author[55]{Ann Schmiedekamp}
\author[55]{Carl Schmiedekamp}   
\author[56]{Madeline Shepley}
\author[57]{Deanna Shields}
\author[58]{Brooke Skelton}
\author[59]{Don Smith}
\author[60]{David Sukow}
\author[61]{John B.\ Taylor}
\author[62]{Elise Weaver}
\author[63]{Michelle Wooten}
\author[64]{Tiffany Stone Wolbrecht}
\author[65]{David Yenerall}

\affil[1]{Las Cumbres Observatory}
\affil[2]{InSTaR}
\affil[3]{University of North Carolina at Chapel Hill}
\affil[4]{GLAS Education}
\affil[5]{Stanford Online High School}
\affil[6]{University of Saskatchewan}
\affil[7]{Deakin University / IAU OAE / MPIA}
\affil[8]{NASA Astrophysics Data System}
\affil[9]{Juniata College}
\affil[10]{Ad Astra Research}
\affil[11]{Robeson Planetarium}
\affil[12]{California Baptist University}
\affil[13]{Coastal Carolina University}
\affil[14]{Elon University}
\affil[15]{Brigham Young University–Idaho}
\affil[16]{Spelman College}
\affil[17]{Howard Community College}
\affil[18]{Georgia College \& State University}
\affil[19]{Wake Forest University}
\affil[20]{University of Washington}
\affil[21]{University of North Carolina at Chapel Hill}
\affil[22]{Western Carolina University}
\affil[23]{Vanderbilt University}
\affil[24]{Saint Anselm College}
\affil[25]{Mesa Community College}
\affil[26]{Vera C. Rubin Observatory (LSST)}
\affil[27]{Coastal Carolina University}
\affil[28]{Rowan–Cabarrus Community College}
\affil[29]{Digitaliseducation.com}
\affil[30]{Elon University}
\affil[31]{Hampden–Sydney College}
\affil[32]{Mitchell Community Schools}
\affil[33]{City College of San Francisco}
\affil[34]{AAVSO}
\affil[35]{University of Nevada, Reno}
\affil[36]{University of Nebraska–Lincoln}
\affil[37]{Georgia State University}
\affil[38]{Radford University}
\affil[39]{Colorado State University}
\affil[40]{Appalachian State University}
\affil[41]{Brigham Young University–Idaho}
\affil[42]{Oglethorpe University}
\affil[43]{Gettysburg College}
\affil[44]{Rowan College at Burlington County}
\affil[45]{University of North Carolina at Chapel Hill}
\affil[46]{New Mexico Tech}
\affil[47]{University of Hawaii}
\affil[48]{Pearson}
\affil[49]{Durham Academy}
\affil[50]{Yale University}
\affil[51]{Hastings College}
\affil[52]{IPAC/Caltech}
\affil[53]{St.\ Mary’s College of Maryland}
\affil[54]{Moorpark College}
\affil[55]{Pennsylvania State University}
\affil[56]{Olivet University}
\affil[57]{University of Arkansas}
\affil[58]{Perimeter College, Georgia State University}
\affil[59]{Guilford College}
\affil[60]{Washington and Lee University}
\affil[61]{Marvin Ridge High School}
\affil[62]{Appalachian State University}
\affil[63]{University of Alabama at Birmingham}
\affil[64]{Associated Universities, Inc.}
\affil[65]{Georgia State University}

\maketitle

\begin{abstract}
As astronomy enters an era defined by global telescope networks, petabyte-scale surveys, and powerful computational tools, the longstanding goals of astronomy education, particularly introductory “ASTRO101", but equally encompassing both higher and lower level courses, warrant fresh examination. In June 2024, the AstroEdUNC meeting at UNC–Chapel Hill convened 100 astronomers, education researchers, and practitioners to synthesise community perspectives on the purpose, content, and delivery of astronomy education. Beginning with historical vignettes, the meeting’s deliberations were organised into six interrelated themes: (1) Context, highlighting astronomy’s evolution from classical charting to multi-messenger discovery and its role as a connective thread across STEM and the humanities; (2) Content, exploring how curricula can balance essential concepts with authentic investigations and leverage open-source and AI-augmented resources; (3) Skills, arguing that astronomy should foreground scientific literacy, computational fluency, and communication through genuine data-driven inquiry; (4) Engagement, advocating for active-learning strategies, formative assessment, and culturally inclusive narratives; (5) Beyond the Classroom, emphasising scaffolding, universal-design practices, and K–12/community partnerships; and (6) Astronomy Education Research, outlining priority areas for assessing knowledge, attitudes, and long-term outcomes. We provide concrete recommendations for future astronomy education research development, underscoring the need for approaches to education that are authentic while meeting the learning and life goal needs of the students, a vibrant community of practice and robust researcher–practitioner partnerships to ensure that introductory astronomy is pertinent, applicable and inspiring to a broad student population.
\end{abstract}

\section{Introduction}

Over the years there have been various efforts and ongoing debates in the United States to determine and characterize the goals of/for astronomy education. These efforts have included developing standardized curricula (Zeik \& Morris-Dueer, 2005), building assessment tools (Hufnagel, 2001), and developing strategies for introductory astronomy courses (Partridge \& Greenstein, 2003). These efforts have given the community new insights, but have also highlighted the complexities involved.

In the current era of large-scale astronomy projects, developments in technology and the rise of vast amounts of data in astronomy, there needs to be a reassessment of the goals of astronomy education now and into the future. The AstroEdUNC meeting held at the University of North Carolina at Chapel Hill brought together 100 participants (astronomers, astronomy education researchers, and practitioners) to discuss and synthesize the goals of astronomy education in the USA, and in the context of Introductory Astronomy (ASTRO101).

The document begins with some historical snapshots of some key efforts in the path to characterizing the goals of astronomy education. Following this the report is divided into six chapters that formed the overarching themes of the meeting at UNC: Context, Content, Skills, Engagement, Beyond the Classroom, and Astronomy Education Research. These discussions and insights that emerged within these themes were informed by a range of questions that were voted by the participants prior to the meeting. Each subchapter unpacks one question, bringing together various perspectives, personal experiences, and a range of resources. Finally, this report ends with some recommendations for the future of ASTRO101.

Astronomy allows for unique connections. Its history extends back thousands of years in history, touching on cultures around the world. Astronomy is \emph{not} a product of western European culture in the last few centuries even though it is often taught that way with textbooks reinforcing this misconception. It draws on writings and observations from \emph{all} cultures, including Asian, African and Native American cultures. From this background, it is clear that astronomy is open to and benefits from this unique cultural diversity.

\subsection{Historical Vignettes}

The Conceptual Astronomy Project (Zeilk \& Morris-Dueer, 2005) aimed to design an astronomy course around community-based core concepts. They had a group of 3 astronomy colleagues reduce 400 concepts gleaned from an astronomy textbook glossary down to the 200 most important and then had a panel of 18 ASTRO101 instructors rank them from highly essential to not essential. Interestingly, specifically excluded was astronomical technology, such as telescopes. Further refinement created a group of concepts that lead the authors to state “the end result of these filters are the key concepts, then, that can form the core of an ASTRO101 course, based on the responses of a panel of expert teachers”. It is important to note that the first author of this study stated that he drops certain of these concepts from his courses based on his own ideas of what is important.

In 1998, a group of astronomy education researchers convened to develop the Astronomy Diagnostic Test in response to recognition that there was no standardized means of assessing astronomy knowledge in ASTRO101 courses. (Hufnagel, 2001). In 2001, studies were conducted amongst a small group of astronomy instructors to gauge what they felt was important in astronomy. Simultaneously an analysis of astronomy course syllabi found online was conducted. A few themes emerged, with a focus on the electromagnetic spectrum and the size, scale, and structure of the cosmos (Slater et al., 2001). Improving attitudes towards science and motivating a lifelong interest in learning were important goals for many of the educators, although this rarely showed up explicitly on syllabi.

In 2001 two workshops for astronomy department leaders were convened by the American Astronomical Society’s Education Office, to formulate recommended goals and useful strategies for ASTRO101 courses to be reviewed by the wider astronomical community. (Partridge \& Greenstein, 2003). Surprisingly, at neither meeting did the participants propose a detailed set of standards for ASTRO101. The generally agreed upon conclusion was that each instructor was free to decide for themselves which astronomical concepts and quantities were important for students to learn. However, the workshops did create a set of content goals and a set of skills, values and attitudes goals for ASTRO101. In a follow-up paper by Lippert \& Partridge (2004), Partridge describes his implementation of these goals within his own small class and reflects on the fact that he did not like the suggestion that students are given a voice in the curriculum because they may choose topics that he sees are less valuable than what he as a professional astronomer values.  There seems to be a consistent theme of individuals picking and choosing from amongst the consensus documents that arise out of attempts to standardize the astronomy 101 curriculum.

In this document created out of the discussions during the AstroEdUNC meeting, we have tried to present the results of several days of discussions among the 100 participants. Within each section, different people have used their own projects as examples, and of course, this provides a non-exhaustive sampling of projects and programs available within the astronomy education community in the context of ASTRO101. We summarise the key findings within each of the dangerous question categories here.

\subsection{Key findings}

The process and structure of the meeting enabled a diverse range of views and insights to be discussed and debated. Many participants highlighted that building a community, or more specifically a “community of practice” is an important aspect for many of the practitioners who attended the meeting. Astronomy education is a layered community particularly when considering the difference between secondary and post-secondary, and the fuzzy nature (at least to the general layperson) of education-outreach.

The key findings from the meeting are provided below:

\subsubsection{Context}

The nature of astronomy as practiced in this current era provides a rich context for giving students insight into the contemporary practices of the discipline. This is perhaps owing to the proliferation of technological innovation and easy access to research grade infrastructure, which includes but is not limited to: robotic telescopes, large-scale research collaborations, and big data through archives.

While its roots lie in charting celestial motions by eye, modern astronomy weaves together physics, chemistry, biology, data science, and engineering: from decoding stellar spectra with atomic models to designing space-based observatories, from mining massive sky surveys with machine learning to probing astrobiological signatures on distant worlds. Professional astronomers balance theoretical insight, technological innovation, and diverse observations—whether gravitational waves, exoplanet transits, or high-precision space astrometry—to piece together the universe’s grand narrative.

Over the past century and a half, the field has transformed from careful cataloging and timekeeping to a rich interplay of theory and technology. Late-19th-century observers compiled nebula and star cluster lists and pioneered photometry and precision clocks. Early-20th-century breakthroughs—atomic models, the Hertzsprung–Russell diagram, Cepheid distance scales, and Hubble’s extragalactic discoveries—laid the foundations of modern cosmology. Mid-century advances in radio astronomy, space-borne platforms, and the Great Observatories (Hubble, Chandra, Spitzer, Compton) opened the full electromagnetic spectrum, while the detection of exoplanets and gravitational waves in recent decades has inaugurated entirely new subfields like asteroseismology and multi-messenger astronomy.

Despite its specialized tools and methods, astronomy need not—and often should not—be siloed. Its concepts and activities dovetail naturally with physics, chemistry, mathematics, environmental science, and even the humanities. Planetary orbits illustrate Newtonian mechanics in math classes; spectral analysis bridges chemistry labs and astrophysical inquiry; light-pollution studies enrich environmental science. Cross-disciplinary modules whether on citizen-science projects, historical case studies, or data-science pipelines feeding Rubin Observatory’s petabytes—bring astronomy into diverse curricula, while astronomy courses themselves serve as gateways to critical thinking, quantitative reasoning, and research skills that propel students into STEM and beyond.

In embracing its broad scope, astronomy both stands as a distinct science and as a connective thread among disciplines. By integrating astronomical themes into varied courses—and by inviting non-specialists to explore real data, creative reflections, and ethical dimensions of cosmic discovery—educators can harness astronomy’s unique power to inspire wonder, build scientific literacy, and cultivate the next generation of interdisciplinary thinkers.

\subsubsection{Content}

The role of content in the teaching and learning of any discipline is complex as it raises questions about the extent to which it influences deep learning. Content is one layer of a larger ecosystem of factors that enable the progression from novice to expert in any discipline. The challenge is how to balance content with the diverse range of skills that equip students for the 21st century and beyond. Some of the key considerations that would inform how to structure the teaching and learning to ensure a balance between content and the range of skills is to take into consideration student interests and prior knowledge. Using a backward design approach to course development together with a flexible curriculum can ensure that there is a balance between content and various authentic investigations which help develop student skills.

One approach would be to ensure that content is used as a way to support curiosity and develop scientific literacy. One aspect of any course is taking in consideration the role of textbooks in teaching and learning, the costs associated with some textbooks (unless using open source) creates barriers of accessibility. There was a general consensus that textbooks do not determine whether a course is able to “achieve its goals for students”, the more important aspect is how the teaching and learning is structured. If the aim is to engage students in critical reading then carefully curating targeted reading material is perhaps a better approach.

Finally, in this era of “AI” grounded in Large Language Models (LLM), it is vital to consider what role these “tools” play in “transferring” content to students, and perhaps completely rethinking the usefulness of textbooks. LLMs also have permeated the assessment space, and as such educators are tasked with designing assessment tasks that do not merely evaluate the “facts of astronomy” rather they encourage students to think critically, and provides them with an opportunity to use these tools appropriately and effectively in the learning.

\subsubsection{Skills}

The general consensus of the meeting was that skills represent the fundamental learning goals for introductory astronomy courses, and that the content is a fun and inspiring vehicle through which to teach the skills that will be important in the workforce these students are entering. The skills that can be taught through these courses align well with both the tenets of scientific literacy and 21st century skills as defined by the National Academies (Defining Deeper Learning, 2012). Implementing a curriculum in which students access professional equipment (a global telescope network) as well as use archival data from professional ground- and space-based telescopes adds authenticity to their learning, while giving them a sense of ownership over their work. An important point of discussion ensued around the different learning goals for astronomy majors, other STEM majors and non-STEM majors. While these groups do have different specific learning needs, the 21st century skills and scientific literacy apply to all of them, and classes can be tailored to populations within them, in terms of their preparedness and what they need to learn to be successful going forward.

Building on this skill focus, authentic, data-driven investigations—whether gathering images from remote telescopes or mining archival surveys—replace cookbook labs and passive lectures. Easy access to public telescope networks and large datasets means students can experience genuine scientific inquiry: defining their own questions, making critical choices in analysis, and wrestling with ambiguity. At the same time, integrating structured writing and oral assignments (from lab reports and code documentation to blog posts and presentations) teaches students to communicate precisely and persuasively, a competency prized both within and beyond STEM.

Finally, these pedagogical choices serve all learners—future astronomers, other STEM majors, and non-STEM students alike—by equipping them with transferable “21st-century” skills: problem-solving, teamwork, ethical standards, computational literacy, and resilience. Moreover, fostering a sense of ownership—through engagement with professional-grade tools, authentic decision-making, and opportunities to share or even publish one’s work—drives motivation and closes self-efficacy gaps. In sum, a skills-centered, authentic, and communicative approach to ASTRO 101 not only deepens scientific understanding but also prepares students for diverse future pathways.

\subsubsection{Engagement}

Research in science education shows that knowledge, affect (attitudes and self-efficacy), and inquiry skills evolve together rather than in isolation. Cultivating genuine curiosity—through open-ended investigations, socio-scientific contexts, and opportunities for moral and collaborative reflection—can fuel students’ motivation to learn and stick with STEM. While it’s tempting to view content mastery and positive attitudes as a binary choice, effective instruction weaves them together: modeling inquisitive questioning, structuring hands-on activities that prompt wonder, and giving students ownership over their learning journeys.

True “lasting learning” emerges when students actively construct knowledge rather than passively receive facts. Interactive engagement techniques (think-pair-share, lecture tutorials, real-data labs) dramatically outperform traditional lectures in promoting retention, especially when paired with attention-grabbing hooks—cinematic clips, vivid graphics, relevance to everyday technologies—and multimodal delivery. Embedding formative assessments, smartphone-friendly simulations, and opportunities for students to predict, test, and reflect helps solidify concepts in long-term memory. Humanizing science—by spotlighting the variety of astronomers and connecting cosmic questions to students’ lives—further anchors learning in personal meaning.

Finally, both authentic research experiences and classroom-based active learning are essential. Authentic projects using real, messy datasets immerse students in the iterative nature of science—designing mini-experiments, grappling with uncertainty, and learning from mistakes—building self-efficacy and scientific identity. Meanwhile, well-scaffolded active-learning strategies foster engagement, close achievement gaps, and refine higher-order thinking. To make these approaches sustainable, we must bridge the gap between education research and everyday teaching: developing accessible, universal digital tools; sharing proven practices through professional communities; and adapting innovations to diverse classroom constraints. Together, these strategies can transform astronomy instruction into a catalyst for enduring understanding, curiosity, and equitable participation in science and beyond.

\subsubsection{Beyond the Classroom}

Instructors can’t assume incoming students arrive with all the background skills needed for an introductory astronomy course, so effective scaffolding must begin by making course logistics and expectations explicit (syllabus quizzes, rubric walkthroughs, clear policies on grading and accommodations). Beyond astronomy content, students often need support brushing up on math (scientific notation, algebra, graph reading), basic physics concepts (units, motion, density), digital literacies (file management, spreadsheets, simulations), and academic habits (note-taking, time management, collaboration). Embedding low-stakes practice assignments, modeling step-by-step problem solving, linking to tutorials (Khan Academy, YouTube), and incentivizing office-hour use or peer tutoring all help bridge gaps without lowering standards.

At the same time, many hidden expectations—unspoken prerequisites in technology access, study skills, prior science exposure, and even cultural familiarity with “college-style” learning—can erect real barriers for learners. Economic pressures, caregiving responsibilities, health or disability concerns, and “imposter” feelings often go unaddressed unless instructors deliberately adopt universal-design practices: flexible deadlines, clear and simple language, frequent formative checks, and easily navigable support resources (academic, IT, health, and counseling services). Early-semester surveys or information sheets can surface students’ needs so that advocacy systems (ombudspersons, learning centers) can step in before struggles compound.

Finally, attracting and retaining students requires both outreach and community building. Recruitment starts long before college—K–12 partnerships, planetarium events, family-centered programming, and visible role models help spark interest and signal belonging. On campus, robust support structures (academic counseling, financial aid for hardware and textbooks, childcare solutions) and micro-level efforts (study groups, learning communities, mentoring by advanced peers and faculty) foster the social and practical networks that keep students engaged. When instructors model healthy work–life balance, facilitate student-led clubs and service-learning, and collaborate with administrators to remove barriers, they create an environment where all students can thrive in astronomy and beyond.

\subsubsection{Astronomy Education Research}

Astronomy education research increasingly recognizes that students’ mastery of facts and problem-solving skills cannot be disentangled from their attitudes, confidence, and career aspirations. Curiosity sparks engagement, which builds literacy and shapes attitudes; positive attitudes and self-efficacy then drive deeper cognitive processing and sustained interest in STEM pathways. To capture these dynamics, researchers must define and measure content knowledge, thinking processes, attitudes, self-beliefs, and career intentions—using low- and high-stakes assessments, surveys, interviews, and concept inventories—and tailor studies to different audiences (from non-majors in ASTRO101 to K–12 learners).

Equally important is deciding where to focus interventions. Career interests often form by fourth grade, yet astronomy is squeezed into limited elementary and middle-school science time. Strengthening teacher preparation, integrating astronomy with literacy and math, and building vertical teaming from K through grade 8 can harness astronomy’s “gateway” potential. At the high-school and collegiate levels, supporting instructors—through authentic data use, active-learning training, and feedback networks with researchers—ensures that evidence-based strategies are implemented effectively.

Finally, research must look beyond the formal ASTRO101 classroom to the broader ecosystem of informal learning—from planetariums and outreach programs to social-media influencers—that shapes students’ preconceptions and curiosities. By mapping the veracity of these sources, employing participatory action research to uncover alternative conceptions, and tracking long-term outcomes, educators can design holistic interventions that correct misconceptions, reinforce accurate scientific models, and foster lifelong engagement in astronomy. Across every stage, strong partnerships between researchers and practitioners are essential to translate insights into equitable, scalable improvements in STEM learning and career pathways.

\textbf{References}

Aristeidou, M., \& Herodotou, C. (2020). Online citizen science: A systematic review of effects on learning and scientific literacy. Citizen Science: Theory and Practice, 5(1), 1-12.

Defining Deeper Learning and 21st Century Skills | National Academies. (n.d.). Retrieved May 18, 2025, from

Hufnagel, B. (2002). Development of the astronomy diagnostic test. Astronomy Education Review, 1(1), 47-51.

Lippert, N., \& Partridge, B. (2004). To Hear Ourselves as Others Hear Us. Astronomy Education Review, 3(1).

Partridge, B., \& Greenstein, G. (2003). Goals for" ASTRO101": Report on workshops for department leaders. Astronomy Education Review, 2(2).

Prather, E. E., Rudolph, A. L., \& Brissenden, G. (2009). Teaching and learning astronomy in the 21st century. Physics Today, 62(10), 41-47.

Tim Slater, Jeffrey P. Adams, Gina Brissenden, Doug Duncan; What topics are taught in introductory astronomy courses?. Phys. Teach. 1 January 2001; 39 (1): 52–55.

Slater, S. J. (2014). The Development and Validation of the Test Of Astronomy STandards (TOAST). Journal of Astronomy \& Earth Sciences Education, 1(1), 1-22.

Waller, W. H., \& Slater, T. F. (2011). Improving introductory astronomy education in American colleges and universities: A review of recent progress. Journal of Geoscience Education, 59(4), 176-183.

Zeilik, M., \& Morris-Dueer, V. J. (2005). What Are Essential Concepts in ‘Astronomy 101’?: A New Approach to Find Consensus from Two Different Samples of Instructors. Astronomy Education Review, 2(3), 61.

\section{Context}

\subsection{What is astronomy? Is it ‘what professional astronomers do’? Is it anything that can be observed with the telescope?}

\begin{flushright}
  \itshape
  Brian Kloppenborg\\
  Noah Kearns
\end{flushright}

Astronomy has a unique draw and ability to inspire (Bode et al., 2008). This widespread general interest has led to the popular belief that astronomers spend all their time looking through telescopes in observatories (Kovalenko, 2019).  Astronomy is more than what can be observed through a telescope; astronomy is an inherently interdisciplinary science dedicated to understanding the universe and our place within it (Berea et al., 2019; Kegel, 1987).  Astronomy is a very dynamic science that constantly evolves.

We began by using stories to describe the universe, progressed through using telescopes ground by hand, and have moved on to detecting gravitational waves using lasers. Astronomy started as and, in some ways, still is an observational discipline. The field was born by watching the skies and the motion of the planets against the stars.  Astrophysics was born when physicists began trying to describe the observations by creating physical models to describe this motion.  Astronomy and physics continue to have a special relationship today where observations are described by physics and astronomy makes observations to verify physics predictions.

As we learn about our place in the universe, we are realizing that we need to consider the results of even more disciplines. Important questions such as what is required for life to evolve on other planets?  Is there a definitive chemical signature of life?  What would the spectra of these chemicals look like?  These questions address our place in the universe but require astronomers to incorporate research from fields that are not traditionally associated with astronomy. Astronomers are using advances in the fields of data science to help analyze and interpret the huge amounts of data that is available to us from observatories around the world.   Astronomy also requires an understanding of engineering to help develop the instruments that will allow these discoveries to happen.

All these diverse tools and approaches have the same goal: trying to describe the story of the universe and our place within it.  Astronomy continues to be a dynamic and interdisciplinary pursuit that has evolved alongside humanity. Astronomers collaborate across diverse fields, from physics to data science and engineering , to understand the cosmos and our place within it.

\textbf{References}
Berea, A., Denning, K., Vidaurri, M., Arcand, K., Oman-Reagan, M. P., Bellovary, J., ... \& Lupisella, M. (2019). The Social Sciences Interdisciplinarity for Astronomy and Astrophysics--Lessons from the History of NASA and Related Fields. arXiv preprint arXiv:1907.07800.

Bode, M. F., Cruz, M. J., Molster, F. J., \& Infrastructure Roadmap Working Group (Eds.). (2008). The ASTRONET Infrastructure Roadmap: A strategic plan for European astronomy. ASTRONET.

Kegel, W. H. (1987). Astronomy, an Interdisciplinary Science. Universitas, 29(3), 157–166.

Kovalenko, N. (2019). Basic astronomy: Common misconceptions and public beliefs according to the audience survey at Kyiv Planetarium. EPJ Web of Conferences, 200, 01023.

\subsubsection{How has astronomy changed over the past ~150 years?}

\begin{flushright}
  \itshape
  Brian Kloppenborg  
\end{flushright}

Over the past 150 years, astronomy has evolved from a discipline focused on observation and cataloging to one characterized by a dynamic interplay between theory, technology, and observation. From the foundational work of cataloging celestial objects to the revolutionary discoveries of the early 21st century, such as the confirmation of gravitational waves and the exploration of exoplanets, astronomy has continually pushed the boundaries of human knowledge and understanding of the cosmos. Each era has seen pivotal moments, from the theoretical breakthroughs of the early 20th century to the technological advancements and space exploration missions of the late 20th and early 21st centuries. As we look to the future, the field of astronomy remains poised for further discoveries and insights into the nature of the universe.

Approximately 150 years ago, much of astronomy focused on the creation of catalogs, precision timekeeping, and development of new and better instrumentation. One of the most important catalogs published from this era was the New General Catalogue of Nebulae and Clusters of Stars (Dreyer, 1888) that was published in 1888. This catalog documented a wide array of diffuse objects which are now recognized as galaxies, star clusters, and emission nebulae. It expanded considerably on the Messier Catalog published some 100 years before. Many researchers also focused on making improvements to instrumentation developed a few years earlier. For example, instrumentation and techniques used for astronomical photometry advanced considerably []. In addition to scientific catalogs, many professional observatories in the late 1800s also provided a very practical service to their local communities by serving as official timekeepers. It was very common for most observatories of the era to have a transit telescope which permitted them to observe the passage of stars across the meridian which served to synchronize the observatory’s local clocks (King, 1976).

In the early 1900s, the field of astronomy underwent remarkable advancements. Theoretical developments in atomic theory including the Rutherford Model in 1911 and Bohr Model in 1913 provided critical insights into atomic structure, which permitted us to explain the emission and absorption spectra of stars. Contemporaneously, both Hertzsprung (Hertzsprung, 1911) and Russel (Russell, 1913) plotted stars on an absolute magnitude vs. photometric color which provided our first insights into stellar evolution. Shortly thereafter, Henry Draper published the first catalog of star positions and stellar classifications, which still serves as one of the most important catalogs for students and amateur astronomers. Likewise, Henrietta Leavitt's 1912 paper (Leavitt \& Pickering, 1912) firmly established the relationship between the period and luminosity of cepheid variables which revolutionized our ability to measure cosmic distances.

The rapid pace of astronomical developments continued into the 1920s. It started with the Great Debate between Harlow Shapley and Heber Curtis questioning the nature of the scale of the universe. These questions were resolved a few years later in 1923 when Edwin Hubble identified a Cepheid variable in M31 (Hubble, 1925; Hubble, 1929) which confirmed that galaxies were indeed far beyond the extent of the Milky Way. Advancements in theory, particularly the Saha Equation (Saha, 1921) provided a physical explanation to the classification of stars during the previous decade. A few years later in 1932, an entirely new observational field of radio astronomy was born when researchers at Bell Labs accidently measured a low, mysterious noise (Penzias \& Wilson, 1965) which later proved to be remnants of the Big Bang.

The time period from 1950 - 2000 saw such a large number of advancements that it is difficult to summarize here. The field of stellar polarimetry owes its existence to experiments by William Hiltner (Hiltner, 1949) and John Hall in 1949 (Hall, 1949). Gravitational waves, first predicted by Einstein in 1912 (Einstein, 1916) were confirmed through observations of a pulsar using radio astronomy in 1974 by Russell Hulse and Joseph Taylor (Hulse \& Taylor, 1975). Similarly, technological developments from the space race of the 1950s - 1960s lead to the launch of many sounding rockets to study earth’s atmosphere and ionosphere. These programs were rapidly followed by the launch of a multitude of Near Earth and Deep Space programs whose success ultimately paved the way for political lobbying that enabled the Great Observatory Programs between 1990 - 2003 (Harwit \& Neal, 1986). These observatories (Hubble Space Telescope, Chandra Space Telescope, Spitzer Space Telescope, Compton Gamma Ray Observatory) expanded our knowledge of the universe by observing across the electromagnetic spectrum from gamma rays to the far-infrared and remain among the most impactful and successful space programs to date. The century closed with the discovery of the first planet outside of our solar system in 1995, Pegasus 51 b, using the radial velocity method (Mayor \& Queloz, 1995, Nature, 378 (6555): 355–359. :). In 1999, HD 209458 b, which was discovered using the radial velocity method was identified as a transiting exoplanet (Charbonneau, D., et al. 2000, ApJ, 529:L45–L48). Thousands of exoplanets are now known (NASA, 2024).

A series of technical and theoretical advancements in the first quarter of the 21st century lead to the creation of three new fields of astronomy and significant advancements in others. Foremost, the fields of asteroseismology had been in development for many years, but the convergence of theory, greater computing power, and speciality spacecraft lead to its rapid development. The high-precision, reduced data products from spacecraft like CoRoT, Kepler, MOST, BRITE, and TESS continue to play an essential role in driving new discoveries in this field. Likewise, decades of meticulous research and advancements in instrumentation and analysis techniques has turned gravitational wave astronomy into a burgeoning field. Although the field of astrobiology can be traced to the discovery of extremophiles in the 1960s (Brock \& Brock, 1968), many of its recent advancements have been enabled by the plethora of exoplanets discovered by Kepler, the Mars Exploration Rovers, detection of organic molecules on the moons of Saturn, and ultra-high resolution spectrographs.

Over the course of the 20th century and beyond, measurement techniques in astronomy moved from very inefficient long exposure imagery on photographic plates to encompassing a variety of methods of scientific measurement such as:

Astrometry: Measurements of position progressed from photographic-plate measurements and manual cataloging to large-scale CCD surveys that dramatically improved positional accuracy. The launch of ESA’s Hipparcos satellite in 1989 marked the first space-based astrometric catalog with milliarcsecond precision, overcoming atmospheric limitations. Gaia now delivers microarcsecond-level parallaxes and proper motions for over a billion stars.

Photometry: Robotic telescopes and fully automated data reduction pipelines are commonly available to both professionals and amateur astronomers alike. Many sky surveys exist (e.g. ASAS, ASAS-SN, ZTF) and many large-scale sky surveys are coming online (i.e. Vera Rubin Observatory’s LSST and the Argus Array). Students and amateur astronomers are eagerly filling in the remaining observational parameter space..

Spectroscopy: There are a small number of robotic, multi-target spectrographs conducting sky surveys; however, many needs of spectroscopists are filled using remote or service-mode observations. Students and amateur astronomers have access to a variety of commercially available spectrographs ranging from R~200 to R~30,000.

Polarimetry: In radio astronomy, polarimetry is a commonly practiced field; however, at optical wavelengths, it remains relatively obscure with only a few professional facilities even offering broadband or spectro-polarimetric data. A few students and amateur astronomers have experimented with polarimetry, but it is not very well supported.

Interferometry: In radio astronomy, interferometry (e.g. VLBI) is a common practice. At optical wavelengths, interferometry is less common. There are a few professional observatories that produce speckle (e.g. Keck) and long-baseline interferometric data products (e.g. CHARA, VLTI). There is a small community of practice in the USA that teaches students how to do double star observations using speckle interferometry.

\textbf{References}

Brock, T. D., \& Brock, M. L. (1968). Measurement of steady-state growth rates of a thermophilic alga directly in nature. Journal of bacteriology, 95(3), 811-815.

Dreyer, J. L. E. (1888). A New General Catalogue of Nebulæ and Clusters of Stars, being the Catalogue of the late Sir John FW Herschel, Bart, revised, corrected, and enlarged. Memoirs of the Royal Astronomical Society, Vol. 49, p. 1, 49, 1.

Einstein, A. (1916). Näherungsweise integration der feldgleichungen der gravitation. Sitzungsberichte der Königlich Preußischen Akademie der Wissenschaften, 688-696.

Hall, J. S. (1949). Observations of the polarized light from stars. Science, 109(2825), 166-167.

Harwit, M., \& Neal, V. (1986). The great observatories for space astrophysics. The great observatories for space astrophysics.

Hertzsprung, E. (1911). Uber die Verwendung photographischer effektiver Wellenlaengen zur Bestimmung von Farbenaequivalenten (No. 63).

Hiltner, W. A. (1949). On the Presence of Polarization in the Continuous Radiation of Stars. II. Astrophysical Journal, vol. 109, p. 471, 109, 471.

Hubble, E. P. (1925). Cepheids in spiral nebulae. Pop. Astr.; Vol. 33; Page 252-255, 33.

Hubble, E. P. (1929). A spiral nebula as a stellar system, Messier 31. Astrophysical Journal, 69, 103-158 (1929), 69.

Hulse, R. A., \& Taylor, J. H. (1975). Discovery of a pulsar in a binary system. Astrophysical Journal, vol. 195, Jan. 15, 1975, pt. 2, p. L51-L53., 195, L51-L53.

King, H. C. (1976). Instrumentation of the nineteenth and early twentieth centuries. Vistas in Astronomy, 20, 157-163.

Leavitt, H. S., \& Pickering, E. C. (1912). Periods of 25 Variable Stars in the Small Magellanic Cloud. Harvard College Observatory Circular, vol. 173, pp. 1-3, 173, 1-3.

NASA (2024). https://science.nasa.gov/exoplanets/discoveries-dashboard/

Penzias, A. A., \& Wilson, R. W. (1965). A Measurement of Excess Antenna Temperature at 4080 Mc/s., 142: 419–421.

Russell, H. N. (1913). " Giant" and" dwarf" stars. The Observatory, Vol. 36, p. 324-329 (1913), 36, 324-329.

Saha, M. N. (1921). On a physical theory of stellar spectra. Proceedings of the Royal Society of London. Series A, Containing Papers of a Mathematical and Physical Character, 99(697), 135-153.

\subsubsection{Must Astronomy be taught in isolated Astronomy courses? How does astro fit into other STEM disciplines? Is astronomy actually a gateway to other STEM disciplines?}

\begin{flushright}
  \itshape
  Angela Osterman Meyer\\
  Lisa Storrie-Lombardi
\end{flushright}

Astronomy is often considered to be one of the oldest sciences and as such has a strong synergy to various disciplines. From an educational perspective, Astronomy is taught at all levels of education, either as an independent course or as part of various courses (Salimpour et al., 2021). There are many activities from astronomy that can be integrated into other courses. Some examples of places where astronomy has been incorporated into other courses given in the AstroEdUNC meeting include::

A module on light pollution for an undergraduate Environmental Studies class (Michelle Wooten -University of Alabama at Birmingham):. Light pollution affects human and environmental health, creating a strong connection between the two subjects. Dr. Wooten provided an Environmental Studies professor guidance on how to participate in the citizen science Globe at Night project (https://globeatnight.org/).

Kepler’s Laws for a high school AP European History (Angela Osterman Meyer Durham Academy): In a course examining the scientific revolution/renaissance, astronomy provided the context for a discussion of Kepler’s Laws Dr. Meyer and the AP teacher recorded a conversation for students sharing their respective expertise and perspectives.

Spectroscopy: importance to both Astronomy and Chemistry. For example, studying the spectrum of sodium in the lab and then showing the sodium line in the Sun’s spectrum.

Connecting Astronomy to Physics and Mathematics. As Lei Zhang (Winston-Salem State University) noted, Newton’s law of universal gravitation is key to understanding why and how the Moon moves around the Earth and how the Earth moves around the Sun.  Mathematics classes can be enriched through using planetary orbits as examples.

Reading and Mathematics: Elementary teachers must often focus the vast majority of their time on reading and mathematics. Astronomy can be incorporated as the topic of readings presented to students.

The question can also be flipped: How can tools and activities from other courses be meaningfully integrated into Astronomy?

Dr. Meyer shared an example of utilizing a , and having students listen to and discuss “Whitey on the Moon” by Gil Scott Heron.

Another conference participant described broadening a final project in astronomy to give students the chance to create art, poetry, or creative writing to show their understanding of the astronomy content. In another example, students wrote a poem about Laika, the famous space dog, or an essay about how a deeply religious person might perceive an astronomy class.

One participant noted that the way of doing and teaching science typical of Astronomy is not the only way in which science is done. Another way to build connections across disciplines and/or to frame instruction within Astronomy is to acknowledge or apply methodologies from diverse cultures and different times, to examine hard science vs. soft science methodologies, and to make connections to computer science.

While Data Science is not usually a course per se, this field came up during many of the question-specific discussions. Some participants expressed the idea that astronomy IS data science. Astronomers and astronomy students must create meaning from data. Astronomers have always done surveys to understand different classes of objects but historically the collection of the data was decoupled from the analysis, it was a serial process. In the last 30 years, the dramatic increase in the availability of computing power, data storage, and mission data archives with science-ready data available to anyone, enables people to engage with astronomy without ever visiting a telescope. We are building facilities, e.g. the Vera C. Rubin Observatory (20 TB of data every 24 hours, (Rubin Observatory, 2024)) or the Square Kilometer Array (0.5 to 1TB per second, SKA Observatory, 2024), where the data management system is of the same scale in cost as the telescope. Scientists and students will never deal with the full original images. While astronomy has always been a data intensive science, the study of data to extract meaningful insights (Data Science) - is itself a multidisciplinary approach to data analysis that combines mathematics, computer science, statistics, and AI to analyze large data sets.

Another program example is the USA Sky Partners from LCO. Through this program, undergraduate and high school instructors and students use the LCO telescopes to collect and analyze their own data on double stars and share their results through publication in the Journal of Double Star Observations. The research process includes accessing some of the large data sets, such as Gaia to look at parallax and proper motion values.

At the K-12 level, a number of participants noted that there are NGSS astronomy standards. These used to fit into Earth and Space Science courses. However, an increasing number of school systems and states no longer require Earth/Space Science to graduate high school, or have dropped such courses entirely (California Department of Education; Benow \& Hoover, 2015). Many school districts (not all) do require Physics, so teachers can choose to integrate Astronomy here, but for many teachers this is a significant challenge, not least of all because many K-12 science teachers do not have degrees in the subjects they teach.

Ardis Herrold’s (Rubin Observatory) work as an education specialist is an example of how to facilitate Astronomy in K-12 classrooms. Under her leadership, Rubin Observatory is developing, testing, and making freely available online investigations across a variety of astronomy topics that students from middle school through undergraduate level can complete within two hours. Teachers who are not experts in astronomy can leverage educational materials specifically targeted at K-12 and undergraduate audiences that are made available on the websites of federally funded and private observatories and organizations.

As Rubin Observatory and other new astronomical observing facilities become available on the ground and in space, the scope and complexity of the scientific questions we can address increases and brings greater connections between astronomy and other disciplines. For example, observations with the James Webb Space Telescope, launched in December 2021, are used to study all phases in the history of the Universe. JWST has substantially enhanced capabilities for measuring physical and chemical properties that help us investigate the potential for life in planetary systems. It drives new research in Astrobiology and Astrochemistry. Bringing research opportunities into Astronomy courses facilitates connections to other STEM disciplines. Research has shown that developing research skills and the confidence to do research, predicts student aspirations for research careers (Adedokum et al. 2013). This result highlights one way in which Astronomy can be a gateway to other STEM disciplines.

An example of incorporating astro101 into the wider world and underserved communities is the LCO Global Sky Partners. Through this program, approximately 35 organizations from around the world have access to the LCO telescopes, and develop projects tailored to their local community’s needs.

Many conference participants noted that amateur astronomy organizations like the AAVSO and events such as star parties can also provide a gateway to STEM, especially for students who otherwise do not have school-based opportunities in science. Sometimes this is where students discover that they are interested in STEM in the first place.

The question of whether or not Astronomy is a gateway to other STEM disciplines came up often in our AstroEdUNC discussions. Astronomy often has a reputation as the “less scary” science, and this can help to bring in students. In the future research strand we wondered whether or not astronomy education research has answered this question or what data would be needed to do so. Within the context strand, participants noted that Astronomy certainly can be a gateway, and that more broadly Astronomy can promote and support scientific literacy for all of our students regardless of level or major.

In Astronomy and Astrophysics in the New Millennium published by the National Research Council (2000), chapter 5 on the Role of Astronomy Education is devoted to how “the astronomical community has the potential to add significantly to the continuing effort to strengthen science education and improve public science literacy.” Recent work by Hirst Bernhardt and Bailey (2024) examined this question at the K-12 level and provides support of “astronomy as a gateway” and of astronomy as a transdisciplinary topic. One conference participant noted Astronomy is also a gateway to critical thinking skills and the process of developing mental models about the world around them. This theme is echoed in section 3.2 of the Astro2020 decadal survey report, “Pathways to Discovery in Astronomy and Astrophysics for the 2020s” – “Astronomy research continues to offer significant benefits to the nation beyond astronomical discoveries. These discoveries capture the public’s attention, foster general science literacy and proficiency, promote public perception of the value, legitimacy, and integrity of science, and serve as an inspirational gateway to science, technology, engineering, and mathematics careers.” This was also the focus of a parallel session of the .

\textbf{Conclusion}

We conclude that there are many ways and many benefits of incorporating Astronomy into other STEM and non-STEM disciplines. Astronomy is not the only possible gateway into STEM disciplines, but it often can be based on students’ interests. The physics, mathematics, and observational skills emphasized in Astronomy are great skills for our many students who go on to other STEM disciplines.

\textbf{References}

Benbow, A., \& Hoover, M. (2015). Earth and Space Sciences Education in the US Secondary Schools: Key Indicators and Trends. American Geoscience Institute, Alexandria, Virginia, USA, 10.

California Department of Education (2024, September 1)

Rubin Observatory (2024, August 29). Data management. https://www.lsst.org/about/dm

SKA Observatory (2024, August 29) Handling a deluge of big data.

Adedokun, O.A, Bessenbacher, A.B., Parker, L. C., Kirkham, L.L., and Burgess, W. D., 2013, Research skills and STEM undergraduate research students’ aspirations for research careers: Mediating effects of research self-efficacy, Journal of Research in Science Teaching, 50, 940

National Academies of Sciences, Engineering, and Medicine. 2001. Astronomy and Astrophysics in the New Millennium. Washington, DC: The National Academies Press. .

National Academies of Sciences, Engineering, and Medicine, 2021, [Astro2020] Pathways to Discovery in Astronomy and Astrophysics for the 2020s. Washington, DC.The National Academies Press. .

Hirst Bernhardt, C., \& Bailey, J. M. (2024). Space for all: a multinational study on the status of astronomy education. International Journal of Science Education, 1–24.

Salimpour, S., Bartlett, S., Fitzgerald, M. T., McKinnon, D. H., Cutts, K. R., James, C. R., Miller, S., Danaia, L., Hollow, R. P., Cabezon, S., Faye, M., Tomita, A., Max, C., de Korte, M., Baudouin, C., Birkenbauma, D., Kallery, M., Anjos, S., Wu, Q., … Ortiz-Gil, A. (2020). The gateway science: A review of astronomy in the OECD school curricula, including China and South Africa. Research in Science Education, 51. https://doi.org/10.1007/s11165-020-09922-0

Reference: “The Moon Landings: History, Politics, and Social Responsibility in Science” by Melanie R. Nilsson. Case copyright held by the National Center for Case Study Teaching in Science, University at Buffalo, State University of New York. Originally published April 7, 2020.

\section{Content}

\subsection{What content do you most hope that students will take with them after your course ends? What content do you think students actually do take with them after your course ends?}

\begin{flushright}
  \itshape
    Kelly Kosmo O’Neil \\
    Yashashree Jadhav \\
    Bradley W. Peterson \\
    Adam McKay \\
    
\end{flushright}

\textbf{Introduction}

A common fallacy in teaching is that the more topics we cover, the more students take away. However, more realistically, there is a limited number of things students actually take away from courses in the long run. Therefore, traditional methods of designing courses have focused on identifying what content is most important first, often following the outline of a specific textbook . However, the skills that students take away from a course can be more impactful than the specific content knowledge that they gain. This becomes increasingly true with the ready availability of Wikipedia, generative AI, etc., which can instantaneously provide facts – or misleading information. It therefore becomes increasingly important that students have skills that allow them to evaluate this information and put it into context.

In providing guidance to instructors designing introductory astronomy courses, we suggest that the traditional approach be replaced with one informed by backward design, where we first identify the key learning outcomes that are most important for students to achieve (Slater \& Adams, 2003; Gauthier, 2020). Content is then selected based on how well it supports those outcomes.

The goal of this document is to help astronomy educators learn to curate content to match learning outcomes – namely, defining higher-level, skill-based learning outcomes and determining the key content necessary to achieve these outcomes. An accompanying chapter details the skill-based objectives that are considered most important for introductory astronomy students to gain. Here, we identify four broad categories of skill-based objectives and for each, we provide a non-exhaustive list of examples of content that can be used to achieve those objectives. Those objectives differ for different audiences (major versus non-major). In addition, we also outline a more comprehensive list of topics that can help students pursuing a major/minor in astronomy prepare for forthcoming courses in astronomy.

\textbf{Aligning Content with Learning Outcomes}

Below, we outline four (non-exhaustive) examples of key skills-based learning outcomes astronomy educators may have for their introductory astronomy classes. All of these categories are aimed at increasing scientific literacy, which benefits students regardless of their major. Scientific literacy encompasses the ability to understand scientific concepts and processes (Laugksch, 2000). It is crucial for making informed decisions and for engaging in discussions about scientific issues, and includes both quantitative skills and qualitative skills (PISA, 2003).

Which of these learning outcomes is most important depends on the students being served, for instance whether they are non-STEM non-majors, STEM non-majors, or students intending to pursue a career in astronomy.  This will in turn influence the content that should be included in the course. For each of the skill categories, we provide a non-exhaustive list of suggestions of content that can be used to develop these skills.

\textbf{Engendering Curiosity}\textbf{
}Topics that best engender curiosity may not necessarily be the most fundamentally important, but instead are the ones that will make the students engage and trigger a “wow!” factor, which will stick with them since it resonates with their prior interests.  This is important for all people, but very important for non-majors, who do not necessarily need a set knowledge base, but rather will engage most effectively with the most interesting material.  These are also often the topics that will appear in popular media in the coming years and decades. These can include:

\begin{itemize}
  \item Exoplanets (background content: Kepler’s laws, the nature of light/electromagnetic radiation)
  \item Black holes (background content: stellar evolution, basics of gravity or General Relativity)
  \item Cosmology (background content: light/electromagnetic radiation, Hubble’s Law)
  \item Gravitational waves (background content: black holes, neutron stars, basics of General Relativity)
  \item Solar system exploration of astrobiologically-interesting bodies (e.g.\ Europa and Titan) 
\item (background content: structure of the Solar System, geology)

\end{itemize}

Focusing on these topics can help engender a positive attitude toward astronomy that can last well beyond the end of the course.

\textbf{Develop Quantitative Skills}\textbf{
}Quantitative reasoning involves, in part, the ability to apply mathematical concepts to interpret and analyze data. This skill is essential for solving problems and making data-driven decisions in various scientific contexts.

\begin{itemize}
  \item Read and interpret graphs
  \item Kepler’s 3rd law
      \item (background content: general planetary motion)
  \item HR Diagram
      \item (background content: stellar evolution; classifications of stars by spectral type or temperature and luminosity or magnitude)
  \item Hubble’s Law
      \item (background content: discussion of light and spectra to understand recessional velocity; use of standard candles to measure distances)
  \item Estimation and evaluation of results
  \item Order of magnitude approximations
  \item Size and scale of astronomical objects
  \item Units and scientific notation
  \item Ratios, proportionalities, power laws, and scaling
  \item Inverse square law
  \item Surface gravity/weight
  \item Stefan–Boltzmann law and luminosity of stars
\end{itemize}

\textbf{Develop Critical Thinking}\textbf{
}Critical thinking requires students to understand concepts that may or may not include numerical data. It emphasizes the comprehension of relationships, patterns, and logical arguments (Ennis, 2018; Framework, 2018).

Critical thinking may be broadly defined as the ability to understand an idea or problem and use evidence to take a position or develop a hypothesis (Framework, 2018). Students should then be able to identify the consequences of their position or hypothesis and analyze their assumptions based on those consequences. Critical thinking can help students assess the validity of information sources, learning to interpret scientific news or articles and discern whether they are reputable.

Topics that lend themselves to critical thinking include, but are by no means limited to:

\begin{itemize}
  \item Evidence that the Earth is round
  \item The Galileo Affair (developing a heliocentric view of the Solar System)
  \item The Great Debate (is our Galaxy the whole universe?)
  \item Qualitative explanation for the observational evidence that there is dark matter in the Milky Way based on the Milky Way’s rotation curve  (this also requires background knowledge of Kepler’s laws and gravity)

  \item Accelerating expansion of the universe, dark energy, and Hubble tension
\end{itemize}

\textbf{Building STEM Identity}\textbf{
}This can be accomplished in two ways. The first concerns not necessarily what content is taught but rather how it is taught. Activities that allow students to participate in the scientific process will advance this goal, so any of the topics listed above can contribute, as long as it is presented in the appropriate way. For example, the OPIS! Curriculum (Reichart, 2021) allows students to collect data using robotic telescopes for use in labs. Zooniverse offers numerous citizen science opportunities. There is also high-quality astronomy data available on the web that may be used for introductory astronomy labs (see, e.g., Rebull 2024). The second approach emphasizes the history of astronomy, and highlights the contributions of various people to our knowledge of astronomy, in particular those contributions from people identifying with underrepresented groups, including:

\begin{itemize}
  \item Global Astronomy (contributions from underrepresented groups)
  \item Discovery of Zero: Aryabhatta
  \item Ancient megaliths around the world: Harappan, Indus Valley, Egyptian, Mayan, etc.
  \item Calendars: Lunar, lunar asynchronous, Zodiac vs.\ Nakshatra, solar
  \item Constellations and star names
  \item Astronomical calculators: Jantar Mantar, celestial spheres, astrolabe
  \item Ancient observations of eclipses, supernovae, comets, etc.
  \item Global astronomers
  \item Women in Astronomy
    \begin{itemize}
      \item Harvard computers (Williamina Fleming, Annie Jump Cannon, Henrietta Leavitt, Cecilia Payne Gaposchkin, et al.)
      \item Caroline Herschel
      \item Jocelyn Bell Burnell
      \item Vera Rubin
      \item Mary Somerville
      \item Female astronauts (Sally Ride, Valentina Tereshkova, Kalpana Chawala, Sunita Williams, Mae Jemison, etc.)
    \end{itemize}
\end{itemize}

In addition to the above goals, there is also a critical knowledge base that students majoring or minoring in astronomy/astrophysics should build. Below, we outline a more comprehensive (though still not exhaustive) list of topics that are important for majors or minors. The specific content included in each class will depend on which other courses are offered at a given university. The items in bold are the topics that we consider to be most important in order for students to progress to future courses in astronomy. 

\begin{itemize}
  \item Celestial Sphere and Coordinate systems
  \item Telescopes and instrumentation
    \begin{itemize}
      \item Design
      \item Types
      \item Detectors
    \end{itemize}
  \item Newton’s Laws and Gravity
  \item Kepler’s laws / orbital dynamics
  \item Tidal forces
  \item Radiation and Spectra
    \begin{itemize}
      \item EM Spectra
      \item Atomic structure and transitions
      \item Absorption and Emission
      \item Blackbody radiation
      \item Radiative transfer
    \end{itemize}
  \item Introduction to Planetary Astronomy
    \begin{itemize}
      \item Solar System planets
        \begin{itemize}
          \item Formation and composition
        \end{itemize}
    \end{itemize}
  \item Modern Physics
    \begin{itemize}
      \item Nuclear reactions
      \item Relativity
    \end{itemize}
  \item Evolution of Stars
    \begin{itemize}
      \item HR diagram
      \item Hydrostatic equilibrium
      \item Stellar remnants (compact objects)
      \item Lifecycle
    \end{itemize}
  \item Exoplanets
    \begin{itemize}
      \item Methods of Discoveries
      \item Properties of known exoplanets
    \end{itemize}
  \item Our spot in the galaxy: the Milky Way and its surroundings
    \begin{itemize}
      \item Local Group, Virgo Supercluster, Laniakea
    \end{itemize}
  \item Galaxies
    \begin{itemize}
      \item Classification
      \item Structure and properties
      \item Mergers
    \end{itemize}
  \item Extragalactic and Cosmology
    \begin{itemize}
      \item Black holes (stellar and supermassive)
      \item Evolution of galaxies
      \item Hubble’s law
      \item Large scale structure
    \end{itemize}
  \item Multi-messenger Astronomy
    \begin{itemize}
      \item Gravitational Waves
      \item Computational techniques
    \end{itemize}
\end{itemize}

\textbf{Conclusion}

When considering the content we want introductory astronomy students to take away and the realities of what students retain from our courses, we propose content should be driven by backward design, and  we advocate for flexibility with content within the curriculum based on the student population. The largest distinction here is often majors vs. non-majors, but other distinctions that vary between institutions may be relevant as well (for instance, learning goals at a small liberal arts college may be different than for a large research university, and the course design should reflect that).

\textbf{References}

Ennis, R. H. (2018). Critical Thinking Across the Curriculum: A Vision. Topoi, 37(1), 165–184.

Gauthier, Adrienne J. in Astronomy Education, Volume 1: Evidence-based instruction for introductory courses. Edited by Impey, Chris \& Buxner, Sanlyn. IOP Astronomy (2020).

Laugksch, R. C. (2000). Scientific literacy: A conceptual overview. Science Education, 84(1), 71–94.

The PISA 2003 Assessment Framework: Mathematics, Reading, Science and Problem Solving Knowledge and Skills. (2003). OECD Publications, 2, rue Andre-Pascal, Paris CEDEX 16, France.

Reichart, Daniel E. The Physics Teacher, 59, 728 - 729 (2021).

Rebull, L. M. Physics Today, 44, (2024).

Slater, Timothy F. \& Adams, Jeffrey P. Learner-Centered Astronomy Teaching: Strategies for ASTRO101. Prentice Hall (2001).

A Framework for K–12 Science Education | Next Generation Science Standards. (n.d.). Retrieved December 28, 2018, from

\subsection{An example of designing a course for non-majors specifically for science literacy and curiosity rather than content. }

\begin{flushright}
  \itshape
    Patricia Craig \\
    Ian Hewitt
\end{flushright}

The vast majority of students who take ASTRO-101 in the United States are non-majors and this will likely be the last science course they ever take (Partridge \& Greenstein, 2003). Therefore, we want to introduce them to science using astronomy as a framework to elicit scientific curiosity in general. This is important because scientific inquiry is fundamental to any science course, and life in general (Ke et al, 2021). Without this basic understanding, students are just passively memorizing and regurgitating facts, not thinking critically about the world around them. In today’s information-inflamed world especially, students need to have the ability to make sense of news stories that they hear about astronomy/planetary science (e.g., water has been found on a planet) and critically analyze information to assess its validity. In addition, engendering curiosity can lead to a positive effect on lifetime learning (Loy 2018).

Of the physical sciences, astronomy is the more approachable science because certain aspects can be experienced by anyone who looks up at night. One can make observations on their own with no equipment (Bara, 2013; Vollmer, 2025), with the increasing access to robotic telescopes students can collect their own data, access to archival data is free, and basic astronomy concepts can be comprehended with little to no mathematics. There can be student biases against other science/lab classes because of the impression that class will be stringent, unforgiving, and will require complex equipment/procedures. Moreover, it is relatively easy to make the content of an introductory astronomy course matter to the student by allowing them to take/generate their own data, creating a sense of ownership (“I created this data; it is mine and I did science with it!”). This also allows the student to have a sense of confidence and accomplishment in their work. Additionally, there is an inherent “safety” in doing astronomy in a hands-on way, without the risk of a severe chemical burn (chemistry) or viral infection (biology). This ties back into the idea that astronomy is the most approachable of the physical sciences.

Finally, students, like all people going back to ancient times, have a natural curiosity about the night sky and by their very nature are drawn to the big questions in the Universe. Questions like “What is out there?”, “Where did it all come from?” and “Are we alone?” have been part of the human psyche for generations. This interest provides a hook to begin an introduction to the processes that we use to tackle questions big and small.

For the reasons described above, a non-major ASTRO-101 course focussed on the broader scientific literacy development of the student may consider the following Student Learning Outcomes.

\textbf{Student Learning Outcomes:}

\begin{itemize}
  \item Explain the scientific process/thought
  \item Summarize the processes and practices of science beyond the middle-school science fair project construct
  \item Identify scientific evidence
  \item Explain what is and what is not scientific evidence (e.g., anecdotal evidence)
  \item Demonstrate critical thinking skills
  \item Understand potential pitfalls in scientific reasoning
  \item Identify logical fallacies in arguments
  \item Apply problem-solving skills
  \item Formulate a coherent scientific hypothesis based on empirical evidence
  \item Formulate a coherent argument for testing a scientific hypothesis
  \item Be able to interpret scientific data using methods like graphical representations or statistics
\end{itemize}

The content of the course should be astronomical in nature, obviously, but should directly support the learning outcomes of the course. A comprehensive understanding of astronomy can be less important than a solid foundation of how science works to explain the natural world. There could be minimal focus on memorization of basic facts and figures (e.g., the mass of the Sun or the size of Jupiter) and instead a focus on a general understanding (e.g., that they are larger than Earth and how we know these things). However, some students may be taking the course to learn such basic facts and figures.

Course elements should include experiential components where the students work with actual data, either gathered by a professional or themselves. This has the effect of building skills and further fostering a sense of ownership of the material. This experiential learning should be combined with reflection and assessment components by the students (“Do my results make sense? If not, why not?”). In addition, group work should be used so that students can collaboratively seek solutions (another pillar of modern science).

\textbf{Conclusion}

An astronomy course for non-majors should focus on improving scientific literacy, specifically including an understanding of the nature of the scientific enterprise and analytical thinking, and provide transferable skills. Incorporating the collection of one’s own data through remote telescopes, or using archival data for analysis, can provide students with a sense of ownership leading to enhanced confidence in their abilities.

\textbf{References}

Bará, S. (2013). Naked-eye Astronomy: Optics of the starry night skies. ETOP 2013 Proceedings (2013), Paper EWP24, EWP24.

Ke, L., Sadler, T. D., Zangori, L., \& Friedrichsen, P. J. (2021). Developing and Using Multiple Models to Promote Scientific Literacy in the Context of Socio-Scientific Issues. Science \& Education, 30(3), 589–607.

Loy, D. E. (2018). Cognitive Curiosity and Third Age Lifelong Learning (Doctoral dissertation, Regent University).

Partridge, B., \& Greenstein, G. (2003). Goals for “Astro 101:” Report on Workshops for Department Leaders. Astronomy Education Review, 2(2), 46–89.

Vollmer, M. (2025). Naked eye celestial objects and phenomena: How far can we see at night? European Journal of Physics.

\subsection{What content knowledge do you assume that students will have coming into your course? How important are your prerequisites (for STEM majors, for non-STEM majors)?}

\begin{flushright}
  \itshape
    David Sukow \\
    Raúl Morales-Juberías
\end{flushright}

In this section, we summarize responses regarding instructor assumptions for content knowledge, skills, and prerequisites for students taking astronomy courses for general audiences and STEM majors. ASTRO101 courses generally have few, if any, expectations of content knowledge, but do rely on basic quantitative proficiency, akin to 4th grade mathematics proficiency and some basic elementary school science models.  Deficiencies in mathematics are identified as a problem that is growing worse. Several ideas arise to deal with this, including pre-assessments, just-in-time teaching, and institutional resources for academic and remedial support. Courses aimed at STEM majors are more easily managed through prerequisites, if well thought out, consistently followed, and administratively supported.

\textbf{Typical content and competency assumptions }

For intro courses like ASTRO101, instructors usually assume that the students bring with them some fundamental content knowledge of mathematics (e.g.,4th grade level mathematics), and also some fundamental mental models about the natural world and how it works (like the earth orbiting around the sun and it being round), regardless of whether the students are STEM majors or not. In terms of academic competencies, most instructors would also assume that the students have basic skills like being able to communicate effectively (basic reading and writing). For more advanced courses, in addition to the basic competencies mentioned before, instructors often have assumptions that are usually based on the curriculum plan for the specific major/degree. If a degree plan/curriculum exists and it is well thought out, each course should have prerequisites that have been designed to provide the necessary scaffolding for the students to be successful in the different courses that they take.

\textbf{From expectations to reality}

For introductory astronomy, there are fewer expectations regarding content knowledge. This is important for access, STEM recruitment, and sharing astronomy as widely as possible.  However, success in ASTRO101 often hinges on a pre-existing level of mathematical competency, such as algebra and geometry. Possessing this competency depends on access to a robust curriculum in preparation, as well as the time that has elapsed since these skills were exercised (whether college-bound students who didn’t go beyond high-school sophomore (age 15/16) mathematics courses, or college seniors with majors in which mathematics is not regularly employed).

Anecdotally, this is the area where increasing numbers of students have weaker preparation and lower confidence. Although the most striking deficiencies that were described are not likely to be representative of the student population as a whole, there was a clear sense that this is an area that is becoming increasingly problematic. This effect has been attributed to a variety of factors, most of them systemic and difficult to fix. Policies such as No Child Left Behind, which emphasize reading and mathematics, have come at the expense of time allocated for science instruction in elementary (age 5-10 years) education (Griffith \& Schkarmann, 2008). COVID-related learning losses are propagating through highschool and college populations (Dorn et al, 2021; Schneiders \& Bobek, 2023).  Factors such as low teacher salaries and insufficient access to training have diminished a well-trained K-12 teacher workforce with expertise in math and science (Sutcher et al., 2019).  These are long-term effects that require societal, policy, and structural changes, which happen slowly if at all. In short, educators expect this problem to last, and probably get worse.

For STEM students, departments and institutions have more control over setting appropriate prerequisites. Sometimes, the challenge is to adhere to institutional policies, in the face of requests for exceptions by students, parents, or administrators, who might advocate for a waiver without fully appreciating the level of mastery required. To make the case that waiving prerequisites is not beneficial to students, it is vital that prerequisites be carefully considered in terms of the needs of subsequent courses, and that they be consistently and fairly applied.

Beyond prerequisites and academic competencies, educators report that students are becoming less familiar with the study habits and independence required for college-level work (Gabriel, 2023). It is not unusual for students to think that attending class or watching asynchronous videos is enough; many are not aware they must learn how to learn independently. Instructors and institutions should expect to address these needs proactively.

\textbf{Closing the gap: remedies and resources}

There are different tools and resources available to institutions and instructors to address the discrepancies observed between instructors’ expectations and students' readiness.

Instructors can use, for example, assessment tests at the beginning of a course to test the students' readiness in the expected competencies. These tests should be designed with the goal of actually measuring the gap between the students' knowledge and the instructor’s expectations. Capitalizing on student agency, students can become partners in this process by being responsible to craft specific plans that will help them achieve the necessary proficiencies (Zeiser et al, n.d.). In order to minimize any negative perception that the students may have about being tested at the beginning of a course, instructors should clearly communicate to the students that the purpose of using such a tool is to create a more meaningful learning experience for them. Computer-based, progressive tests, that increase with difficulty, might also help gauge student readiness and empower students by reminding them what they do know. Such a test would not ask dozens of difficult questions beyond each student's grasp; it would stop shortly after reaching a student's limit. Alternatively, if tests are received negatively, the instructor could choose to introduce and review specific skills as they arise naturally in the course, with an invitation to review further outside of class if needed.

Minor differences detected through such assessment tools could be easily addressed by spending some time at the beginning of the course providing some review material to build the missing competencies before moving on to the main course material. If assessment tools like pre-tests indicate major gaps in the students' knowledge that cannot be addressed within the context of the course due to limited time and resources, then this would be indicative of systemic problems that would need to be addressed at a departmental or institutional level. In the case of introductory courses, this could involve curating remedial modular curricula or creating entire courses that then would become prerequisites. In the case of courses for the major, it could involve revisiting the course prerequisites and/or redesigning the course curriculum to make sure that the gaps are being adequately covered in previous courses. However, these approaches could well create financial and staffing challenges for many institutions. Similarly challenging, and important, is a focus on recruiting and training pre-service and in-service K-12 teachers in mathematics and science, and to ensure their efforts and expertise are compensated and rewarded.

There are practical approaches that instructors can also use to level the playing field of a diverse cohort of students, for example, adopting a first principles approach that presents the material to the students in a unified fashion. This also speaks to the importance of standard best practices in teaching, such as clear expectations for the course, learning objectives, and assignments. Communicating these expectations with students can be done effectively through the use of rubrics that the instructor can use to emphasize what is important and that students can use to guide their learning.

Other tools that are usually available to instructors and students are support offices on campus. Many institutions have developed offices, for example, tutoring centers or writing centers that are aimed at addressing students' lack of competencies that may be hindering students' success. Instructors should be encouraged to work with these offices referring students and creating a support network for them. Some institutions even have offices that are dedicated to solving systemic problems that may affect students' success such as economic hardships and food insecurity. Instructors should aim to create an environment where students feel comfortable asking for help in these areas as well.

\textbf{Conclusion}

While introductory astronomy generally does not have prerequisite specific content knowledge, other than mental models of simple systems such as the Sun-Earth-Moon system, students are coming into college less prepared in math and study skills than they used to be. To address these issues, there should be an adherence to prerequisite policies, and these prerequisites should be carefully thought through. Formative assessments could be used to help students and teachers determine where they need to catch up on material and skills. In addition, campus support offices, such as tutoring centers, can be utilized to help students be successful.  Compounding the problems, low teacher salaries and lack of teacher-preparation contribute to a diminishing well trained educator workforce. This requires systemic change to alleviate the problem.

\textbf{References}

Dorn, E., Hancock, B., Sarakatsannis, J., \& Viruleg, E. (2021). COVID-19 and education: The lingering effects of unfinished learning. McKinsey \& Company, 27.

Gabriel, K. F. (2023). Teaching Unprepared Students: Strategies for Promoting Success and Retention in Higher Education. Routledge.

Griffith, G., \& Scharmann, L. (2008). Initial impacts of No Child Left Behind on elementary science education. Journal of elementary science education, 20(3), 35-48.

Schnieders, J. Z., \& Bobek, B. L. (2023). Supports Key to the College Preparation of Students from the Covid Cohort. ACT Research. ACT, Inc.

Sutcher, L., Darling-Hammond, L., \& Carver-Thomas, D. (2019). Understanding teacher shortages: An analysis of teacher supply and demand in the United States. Education policy analysis archives, 27(35).

Zeiser, K., Scholz, C., \& Cirks, V. (n.d.). Maximizing Student Agency: Implementing and Measuring Student-Centered Learning Practices.

\subsection{If we move toward investigation driven approaches, what content are we getting rid of?}

\begin{flushright}
  \itshape
    Donovan Domingue \\
    Don Smith
\end{flushright}

As the community of astronomy educators considers inclusion of active learning, investigation-driven approaches and online data collection for authentic research experiences, many concerns have been expressed about classroom time management. One approach that may be taken is to actually remove some of the content that instructors have become accustomed to inclusion in the syllabus. What topics should we remove? The challenge with answering this question is that there is not a set list of what content should be \emph{included} in an ASTRO101 class, i.e., a curriculum. Therefore, it becomes doubly difficult to make a list of what topics should be \emph{excluded}.

Major topics for an introductory astronomy class that were mentioned in the group discussion ranged from the Big Bang to stars and stellar evolution to naked eye observables (such as seasons, Moon phases, and azimuth of sunset) to planetary exploration/rocketry to aurora. During the discussions, topics that some colleagues considered critical or interesting were considered irrelevant or boring by other colleagues. There was absolutely no consensus on a core set of topics that MUST be included in an ASTRO101 course, although some high school instructors may be constrained by state standards such as the NGSS.

That said, several people expressed an interest in shaping the course content based on student interest: topics that engage student curiosity can be expanded upon and allocated more time, while topics that students find simple or boring can be omitted or allocated less time. This approach requires real-time assessment of the specific students in the specific class, and demands flexibility of the instructor. Large classes or courses that need much advanced planning would find this approach challenging or impractical.  One colleague suggested providing a range of possible topics for exploration, so students could tailor their experiences to their own interests.  Then each student's content for ASTRO101 would be different.  Some suggested moving less important topics to hybrid delivery and letting students explore on their own.  But will they actually bother?  Is it okay if they don't?  Some suggested that a reorganization of topics away from a chronological or topical grouping into a grouping that is motivated by the investigation itself (empirical questions leading to interpretive questions) might allow topics to be "covered" that otherwise might seem necessary to skip.

During the discussion it was said that perhaps a move toward investigation driven approaches does not necessarily mean that there is much loss of total content, It would merely involve a reorganization. It would mean getting rid of a coherent, logical (chronological?) development of what we understand about the Universe. Some important and interesting historical case studies may be neglected. In that case, a set of questions that come up naturally in the course of an investigation would be the replacement. For example, suppose an instructor chooses to do a project on star clusters. One approach would be starting with basic questions (What kinds are there? Where are they in the sky? How do they look?); leading to methodological questions (How do we observe the colors of stars? How do we process the images?); leading to interpretive questions (What do the colors mean? What additional information is encoded in the stellar population? What could be throwing things off?). The investigative/research process is a lot like the JITT (Just-In-Time Teaching) approach in which students learn what they need to learn, when they discover that they need to learn it (Novak, 2011). In this way, with well-crafted investigations instructors would  find themselves retaining a lot of content (and discovering what is not as important in the process).

It seems like "history" is challenging to justify including in inquiry-based/active-learning classes.  Yet there is evidence that for students from underrepresented groups, learning about representatives of their identity groups that have made significant contributions to Astronomy can increase engagement and a sense of belonging (Malcolm-Piqueux, 2024). Perhaps it is important to not jettison history entirely. A widely employed investigative approach that is based on history has been the “Reacting to the Past” model (, Hagood et al, 2018)

\textbf{Conclusion}

To address the challenges posed in this section,  a framework is proposed for the consideration of included and excluded topics. This involves a three step approach.

1) Instructors should identify their goals for the course considering investigative experiences they wish to implement. Goals may include the investigation of topics such as the expansion of the universe, the phases of the moon, or other phenomenon-based inquiry.

2) After a list of goals has been established, instructors can identify the required topics that scaffold the understanding of the goal topics.

3) Instructors should evaluate each topic that has traditionally been included against their list of required scaffolding topics to achieve the course goals, this comparison will provide potential topics to eliminate in a more surgical and strategic manner rather than running out of time to complete important instruction usually occurring near the end of the semester.

\textbf{References}

Jennifer Delgado; Less content is more in introductory astronomy. Phys. Teach. 1 January 2024; 62 (1): 70–71.

Hagood, T. C., Watson, C. E., \& Williams, B. M. (2018). Reacting to the past: An introduction to its scholarly foundation. Playing to learn with reacting to the past: Research on high impact, active learning practices, 1-16.

Hanley, Jennifer, "“Reacting to the Past:” How to Use and Assess Role Playing Games in American History" (2017). Arts and Humanities. 

John L. Safko; Structure for a large enrollment, self‐paced, mastery oriented astronomy course. Am. J. Phys. 1 July 1976; 44 (7): 658–662.

Malcom-Piqueux, L. (2024). Detecting the Signal Amidst the Noise: Ambient Exclusion as a Barrier to Advancing Diversity in Physics and Astronomy. In An Astronomical Inclusion Revolution: Advancing Diversity, Equity, and Inclusion in Professional Astronomy and Astrophysics. IOP Astronomy.

Novak, G. M. (2011). Just-in-time teaching. New Directions for Teaching and Learning, 2011(128), 63–73.

Tim Slater, Jeffrey P. Adams, Gina Brissenden, Doug Duncan; What topics are taught in introductory astronomy courses?. Phys. Teach. 1 January 2001; 39 (1): 52–55.

Philip M. Sadler; Choosing between teaching helioseismology and phases of the moon. Phys. Teach. 1 December 2001; 39 (9): 554–555.

Jay M. Pasachoff; What should students learn?. Phys. Teach. 1 September 2001; 39 (6): 381–382.

Daniel Caton; What should students remember?. Phys. Teach. 1 September 2001; 39 (6): 382–383.

Mark Carnes, Minds on Fire: How Role-Immersion Games Transform College, Harvard University Press, 2018.

Elaine Seymore, ed.  Talking about Leaving Revisited: Persistence, Relocation, and Loss in Undergraduate STEM Education, Springer, 2019.

\subsection{Could we do away with textbooks? Or do we need them, but not as much, just background resources? Mastering Astronomy, gaming sites are they valuable?}

\begin{flushright}
  \itshape
    Imad Pasha \\
    David Yenerall
\end{flushright}

Predominantly, participants felt that textbooks held inherent value, albeit primarily as reference and resource sources as opposed to being highly integrated into the course (i.e., teaching to a textbook; significant required readings). However a survey of over 2000 students across 150 campuses found that ~65\% of students did not purchase the textbook for their courses, primarily due to costs, as the survey also found that students that did \emph{not }purchase the textbook nearly-universally felt concern that not having the book would affect their grade (Senack, 2014) . Conference participants presented anecdotal evidence that “hitting the books” (i.e., studying) is seen by the current generation of students as primarily watching videos or reviewing lecture slides, possibly contributing to an under utilization of textbooks as an appropriate resource. This is borne out in more formal analyses which have found that faculty have difficulty engaging students in reading assignments, and when students \emph{do} read, they do not read critically. The lack of critical reading hampers comprehension, participation in lectures, and ultimately their grade point average (Becker, 2023).

Other anecdotal evidence from participants indicated that students did value having a single authoritative source of information for a course; they reported feeling “adrift” without a text to provide structure and coherence to the subject matter. “A collection of ‘best’ readings on different topics, or multiple classic perspectives on the same topic, only confused them rather than enriching their understanding,” according to one participant. These various sources point to a potential interpretation that questions about the efficacy of textbooks needs to be paired with a discussion of student trends and attitudes towards reading in general. Attempting to reduce the amount of overall text (e.g., with sets of smaller readings) may not immediately fix problems associated with getting students to read textbooks.

Along a separate thread, there seems to be a general sense that the currently available set of textbooks for introductory astronomy are lacking in various regards; primarily in being too broad. Additionally, the associated tools for managing homeworks (e.g., Mastering Astronomy) have prompted frustration for students and instructors alike. One participant mentioned using the popular science book as their “textbook” for a course with success, so it is possible that alternatives to the traditionally published textbooks for the subject could be explored. The alternative textbook avenue also provides opportunities for ensuring that the barrier for economically-disadvantaged or disabled students is minimized; e.g., in the Bad Astronomy case, there were audiobook options and the ability to access the book for free or at low cost.

A major thread of discussion surrounded what textbooks might be replaced with, if at all. Some folks mentioned the Open Education Resources (OER), specifically OpenStax (Fraknoi et al., 2016), but others found that it was often easier to “write your own textbook” (or content sections) as needed (or highly adapt OER), which is more customizable but also more work.

Two techniques which can be used in tandem which were discussed in the context of student-content interaction were “just in time teaching” (JITT, Novak, G et al., 1999) and investigation driven classrooms (Duncan \& Chin, 2021). JITT focuses on ensuring that instructors have up-to-date information about student learning by having students complete pre-class assessments. Instructors can then use those assessments to make modifications to their classroom time to better address noted misconceptions or missing knowledge. Investigation driven classrooms focus on utilizing classroom time for active learning exercises and activities, which are known to help with retention and self efficacy (National Research Council, 2000; Hake, 1998; Paulson, 1999;Udovic et al., 2002). One consideration when utilizing class time in this way is that time for processing and gathering prerequisite information generally moves to outside the classroom, which can sometimes make it difficult to know whether information is being parsed. JITT assessments can thus pair well with this mode of teaching, allowing time to be spent on investigations while still ensuring needed preliminary information is being learned. Lanos et al. (2021) found that JITT paired with the use of gamification platforms is an effective way to increase student comprehension. JITT and gamification are also effective methods of formative assessment that are useful to identify and address incomplete or flawed foundational content knowledge. The major challenge with this approach is that changing investigations or lecture material at the last minute can be a considerable effort.

One participant highlighted that the just-in-time content delivery that many people mentioned resonated with them, especially for some kinds of content. For example, in Astrobiology, they have a unit on the habitability of Jupiter's moons, both for heretofore undiscovered forms of extraterrestrial life and for potential human colonization. For that unit and during that class discussion, it is important for students to know the names of Jupiter's moons (so that they don't mix them up with moons of other planets) and various other facts about magnetic field strengths and various other physical phenomena. However, the instructor does not need or expect them to remember those sorts of things outside of that unit or outside of that discussion. On the other hand, in the astrophysics class, there are some kinds of content that are inherent and important in so many parts of the curriculum, that the instructor assumes those ideas are always at the students' fingertips. For example, the treatment of cosmology would not make sense to a student who did not have a whole bunch of pre-existing physics and mathematics knowledge. As these examples illustrate, the use and relevance of content depends on context, and audience, and a whole bunch of other factors.

In this form of classroom organization, there is still space for textbooks, as a source of information that students are instructed to engage with during the pre-class assignments. In recent times, it has been found (anecdotally) that uptake is higher when the information is made available as, e.g., YouTube videos or MP3 files/podcasts (Just-in-time teaching). It is possible that new/additional textbook formats, in which a “text” on a subject is presented in a podcast or video format, would help bridge these requirements, regardless of the teaching techniques being used in the classroom. That said, whether the underlying trend reflects a general regression in reading acumen among students is a concern we may not wish to propagate by removing reading from our courses. Nationally, educators have reported that students are averse to working outside of the classroom and have poor reading and comprehension abilities (Haysom, 2024). David Brown of Wake Forest University attempts to mitigate this with a pairing of pre-lecture readings with a “” assignment in which students send the most confusing element of what they read that week; these are compiled and contribute to the material covered in the class. That type of framework may encourage students to read a textbook as they have to engage with and respond to it critically as part of a graded assignment.

\textbf{Conclusion}

Ultimately, we find that textbooks as student resources are not the primary factor governing the success or failure of the course to achieve its goals for students, and their presence will benefit at least some students, as long as appropriate steps can be made to ensure accessibility (both in cost and in terms of disability access).  Core interventions surrounding structure (e.g., active learning, investigations), support structures, and…..., appear to have a larger impact than the use (or non-use) of a textbook. But a clear throughline of the discussions presented is that if you do elect to require a certain amount of critical reading from any source outside of the classroom, then separate interventions may be needed to address the aforementioned struggles with actually engaging students in reading.

\textbf{References}

Becker, K. L., Gilbert, D., \& Bezerra, P. (2023). Promoting College Reading Completion and Comprehension with Reading Guides: Lessons Learned Regarding the Role of Form, Function, and Frequency. Journal of Political Science Education, 20(1), 14–30.

Duncan, R. G., \& Chinn, C. (Eds.). (2021). International Handbook of Inquiry and Learning. Routledge.

Fraknoi, A., Morrison, D., \& Wolff, S. C. (2016). Astronomy (OpenStax). OpenStax.

Hake, R.R. (1998). Interactive-engagement versus traditional methods: A six-thousand-student survey of mechanics test data for introductory physics courses. Am. J. Physics 66, 64–74.

Haysom, H. (2024). Is this the end of reading. The Chronical of Higher Education, volume 70 (19),

J. Llanos, C. M. Fernández-Marchante, J. M. García-Vargas, E. Lacasa, A. R. de la Osa, M. L. Sanchez-Silva, A. De Lucas-Consuegra, M. T. Garcia, and A. M. Borreguero (2021). Game-Based Learning and Just-in-Time Teaching to Address Misconceptions and Improve Safety and Learning in Laboratory Activities. Journal of Chemical Education  98 (10), 3118-3130

DOI: 10.1021/acs.jchemed.0c00878

Just-in-Time Teaching. (n.d.). Retrieved April 14, 2025, from

National Research Council (2000). How People Learn: Brain, Mind, Experience and School, Bransford, J.D., Brown, A.L., and Cocking, R.R., eds. Washington, D.C.: National Academy Press.

Novak, G et al., (1999). Just-In-Time Teaching: Blending Active Learning with Web Technology, Addison-Wesley Professional,

Paulson, D.R. (1999). Active learning and cooperative learning in the organic chemistry lecture class. J. Chem. Educ. 76, 1136–1140.

Senack, E. (2014). Fixing the broken textbook market: How students respond to high textbook costs and demand alternatives. U.S. PIRG Education Fund \& the Student PIRGs. Retrieved from

Udovic, D., Morris, D., Dickman, A., Postlethwait, J., and Wetherwax, P. (2002). Workshop Biology: demonstrating the effectiveness of active learning in an introductory biology course. BioScience 52, 272–281.

\subsection{How do we respond to ChatGPT?}

\begin{flushright}
  \itshape
    Kalée Tock \\
    Deanna Shields \\ 
    Anthony Crider \\
    Qurat-ul-Ann Mirza
\end{flushright}

\textbf{Introduction}

The rise of ChatGPT and other large language models (LLMs) presents a significant challenge and opportunity for educators (Adeshola \& Adepoju, 2023; Lo, 2023). Its capabilities are expanding rapidly, making any statement about its current limitations temporary. Therefore, it is misguided for instructors to focus on what AI cannot do. Instead, educators should teach students to use it appropriately and effectively (Adeshola \& Adepoju, 2023; Lo, 2023). This perspective shifts the focus from basic skills to more advanced competencies, such as techniques for query, follow up, and cross-checking (Tenório \& Romeike, 2023).

Flexibility in teaching is crucial as AI continues to evolve. Educators must be prepared to adapt their policies and assignments in real-time, clearly stating the AI policy for each task.  Students must also be prepared for the AI policy to change over the course of a semester with inevitable changes to the software (see the Sample Syllabus policy in Appendix I).  Above all, it is important for instructors to be transparent about their own use of AI to model the transparency expected from students. 

\textbf{Rethinking the Honor Code}

The sample assignment cover sheet below can be used with any assignment.  Students are prompted to document their interactions with AI as they are working on the assignment, and to “confirm or problematize” at least one such interaction via external research.  The exercise gives students practice asking questions and cross-checking information, which are metacognitive skills in addition to AI-specific skills.  In addition, use of the cover sheet stimulates reflection on the ever-shifting best practices for interacting with AI and promotes open communication between students and instructors on this topic.  However, it requires non-negligible extra time on the part of students to fill out thoughtfully, and significant extra time on the part of instructors to read and respond to.  Therefore, including the cover sheet may necessitate shortening the assignment(s) with which it is used.

\begin{figure}
    \centering
    \includegraphics[width=1\linewidth]{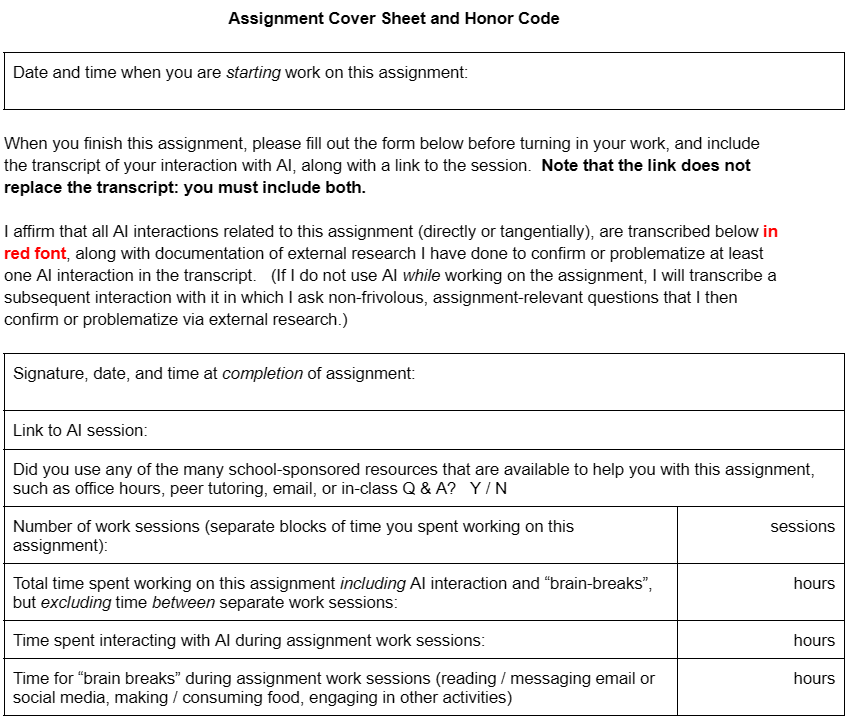}
    \caption{Assignment Cover Sheet}
    \label{fig:enter-label}
\end{figure}

\textbf{Rethinking the Syllabus Quiz}

Many instructors start the semester with a Syllabus Quiz to help students become aware of course policies and structure.  A common source of instructor frustration is misguided expectations that students often exhibit despite having taken a Syllabus Quiz and having agreed to the course policies.  However, students often take several different courses per semester, each of which is structured differently.  Differences between the corresponding syllabi dilute the efficacy of any one Syllabus Quiz.  For example, it is unrealistic to expect students to internalize six different policies for late work, which might differ not only between courses, but also between types of assessments \emph{within} a single course.

If students are prompted to upload syllabus documents for \emph{all} their classes into ChatGPT to pre-train a custom model, then more effective strategies become possible.  First, students can have the AI summarize the similarities and differences between various policies (late work, missing work, class participation, assignment weights, etc.) for the classes they are currently taking.  This might help them to construct a more useful and targeted reference than several lengthy documents that are each organized differently.   In the same vein, ChatGPT can generate a custom Syllabus Quiz that draws upon these similarities and differences, highlighting policies that are unique to the instructor’s own course.

An essential component of any ChatGPT-augmented Syllabus Quiz assignment is to prompt students to check the AI’s summaries and / or questions against the original Syllabus document(s) for their course(s).  Students are intrinsically motivated to do so, since accurate knowledge of course policies factors into their academic success.  The exercise of ground truthing ChatGPT’s content will expose the AI’s limitations (if any), while simultaneously deepening students’ awareness of the policies with which Syllabus Quizzes are designed to acquaint them.

Another advantage of a redesigned, AI-augmented Syllabus Quiz assignment is that in reading students’ responses, \emph{instructors} will become better acquainted with their colleagues’ different approaches.  In addition to the beneficial cross-pollination that might arise from this, instructors might become able to be more genuinely empathetic when their own policies are misinterpreted or mistaken.

\textbf{Rethinking Writing Assignments and Long-Answer Questions}

Taking notes, writing summaries, and drawing connections between content areas is known to be an effective way for students to internalize concepts and build deep understanding.  As is explained in John Bean’s “Engaging Ideas”, writing is an important way of thinking (Bean, 2001).

For this reason, instructors often ask students to write definitions, descriptions, and summaries “in their own words”.  Since it is the \emph{act }of summarizing rather than the summary itself that builds students’ understanding, these sorts of assignments become ineffective when assisted by AI.  Given increasingly widespread student use of AI-assisted writing, alternatives are necessary both for building students’ understanding and for assessing its extent.

\textbf{A. Presentations}

Presenting a topic to a group of peers incentivizes students to develop the sort of understanding that writing an “in your own words” summary would help them to construct, especially when these involve live Question and Answer (Q\&A) sessions.  Using AI to help prepare for a presentation is not counterproductive to the student’s understanding as using it to generate a written summary would be.  In large classes, it can be helpful to split students into small groups for presentations, and / or have them “present” by exchanging pre-recorded videos while reserving the live element for Q\&A.

For live presentations, grading students on the questions they ask of each other in addition to the presentations and Q\&A responses they give provides an opportunity for students to practice asking useful questions while simultaneously ensuring that the Q\&A session includes questions for every presentation.  An instructor could solicit questions live via a Google form and pick one or two to ask the presenter.  The immediacy of an oral presentation can give instructors an important window into the actual extent of a student’s grasp of the material, while simultaneously giving the student practice with scientific communication and presentation skills.

\textbf{B. Model Building }

Constructing and interrogating models is another way to deepen students’ understanding.  For example, a common introductory astronomy assignment involves having students build scale models of the solar system.  This helps students come to terms with the vastness of space, the large differences between the sizes and distances of the planets, and the limitations of common solar system depictions.  Another advantage of such an assignment is that repurposing materials in their environments for the model helps students realize that science is all around them, not confined to a lab or a textbook.  Additionally, the hands-on aspect helps build a bridge between concrete and abstract representations of the same idea.

Beyond the solar system, other student-constructed models can help build conceptual understanding of common astronomy topics. Most of the deliverables for these assignments include a short video (starring the student) that demonstrates the corresponding principle in a hands-on way.  It is important for instructors to watch the videos carefully to assess them, because students’ models sometimes betray misconceptions.  For this reason, adequate instructional and grading resources are needed to implement these types of assignments effectively.  (Students often interpret “creative” assignments such as these as “everyone gets an A” assignments; without careful, individual, time-consuming assessment, this is what they become.)

\textbf{Rethinking the Practice Test}

Building fluency in any discipline requires practice.  Students need multiple opportunities to apply content knowledge in a formative environment where it is safe to take intellectual risks and make mistakes.  In quantitative disciplines, this might involve solving problems that draw upon different dimensions of a topic or in which multiple concepts play into the overall solution.

Instructors often generate practice problems or publish exams from previous years to help students build this fluency.  Sometimes the repository they can provide is insufficient, especially for students who need (or think they need) extra practice opportunities to feel comfortable with complex topics.  Existing course materials, including published practice assessments, can be used to train a custom ChatGPT model to generate new questions.  These can be used by instructors to generate new exams and by students to generate practice opportunities without duplicating the same questions for both groups.   Explicitly instructing the AI to design problems based on a constrained set of training materials can help to ensure that the questions are relevant to the learning goals of the course and do not extend beyond the scope of the intended curriculum, as sometimes happens when students seek out external study resources.

\textbf{Rethinking Exams}

Just as ChatGPT can generate new questions based on course content, it can also answer many of them. Students operate in an environment where time is tight and pressure is high, such that the distinction between “getting a little help” and having ChatGPT do their work becomes blurry.  To clarify the distinction and remove the incentive, assessments such as exams that are intended to be completed \emph{without }consulting AI should enforce, to the extent possible, the removal of students’ access to online help.

However, although exams are useful for assessing a student’s knowledge base and problem-solving capabilities, they are imperfect assessments of skills like scientific communication, data visualization and coding, model building, spreadsheet manipulation, and engagement with authentically messy and ambiguous data. Because of this, and because the artificial environment of high stakes, time-limited exams can be distractingly stressful for students, exam performance sometimes misrepresents what an instructor would consider to be a student’s “true” scientific ability.

To address this, instructors routinely incorporate other types of assessments that explicitly target “20th Century skills”. Often, these assessments resemble work done by professionals in the field, which is intrinsically motivating to students. Especially in the era of ChatGPT, it is critical to leverage students’ intrinsic motivation by foregrounding, to the extent possible, assessments that draw upon it. These sorts of assignments are especially effective when they can be designed to foster and scaffold deep comprehension of scientific principles that are intrinsic to the curriculum.

Examples include student participation in Citizen Science efforts such as Zooniverse, where the course deliverable might involve submitting a screenshot of their work to confirm their active participation, along with a presentation to their peers describing their experience and spreading the word about the project in which they participated. Another Astronomy Citizen Science project in which students can participate is Exoplanet Watch: in this case, student participation can be confirmed by scraping students’ American Association of Variable Star Observers (AAVSO) observer codes from the NASA project website.

\textbf{New AI-specific Skills Assessments}

Additionally, some emerging AI-specific skills are important for students, as future practitioners, to experience and explore.  For example, techniques for obtaining and summarizing background information to write a more helpful query, embedding one or multiple examples within a query to obtain a more targeted response, and pre-training AI models on background information to which a query pertains are increasingly important skills for students to master.  Practice with this can not only improve the student’s fluency in AI query techniques, but also their ability to ask a clear question in general.  As many instructors attest, students’ questions are notoriously incoherent and imprecise, which often makes it challenging to formulate an answer.

As the software continues to evolve, AI-specific skills are a moving target, so that continuing engagement is necessary.  Because of this, instructors may find it helpful to incorporate catch-all AI assessments such as the Assignment Cover Sheet and Honor Code described above into multiple assessment types.

\textbf{Using AI Metacognitively}

One strategy for teaching students to use ChatGPT metacognitively is to ask them to interact with ChatGPT’s answers to their questions: following up on its responses, identifying and correcting errors.  In the following sample assignments, students ask the AI to explain a phenomenon at varying levels of complexity. They then analyze the explanations, identify aspects they do not understand, and seek further clarification.  This is best accomplished using a phenomenon with which the instructor has pre-existing expertise, so that the instructor can identify and address hallucinations or misdirection offered by the AI.

\textbf{Explain it Like I'm Five Assignment}

For this assignment, you'll ask ChatGPT to explain gamma-ray bursts as if you are the following:

\begin{itemize}
  \item About to start elementary school.
  \item About to start middle school.
  \item About to start high school.
  \item About to start a college “ASTRO101” class for non-majors.
  \item About to start graduate school, where you will begin research on gamma-ray bursts.
  \item About to apply for NASA funds to study gamma-ray bursts as a researcher.
\end{itemize}

After you have read those, tell ChatGPT which levels you understood completely and which ones you didn't fully understand, and then ask it to create a custom explanation for you.

Finally, give it the following prompt and answer each question.

"Ask me a series of twenty multiple-choice questions about gamma-ray bursts, one at a time, that incrementally get more difficult from fifth-grade level up to expert to determine how well I understand gamma-ray bursts."

Upload the link to your ChatGPT conversation when you're done.

\textbf{Summarize This Paper Assignment}

For this assignment, you'll ask ChatGPT to summarize a recent research paper about gamma-ray bursts:

\begin{itemize}
  \item It must have been published in Nature, ApJ, or MNRAS in 2023 or 2024.
  \item The final summary should be 150 to 200 words long.
  \item You should have a conversation with ChatGPT and simplify the summary until you understand every word and sentence in the summary.
  \item Ask ChatGPT a series of at least five questions you have about the paper.
\item At the end of your chat, tell ChatGPT what you think is the most important figure in the paper (e.g., Figure 3). It should be the one you'd show on a single slide when discussing the paper.
  \item Upload the link to your ChatGPT conversation when you're done.
  \item Repeat the preceding steps with a custom ChatGPT dedicated to research papers (e.g., Consensus, Scholar GPT, ScholarAI, SciSpace), and compare the results.
\end{itemize}

With the training and cautions instilled by these types of assignments, students (and instructors) can work together to develop effective ways to use AI effectively to streamline a literature search.

\textbf{Using AI for Coding}

Coding is increasingly useful and important in modern astronomy, and many introductory astronomy course instructors struggle with whether and how to incorporate it into their curriculum.  Previous knowledge of coding principles and syntax increases the sophistication of the code that students can use the AI to help write, so that many astronomy courses even have a coding prerequisite.  Other courses strive to give students a taste of coding by incorporating introductory coding tasks into their assessments where feasible.

A few assignments designed to introduce novice students to astronomy coding tasks that they can accomplish with help from AI are linked .  Many of these incorporate Google Colab to facilitate students’ ability to share their code with instructors for troubleshooting help.

\textbf{Appendix: Sample Syllabus ChatGPT Policy}

Between the time I write the syllabus and the time you read it, AI will have evolved. The AI policy, as written, may then be obsolete. This policy is therefore subject to change at any moment.

When you use AI:

\begin{itemize}
  \item You must include a link to your AI log. Test the link to make sure it is publicly available.
  \item You must also include a transcript of your interaction with the AI \textbf{in red font}. This does not replace the link to the log requested above; both must be included.
  \item You must fact-check any AI you use with reputable, independent sources, and cite those sources.
  \item You must significantly revise the AI output in writing your answers to questions. In other words, your work must be written primarily by you.
  \item You must understand the concepts and vocabulary in your writing. The instructor reserves the right to quiz you, orally or in writing, on any of the material that you turn in to confirm your understanding.
  \item You must answer questions using concepts, ideas, and terminology discussed in class rather than those encountered elsewhere, or using new concepts and vocabulary introduced to you by AI.
\end{itemize}

\textbf{Conclusion}

In this quickly evolving world of AI teachers must be flexible and evolve alongside it. We need to rethink our honor codes, syllabi, writing assignments and assessments. Students should learn the value of AI and when it is useful and when it inhibits learning, and AI skills should be integrated into courses.

\textbf{References}

Adeshola, I., \& Adepoju, A. P. (2024). The opportunities and challenges of ChatGPT in education. Interactive Learning Environments, 32(10), 6159–6172.

Bean, 2001.  Bean, John C. Engaging Ideas : the Professor's Guide to Integrating Writing, Critical Thinking, and Active Learning in the Classroom. San Francisco :Jossey-Bass, 2001.

Lo, C. K. (2023). What Is the Impact of ChatGPT on Education? A Rapid Review of the Literature. Education Sciences, 13(4), Article 4.

Tenório, K., \& Romeike, R. (2023). AI Competencies for non-computer science students in undergraduate education: Towards a competency framework. Proceedings of the 23rd Koli Calling International Conference on Computing Education Research, 1–12.

\section{Skills}

\subsection{What matters more in ASTRO 101? Skills or content? Learning by doing, or learning the results of others’ doings? To what degree does this have to be a tradeoff between breadth and depth? Are the gains in skills (and in understanding/appreciating what science actually is) worth the tradeoff?}

\begin{flushright}
  \itshape
    Kevin Healy \\
    Jonathan Keohane \\
    Stephen McNeil
\end{flushright}

Why should a college require their undergraduates to take an introductory laboratory science? This is clearly not because of the importance of learning astronomy, per se, for a student’s future success, or even because it is one of the seven classical liberal arts. Unless we can justify the importance of astronomy for the modern college student, it will be deemed unimportant and thus dropped from the core curriculum to make room for more modern subjects, such as data science.

In practice, however, professional astronomers are data scientists. We analyze images and spectra from either our own smaller telescopes or larger telescopes or we work with large archival data sets. We write computer code to automate data analysis pipelines, and we create tools to show correlations between measurable results. Within the context of a good laboratory astronomy course, our students will be introduced to these critical data analysis skills. However, rather than learning how to use these tools in a simulated business context, our students use them to discover for themselves the answers to some of life’s most interesting questions.

At this meeting, there was a broad consensus that \emph{skills} are more important than \emph{content}.  However, since we did not have a working definition of the term \emph{content}, it was often considered to mean the different topics usually written on a course description. Content is the vehicle through which we teach skills. And we definitely want some basis of content to give context. So maybe "giving students an opportunity to develop skills useful in their future careers competencies is more important than breadth of content"? Of course this gets us back to the breadth vs. depth discussion, which in some ways is the same as skills vs. content.

At another point in this meeting however, there was a  clear pushback against surrendering  \emph{content} for \emph{self-efficacy}. The consensus in that context, among the different discussion groups, was that increases in self-efficacy come about by students mastering skills that are perceived as difficult. We do not care about empty self-efficacy, but, rather, we want students to acquire a realistic measure of their own abilities. Thus, we teach important and difficult skills, and this, in turn, leads to advances in self-efficacy.

It is obviously impossible to cover the whole universe in a semester. Similarly, it is impossible to teach every possible useful skill. Just as with content, we must select the most important skills to teach. A natural solution would be to design the \emph{content} around what observations the students make (or what archival data they access), and the \emph{skills} around what we, as astronomers, value. However, this still makes for a very tall order for a semester-long class with no assumed prior knowledge.

Different students need to learn different things. A student who has a strong mathematical background may need to learn a different set of numeracy skills than someone who has always been weak at math. Similarly, someone who is a strong writer may need something different than someone for whom writing has always been a significant challenge.

Putting these ideas together, we could arrange our introductory college astronomy classes around the two important skills of numeracy and data analysis. To keep expectations manageable for the student, the instructor should pick one primary numeracy skill and one data analysis tool. Then, the instructor should design the course content around students using those two skills.  Ideally, both skills chosen will be ones that few students already have, yet be highly applicable to multiple disciplines.

For example, astronomers solve novel problems by analogy, often using estimates, ratios, proportionalities and scaling with power laws. We study a close-by object in minute detail. Then, we assume that far-away, but otherwise, similar objects are the same.  Eventually, measurements replace our simplistic assumptions, as we learn more and more. This \emph{astronomical scientific method} stresses basic numeracy skills, but that are new to everyone regardless of background.

Similarly, it makes sense to deliberately teach one data analysis tool, such as the spreadsheet program adopted by the College. With more advanced students, a good choice may be a coding environment package, such as Python on a Jupyter notebook server or Google Colab notebook. Whatever tool the instructor picks, it should be relevant to both potential majors and non-majors.

After the instructor has made these defining choices, he or she can select exercises around a theme.The instructor should choose each topic in order to teach the core numeracy and data analysis skills in the context of answering those questions that she or he finds most profound.

\textbf{Conclusion}

The general consensus was that skills should be a main focus, and content can be the vehicle through which we teach the skills. Content can be selected flexibly in order to address the learning needs and interests of the students. Numeracy and data analysis skills are important science literacy and 21st century skills that could anchor an introductory astronomy course.

\subsection{To what degree should we be using real data-based investigations to teach content, vs. more traditional lecture \& lab approaches?}

\begin{flushright}
  \itshape
    Kevin Healy \\
    Jonathan Keohane \\
    Stephen McNeil
\end{flushright}

Traditional astronomy lab activities might use sample data and worksheets or simulations of phenomena. Today, with such ease of access to massive troves of space and ground-based telescope data, along with the ability of students to collect their own data, astronomical investigations can take on a new authenticity.

Traditional lecture and laboratory courses were developed at a time when laboratory exercises required separate physical spaces and were designed as separate courses. Lecture is traditionally a  larger credit course and laboratory a smaller credit  course. Today, technology, such as laptops and portable laboratory equipment,  can be used in a regular classroom space. This change eliminates the physical and pedagogical separation between lecture and laboratory activities and class periods.  With the availability of modern equipment and classroom spaces, those institutions that have a single large lecture course with no laboratory component can incorporate real-data activities into their lecture classes. They can also use remote telescope networks to gather data (Gomez \& Fitzgerald 2017).

Classroom investigations using real data gathering and realistic data analysis provide the best opportunities to give students authentic science experiences (King \& Ritchie, 2012). Real datasets provide students with the starting point for authentic scientific inquiry for any physical or astronomical concept in an introductory course. Students can select their own data from publicly available archives or collect their data using laboratory equipment or from local or remote telescopes. This data-gathering exercise generates a sense of student ownership of the data and of the investigation. With modern data and software tools, students can perform investigations that were difficult or impossible to accomplish by professional astronomers a century ago.

Portable table-top laboratory equipment and physical demonstrations provide students hands-on experience with physics and chemistry concepts such as atomic spectroscopy, the inverse square law, Moon phases, and impact cratering. In addition, access to robotic telescopes by undergraduate students is now much easier and affordable. Small telescopes with CCD and CMOS cameras can capture images with higher resolution and greater sensitivity than was possible with previous generations of larger telescopes using film cameras. The latest generation of small telescopes can be easily used by beginner astronomy students with no formal technical training (Arditti, 2024).

Among the several telescope networks that are free or have relatively low fees are Las Cumbres Observatory (Brown et al. 2013), SkyNet (Reichart et al. 2005), MicroObservatory Telescope Network (Sadler et al. 2001), Falcon Telescope Network (Chun et al., 2018), OurSky (https://oursky.ai/), and The Virtual Telescope Project 2.0 (The Virtual Telescope Project). An assortment of companies sell time on robotic telescopes, such as Telescope Live (Telescope Live) and iTelescope.net (Boley et al., 2019). Many all-sky surveys spanning the electromagnetic spectrum are publicly available. These surveys, and other targeted observations, have been made with large-aperture ground- and space-based telescopes run by national research institutions. In short, the availability of high-quality astronomical data has never been greater.

These data, whether gathered from public archives or collected by astronomy students themselves, can be used as the basis for authentic explorations of a wide variety of astronomical phenomena. Some examples demonstrated at this conference include radio observations of pulsars and neutral hydrogen in nearby galaxies, visible-light observations of the orbital motions of planetary satellites and binary stars, photometry of stars harboring exoplanets, and photometry of stellar populations in open and globular clusters. These kinds of data can also be used by instructors to create figures or graphs for introductory lectures, homework assignments, or exam questions.

Many traditional laboratory exercises give students the impression that science involves following standard procedures to reach a predetermined result. Real scientific research is much closer to hunting in the dark with no prior goal. Often the data is messy and not as straightforward as in a traditionally constructed lab activity. Consequently, scientists and science students have broad choices concerning how they collect data and conduct their research. For students to gain an understanding of real scientific inquiry, they need to keep in mind the big questions at the heart of a research program, develop persistence in the face of problems, master numerical literacy skills to check their results, and effectively communicate results to their peers.

Since its beginnings, astronomy has required mathematics to describe the universe (Williams, 1930). While modern astronomy research involves sophisticated computer software and simulation, students can accomplish much with spreadsheets, graphing tools, and some basic programming. An introductory science course may be the first time a student uses a computer in this way. The skills students develop with these computing tools have broad application, from the workplace to community organizations to home finance. Through exercises involving real data, students can acquire an appreciation for the usefulness of software tools and how computer software plays a major role in modern society.

In real scientific research, promising strategies often lead to dead ends. Novel data analysis techniques may include mathematical or software errors, which need to be identified and corrected. Preliminary results need to be confirmed by the scientific community before a new technique is applied to an entire dataset. Scientists spend a significant amount of time troubleshooting problems. This same skill is valuable for any student, whether they are investigating an astronomical topic in class, or determining the cause of a problem at work, in their community, or in their home. Persistence in the face of problems is a crucial skill that can be developed in an authentic astronomical investigation.

Progress in scientific research is made by placing each discovery in context and confirming that new results are physically plausible and consistent with prior knowledge. Scientists routinely double-check their results along the way with sample calculations. Students who can perform the textbook version of a calculation often do not apply this kind of “sanity check” on their results. Authentic scientific inquiry can train students to think about their findings like a scientist, gaining confidence in their work as they proceed.

Plotting data graphically is another way to put results in context and check for consistency and plausibility. Students in introductory science courses are novices at reading graphical displays of data. Constructing their own graphs and interpreting them can strengthen students' overall comfort with, and their skills at reading, graphical data.

Scientific discoveries are not useful unless they are shared. Scientists are expected to share their results with peers at science meetings and publish research articles in science journals. Some students may participate in these formal activities. All students can build valuable skills in communicating their research methods and results with student peers through writing exercises. As with their numerical and graphical skills, students in introductory science courses are novice writers. The concise form of science communication teaches students to write and speak with specific meaning and convey the important details.

\textbf{Conclusion}

With access to remote global telescope networks, and increasingly easy to use small telescopes, as well as ever larger data sets from ground and space-based telescopes, students are more able now than ever to participate in the science of astronomy. Within the introductory astronomy courses they can carry out authentic activities with real data, whether gathered by themselves or professional astronomers. We argue that participating in this authentic way can improve the learning process, giving them opportunities to think critically and scientifically about what they are learning. These authentic experiences can be extended to communicating the science they are doing, as students prepare graphical representations of their data and share their findings in presentations and written papers.

\textbf{References}

Arditti, D. (2024). Smart telescopes. Journal of the British Astronomical Association, 134(2), 100–101.

Boley, A. C., Bridges, T., Hickson, P., Richer, H., Gladman, B., Heyl, J., ... \& Stairs, I. (2019). Small and Moderate Aperture Telescopes for Research and Education. arXiv preprint arXiv:1910.00227.

Chun, F. K., Tippets, R. D., Strong, D. M., Della-Rose, D. J., Polsgrove, D. E., Gresham, K. C., ... \& Stoll, E. (2018). A new global array of optical telescopes: The falcon telescope network. Publications of the Astronomical Society of the Pacific, 130(991), 095003.

King, D., \& Ritchie, S. M. (2012). Learning science through real-world contexts. Second international handbook of science education, 69-79.

Telescope Live (2024, Sept 1) https://telescope.live/home

The Virtual Telescope Project (2024, Sept 1) https://www.virtualtelescope.eu/

Williams, H. S. (1932). The Great Astronomers. Simon and Schuster. New York.

\subsection{Science is much more than…science. It’s communication.  It’s writing, both formal and informal. To what degree should this play a larger role in how we teach science?}

\begin{flushright}
  \itshape
    Jennifer Lynn Bartlett \\
    Brooke Skelton \\
    Raúl Morales-Juberías
\end{flushright}

\textbf{Introduction}

Astronomy, and science generally, remains unknown unless practitioners communicate their work with other scientists and the broader community (Weingart et al., 2021). Therefore, departments and instructors should integrate the strengthening of their students’ communication skills at all levels. Doing so prepares the students for their future careers regardless of their professed major. The exercise of communicating what they have learned, whether to an instructor or a classmate, solidifies learning and reveals areas of weakness (Bangert-Drowns et al., 2004; Glynn \& Muth, 1994; Rivers \& Straw, 2000). Including “language arts” in a science classroom also shows students the connectedness of knowledge. Working with colleagues in English Departments and writing support programs also builds community and common expectations across an institution. Fortunately, instructors can integrate many reasonable written and oral assignments into their classes without significant revision. For most courses, teaching communication skills simply requires intentionality.

\textbf{Justification}

Writing and communication are a vital career competency for all students, regardless of their future career plans (Shoja et al., 2020). Employers need graduates prepared to communicate clearly in both written and oral settings.  Communicating scientific ideas, whether it be in a persuasive essay, research project, or laboratory report, is excellent preparation for job readiness because it requires precision in both terminology and explanation. It also requires students to digest, internalize, and reflect on the concepts they are learning. In addition, writing assignments can be an effective avenue for novice students to explore the interdisciplinarity (Kegel, 1987) of the concepts covered in astronomy courses.

At the introductory, non-major level, students practice using language as a tool to communicate data-driven arguments.  Professionals in many fields need to convince others—be they colleagues, supervisors, or clients—that they understand the tools they are using, developing, or selling. In addition, individuals in many jobs need to document their work effectively and clearly convey questions and explanations in emails.

For undergraduate astronomy and astrophysics majors, written communication is vital for job readiness as well, both in graduate school and beyond. Writing proposals, constructing scientific papers, and communicating ideas with colleagues are skills that they will use daily. Scientists interact with the scholarly literature regularly and must be able to read and reflect on the work of their colleagues. To be successful, astronomers must also engage in clear and effective oral communications with supervisors, in group meetings, and during conferences.

Even with the increasing frequency of Artificial Intelligence (AI)-generated text, crafting an authentic voice will remain important, especially for crucial communications. Professionals must also brief one another and exchange ideas in situations where they do not have access to such tools.

Science communications is a growing field of its own. It may be the right choice for students who enjoy science but do not want to pursue a research or academic career. Students interested in these opportunities benefit from multiple communication opportunities at every stage of their education. They might also need a more structured technical writing or science communication course, which could be offered in conjunction with English and other science departments. Astronomy majors may also welcome the opportunity to improve their communication skills through a dedicated class, besides the opportunities woven into their standard curriculum.

We believe that scientific writing at all levels is a skill to be developed and honed throughout a student’s career. While details of professional scientific writing might be taught in a specific course, students need to demonstrate their mastery of multiple written formats, from informal emails to outreach blog posts to detailed proposals and to publishable research reports. Even with AI-generated text, students must also be able to express themselves orally, including briefing colleagues, demonstrating competency and confidence in an interview, and selling ideas or products.

\textbf{Implementation Strategies}

Asking an instructor to add communication skills to their already full curriculum may initially seem onerous. However, students are already confronting multiple forms of communication they must digest and must respond to their instructors in different venues. Many programs already have a communication component. However, we argue that intentionally including writing and oral skills, even at the earliest levels, is beneficial to students. Our suggestions apply to courses for both majors and non-majors generally. Someone teaching a majors course should naturally tailor their assignments, formats, and expectations to the common practice of their discipline.

Instructors should present examples of science writing in their courses and hold students accountable for reading and interpreting written work. Brief discussions of what makes passages of textual information more readable and understandable help students hone their own skills. In addition, seeing written communication intended for a variety of different situations assists students in tailoring their own writing. Textbooks, blog posts, press releases, magazine articles for laypersons, and journal articles for professionals can be aspirational examples.

Instructors who lack confidence in their ability to build their students’ communication skills can partner with many campus resources, like writing centers or tutoring sessions. Constructing detailed rubrics empowers both instructors and students to conform to common expectations tied to specific communication styles. It also simplifies assessment for instructors.

\textbf{Problem Set.} If students solve problems or answer questions, instructors should evaluate how they present their work, not just the “right answer.” For calculations, do students restate the problem, draw a picture, define their variables, identify their equations, and proceed methodically? For textual questions, do students explain their response in an organized manner, with correct grammar, and in a suitable tone? If instructors use computer-grading, perhaps a select few problems could be hand-graded and evaluated for their presentation of ideas or presented in class on a whiteboard. By asking our students to show their work, we are ensuring they understand the individual steps. When a student makes mistakes, the instructor can more easily pinpoint the student’s individual area of weakness and help them improve. Whether professionals are explaining a new work procedure or a novel scientific conclusion, their colleagues will have greater confidence in the results if they understand the process.

\textbf{Code documentation.} If students write code for an assignment or laboratory exercise, the instructor should hold them accountable for their documentation, whether as an accompanying README file, a file header, comments, or the cells of a computational notebook. Documenting their code well encourages students to think through the choices they are making and reduces the time they spend debugging their code. When reviewing the documentation, the instructor can better understand the student’s intent and help them develop more efficient and versatile solutions. Students who pursue careers in software engineering, data science, and research will have to adhere to workplace code and documentation standards.

\textbf{Laboratory Report.} If students complete laboratory exercises, they should describe their experiences in a formal report. The instructor could require an abstract, a blog post, a poster, an oral briefing, or a journal article. Their report should convey what they did, what results they got, and how they interpreted their experiments using words, tables, graphs, and illustrations. The instructor can set the standards for the report based on the course level, with more advanced students being held to higher standards and majors students to more professional formats. Asking students to reflect and draw conclusions on their laboratory experience allows them to connect these exercises with lecture concepts. Articulating their results helps students deepen their understanding and identify concepts that are still unclear to them. Students also develop critical reasoning skills through evaluating whether their results are reasonable and assessing sources of error. They can apply these reality checks professionally when reviewing the work or claims of others. Regardless of their professional choices, students will be better prepared to explain themselves and their work in the preferred format of their field.

\textbf{Special Topic Assignment.} If students investigate a topic or news report, they should explain it in a fitting tone and reflect on what they learned. Furthermore, they should consult multiple trustworthy resources and credit their sources. Compiling writings in class journals would allow students to take on the additional roles of editor and reviewer. Depending on the course, such an assignment could range from reading and responding to news articles or press releases through reading and evaluating journal articles. Many formats allow students to showcase their understanding of their topics and their abilities to approach their sources critically. Students could prepare oral or video or poster presentations, podcasts, blog posts, discussion forum contributions, term papers, or portfolios. Instructors can scale these assignments appropriately for their classes and choose formats applicable to either non-majors or majors. Explaining a topic to their peers or their instructor provides students opportunities to engage with an area of interest more deeply, to apply concepts and skills learned in class, and to develop their confidence as scientists or critical consumers of scientific information. An extension of this kind of exercise could be to ask the students to have ChatGPT summarize the topic, and then having the students grade the ChatGPT output. This would be a quick and practical way to make the students critical of AI output. In assisting a student in selecting and developing their topic, the instructor helps the student explore their own relationship to science and identify ways in which science is relevant to their personal experience. Students who do not enter a research-oriented field may still have to do background research, prepare digests, and present reports.

\textbf{Informal Science Communication Assignment.} Because social media plays a strong role in students’ own communications and identities, instructors can tailor a special topics assignment to reflect this reality. Students could identify, critique, and respond to social media posts on scientific topics. TikTok and other forms of social media are ripe for short explanations of fascinating and cutting-edge astronomy that catch the attention of novices of all ages. Providing students with structured opportunities to convey scientific information in more informal settings allows them to explore the strengths and weaknesses of these platforms for conveying complex ideas. Regardless of career choice, students will continue to consume and contribute to evolving social media platforms in both personal and professional capacities. Such assignments remind students they can contribute to the scientific community in ways they may not think of as scientific.

For majors, the effectiveness of all these strategies can be boosted if they are adopted at a departmental level. Doing so ensures that a consistent set of communication expectations persists as students progress through the curriculum. Departmental guidelines also prevent situations where individual instructors may have different expectations. For non-majors, students will be more successful if the format and standards closely align with those required in core courses.

\textbf{Assessment Strategies}

Student efforts to communicate astronomy concepts provide a way of assessing how well they comprehend the course material. They could demonstrate their mastery through a myriad of forms, but such assignments challenge them at every educational stage. However, reviewing student projects, providing meaningful feedback so they improve, and ensuring consistent and objective grading is time-consuming for instructors. Ideally, students should revise and resubmit less successful assignments, which requires more time from both instructors and students.

Reviewing student writing and oral assignments is not a problem unique to the individual instructor, the course, or the department. Institutional resources that support student learning of communication strategies can also assist instructors. By partnering with such organizations, instructors should develop techniques for offering meaningful feedback to students. Providing clear instructions, a detailed rubric, and imitable examples of acceptable work helps students complete assignments successfully.

Limiting revision of assignments for an improved grade to singular, major projects reduces the time commitment for instructors and students. However, the timeline for such assignments should include time for revision before the end of the grading period.

For repetitive assignments, such as problem sets, coding assignments, or weekly reflections, the instructor should encourage students to improve their next assignment better rather than redo an old one. In these situations, the instructor could systematically change the rubric so that he or she expects more of students after completing several assignments than at the beginning of the semester. For instance, if the instructor asks students to use a formal tone and correct grammar when writing a laboratory report, she or he could grade the first few assignments solely on completing the steps in the manual, along with constructive feedback regarding the presentation of this information. By the end of the semester, the grade on a laboratory report should reflect both the scientific concepts expressed and the manner in which they are conveyed.

\textbf{Conclusion}

In our modern interconnected world, clear communication, both formal and informal, is an extremely important tool. To inform colleagues and the public about new advances in science and to fight misinformation, scientists must communicate well. Therefore, exposing students to different methods of science communication is important. Similarly, they need structured opportunities to practice this overlooked aspect of the scientific endeavor. Regardless of a student’s career trajectory, using language effectively is a vital career readiness skill. Consequently, we propose all assignments should include components that involve effective communication of ideas.

A variety of assignments that faculty could implement to help students strengthen their use of language are presented above. Instructors should adjust course expectations to challenge all students to improve their communication skills. Novice science learners to the advanced undergraduate, all benefit from such requirements. Most importantly, students should learn that communication is important in all areas, not just specific assignments or writing courses.

\textbf{References}

Bangert-Drowns, R. L., Hurley, M. M., \& Wilkinson, B. (2004). The effects of school-based writing-to-learn interventions on academic achievement: A meta-analysis. Review of educational research, 74(1), 29-58.

Glynn, S. M., \& Muth, K. D. (1994). Reading and writing to learn science: Achieving scientific literacy. Journal of research in science teaching, 31(9), 1057-1073.

Rivard, L. P., \& Straw, S. B. (2000). The effect of talk and writing on learning science: An exploratory study. Science education, 84(5), 566-593.

Shoja, M. M., Arynchyna, A., Loukas, M., D’Antoni, A. V., Buerger, S. M., Karl, M., \& Tubbs, R. S. (2020). A Guide to the Scientific Career: Virtues, Communication, Research, and Academic Writing. John Wiley \& Sons, Incorporated.

Weingart, P., Joubert, M., \& Connoway, K. (2021). Public engagement with science—Origins, motives and impact in academic literature and science policy. PloS one, 16(7), e0254201.

What is Career Readiness? (n.d.). Default. Retrieved May 8, 2025, from

Kegel, W. H. (1987). Astronomy, an Interdisciplinary Science. Universitas, 29(3), 157.

\subsection{All of these skills matter for the 0.1\% of students becoming astronomers. To what degree do they matter to the larger number of other-STEM students? To what degree do they matter to the even larger number of non-STEM students?}

\begin{flushright}
  \itshape
    Matthew Beaky \\
    Michael Rutkowski \\
    Erika Grundstrom
\end{flushright}

Many students engage in astronomy and astrophysics in higher education, and few will ever join the workforce as “astronomers” or “astrophysicists” (AIP, 2019).   We address how the skills and training provided in lower division (“ASTRO101”) and upper division coursework (traditionally, lecture-based courses with more advanced student participation and learning through exercises in diverse formats) are applicable beyond the “astronomer” occupation.

In a discussion of skills related to what Astronomy students practice, we posit that many of the skills used in astronomy are:

\begin{itemize}
  \item \emph{content agnostic} in that they can be practiced in many different fields
  \item \emph{demonstrable} to future employers (and instructors). For example:
    \begin{itemize}
      \item “I carried out this project\ldots”
      \item “I analyzed these data using these tools\ldots”
      \item “I presented a \{oral or written\} report to \{audience\}\ldots”
    \end{itemize}
  \item \emph{measurable} (noting that the ease of measuring is variable)
  \item \emph{valuable} for creating a better citizen and a better human being
\end{itemize}

For context, we define the workforce into which the students enter.  In the US, there are 165M workers, of which ~10M are employed within the STEM fields, (https://www.bls.gov/emp/tables/stem-employment.htm), as classified by the Bureau of Labor Statistics.  STEM field employment is expected to grow, and faster than non-STEM fields, by a factor ~2x over the next decade.   Within the STEM fields, the divisions in employment across the diversity of STEM, the total employment within the fields, and qualifications necessary to achieve employment are summarized in the interactive “Periodic Table of STEM Occupations” by the BLS (K-12 Student Resources).

In this discussion, we divide the discussion of skills and training that astronomy/physics supports for future employment amongst the lower and upper divisions for simplicity, recognizing that enrollment in ASTRO101 is composed of students who may enter either STEM or non-STEM fields.  It is assumed for brevity that the primary career trajectory for students in upper division courses will end in a career in the STEM fields. This excludes a substantial, growing population of ~20\% astronomy graduates that do not engage in STEM as a career (AIP Statistics, 2024).  The AAS presents a more fine-grained impression of career paths (American Astronomical Society) and is developing and implementing efforts to support non-academic career transitions.

Most students who enroll in ASTRO101 intend to pursue careers outside of STEM. For many, this course represents their only or terminal science class. Through ASTRO101, these students can gain a variety of valuable skills that are applicable beyond STEM careers.

\emph{Analytical Skills}: Students learn to analyze data, identify patterns, and draw conclusions. These skills are applicable in many fields, from business to social sciences.

\emph{Problem-Solving}: ASTRO101 courses often involve solving complex, unfamiliar, or non-intuitive problems, which helps students develop their creativity and the ability to approach and solve unfamiliar problems, a skill useful in any profession.

\emph{Understanding Data}: Students gain experience in interpreting and presenting data presented in various formats, including graphs, tables, and charts. This skill is essential in a world increasingly driven by data.

\emph{Probability and Error Analysis}: Learning to understand and apply concepts of probability and error analysis helps students make informed decisions based on data.

\emph{Written and Oral Communication}: Writing reports and essays helps students improve their ability to communicate complex ideas clearly, effectively, and concisely, while presentations and group discussions help students develop their public speaking and interpersonal communication skills.

\emph{Use of Software}: Students become familiar with software tools for data analysis and presentation, which have value in many non-STEM careers, such as marketing, finance, and management.

\emph{Teamwork}: Group projects and collaborations enhance students' ability to work effectively in teams, a crucial skill in almost any workplace.

\emph{Resilience and Intellectual Independence}: Working through challenging problems helps students develop the ability to persevere in the face of difficulties. They come to understand the limits of their knowledge and learn when and how to ask for help.

\emph{Interdisciplinary Awareness}: Astronomy often intersects with other fields such as physics, chemistry, and environmental science, giving students a more holistic understanding of how different disciplines connect.

By acquiring these skills, students who take introductory astronomy courses are well-prepared to succeed in a wide range of non-STEM careers, as these competencies are highly valued across many professions (Vista, 2020; Whorton et al., 2017). Future STEM professionals also benefit from development of essential skills (communication, teamwork) that may be further developed in advanced astronomy courses.

The skills and training obtained in lower division courses (ASTRO101 and associated laboratory exercises) are reinforced and expanded upon within the upper division, with direct benefit and parallels within the future of employment of the STEM majors.  Specific instances in which this training is applicable within the workforce are highlighted here:

\emph{Collaborative Effort} – Astronomy is increasingly collaborative, in research and in teaching. Employers seek employees with strong oral and written communication skills (Asefer \& Abidin, 2021) and with the ability to work on collaborative teams.

\emph{Professionalism/Work Ethics} – Astronomy is “self-policed” – we do not have “Boards of Ethics”. Upper division courses in research methods can define strong codes of ethics regarding academic honesty. This is highly sought after (Finely, 2023; Hirudayaraj et al., 2021) by employers.

\emph{Computers as a tool} — Ubiquitous use of computers as tools for manipulating, analyzing, and interrogating large datasets requires familiarity with the use of software and programming language(s). Industry uses JavaScript, HTML/CSS, and Python , which are well-supported across astrophysics.

\textbf{Conclusion}

The question of whether the skills learned in an introductory astronomy course are applicable to non-STEM majors and those who will work outside of STEM fields really gets at much of the focus of the discussions during the three days of meetings. Preparing a science-literate citizenry is crucial to addressing global issues such as pandemics and climate change. We need to be preparing students who will become the lawyers, politicians, and poets of tomorrow to participate meaningfully in a democracy, to make informed and scientifically sound decisions throughout their lives. These students will be working in an ever more technical and globally connected world, and the 21st century soft skills that can be taught within an introductory astronomy course will continue to be valuable for all.

\textbf{References}

AIP Statistics (2024, Sept 2). New Astronomy Bachelors and Masters: What Comes Next

Astronomy Degree Recipients: One Year After Degree. (2019, August 1). AIP.

Astronomy-Powered Careers | American Astronomical Society. (n.d.). Retrieved February 26, 2025, from

K‐12: Student Resources. (n.d.). Bureau of Labor Statistics. Retrieved February 26, 2025, from

Asefer, A., \& Abidin, Z. (2021). Soft Skills and Graduates’ Employability in the 21st Century from Employers’ Perspectives: A Review of Literature. 9.

Finley, A. P. (2023). The Career-Ready Graduate: What Employers Say about the Difference College Makes. American Association of Colleges and Universities.

Hirudayaraj, M., Baker, R., Baker, F., \& Eastman, M. (2021). Soft Skills for Entry-Level Engineers: What Employers Want. Education Sciences, 11(10), Article 10.

Vista, A. (2020). Data-driven identification of skills for the future: 21st-century skills for the 21st-century workforce. Sage Open, 10(2), 2158244020915904.

Whorton, R., Casillas, A., Oswald, F. L., \& Shaw, A. (2017). Critical skills for the 21st century workforce. Building better students: Preparation for the workforce, 10.

\subsection{“Ownership” appears to be an important – possibly even critical – ingredient to boosting STEM self-efficacy. But ownership of what? Of these amazing images that they’re collecting? Or of the process of collecting/processing them, making them amazing? “This is mine.” vs. “I did this.” And in either case, how well does this translate to non-image data? What about the analysis and modeling of non-image data? Are astronomical images a one-trick pony with respect to to boosting STEM self-efficacy, or just the beginning of a larger journey?}

\begin{flushright}
  \itshape
    Luisa Rebull \\
    Dan Reichart
\end{flushright}

\textbf{Introduction}

“Ownership” of the project or product(s) of a STEM experience can be a complex thing, and can mean different things for different levels of participants. One unifying element, however, is that a participant has to be engaged in the activity before ‘ownership’ can even be addressed (Hanauer et al., 2012). A student who is racing through an activity “just to get a number” may be unlikely to feel ownership regardless of effort on any other front. But an engaged participant could feel ownership at any of a number of levels.

Working with authentic data and tools makes it easier to ‘hook’ participants into being engaged.  If it seems like they’re working with something simple, they may be less likely to be engaged than if they realize they are working with “the real thing,” e.g., the same telescopes/ data/ tools that professional astronomers use. Moreover, it is critical that they make important decisions about what to do with the data or analysis, e.g., decide significant things in the process of doing whatever they are doing with the data. Making important decisions most frequently can result in feelings of ‘ownership.’ If participants are taking images where they get to decide where to point the telescope, that can result in ‘ownership.’ Working with others’ data, where they get to decide which images they use and how to make 3-color images, can also result in ‘ownership.’  However, if they are blindly following a set of scripted directions, it doesn't matter even if they requested the images; it still doesn’t feel like “their” data.

Slightly more sophisticated cookbook/inquiry isn't necessarily good enough either. “Ownership” doesn’t necessarily arise if it is a situation where the lab participants are struggling to figure out what they know especially when they feel if the instructor is, in essence, saying, “I know the answer but you have to figure out the information I’m hiding from you.” This is an issue that many non-STEM students seem to struggle with at the OPIS! level. Initial excitement over images is tempered by the fact that many expect Hubble-quality images, while others are dismayed by their problems in extracting data from their images. A common problem is that students often don't recognize when an answer they've generated is nonsensical...for example that the mass of their planet is many orders of magnitude larger than the mass of the sun, or that its diameter is larger than its orbital radius! An awareness that you should always question is a part of ownership. How do we instill it? Students feel ownership if they have to make significant decisions that impact the answers they get and maybe the questions they are even asking in the first place.

We have assembled case studies reflective of our own experiences. They come from two contexts: the OPIS! and MWU! undergraduate labs developed at UNC-Chapel Hill and the NASA/IPAC Teacher Archive Research Program (NITARP; see Rebull 2018RTSRE...1..171R), a program where high school teachers work with archival data from some of NASA’s professional astronomy research telescopes.

\textbf{OPIS!/}\textbf{MWU!}

In 2008, University of North Carolina at Chapel Hill decided to move away from their traditional lab offerings (small-telescope experiences at night and planetarium-based experiences during the day, neither of which connected strongly to the lecture curriculum) for their ASTRO 101 / survey-level students, and develop a Skynet-based curriculum (Dan Reichart, personal communication).  Called “Our Place in Space!” (OPIS!), UNC Chapel Hill began offering this in 2009.  Students learn how to acquire images with professional-quality telescopes at remote, professional-quality sites (in Chile, Australia, the United States and Canada), using Skynet’s queue-scheduling system, and then carry out investigations, such as measuring the mass of gas giants using Kepler’s Third Law, measuring the distance to main-belt asteroids by observing them in both northern and southern hemispheres simultaneously, measuring the distance to standard candles such as RR Lyrae in globular clusters, cepheids in nearby galaxies, and Type Ia supernovae in distant galaxies, measuring the rotation curve and mass of the Milky Way with radio observations, and various other authentic investigations.

Almost immediately, we realized that we had struck upon something. Students began posting their images on their social media sites, writing things like “best homework ever!” and “pretty damn awesome!”.  Others took their observing credits and went beyond the curriculum, making RGB combination images of deep sky objects, and posting these on social media as well.

Funded by NSF (Grant N 2013295), we have a team of education researchers studying this now. It is a rich data set, and so far they’ve only skimmed its surface, but the stand-out results are (1) very large gains in self-efficacy, and (2) an elimination of the gender self-efficacy gap, between pre and post surveys (Freed et al. 2024).  Now, they are digging deeper into what’s driving these gains, but preliminary focus-group results point to a combination of ownership and “realness.”  Students clearly feel a sense of ownership over their images.  That they’re using “real” telescopes at “real” sites, that are simultaneously used by “real” astronomers doing “real” science, also appears to be a key ingredient.

Interestingly, anecdotal evidence suggests students’ sense of ownership changes between OPIS! and OPIS!’s follow-up curriculum, “Astrophotography of the Multi-Wavelength Universe!” (MWU!) (Megan Dubay, Dan Reichart, personal communication).  Funded by DoD STEM, students carry out deeper (fainter-reaching) observations, at both optical and radio wavelengths, incorporate archival images from NASA/IRSA, and carry out deeper (physically) investigations, most closely matching the content matter of second-semester astronomy courses, at the interface of the non-major and major level.

Like OPIS! preliminary analysis of pre- and post-surveys also show large gains in self-efficacy, and the “realness” aspect described above still seems to apply.  However, the ownership aspect quickly evolves.  In MWU!, students work in teams, pooling their telescope access to carry out these deeper investigations.  Initially, students want to be the one to submit the observations for the team.  However, by mid-course, no one really cares who submits the observations.  Instead, their sense of ownership has transitioned from images that they themselves observed, to the work product, typically an LRGB + narrowband + infrared composite image.  These layers can be combined in different ways, with both different physical choices and different artistic choices to be made.  Although not required, different students on the same team often opt to make their \emph{own }combination, and that is the one they identify most strongly with (Mae Dubay, Dan Reichart, personal communication).

Although MWU! focuses mostly on what can be learned from color, and multi-wavelength layered, images, some of MWU’s explorations draw upon non-image data (e.g., pulsar timing data at radio wavelengths).  One of MWU’s most advanced tools is Clustermancer, which allows students to separate star cluster members from unrelated field stars, plot HR diagrams, and match isochrones to them, measuring the star cluster’s age, distance and metallicity.  Using this tool, students can often produce better results than what can be found in the professional literature (Daryl Janzen, Michael Fitzgerald, personal communication), and as such, we are adding a “publishing” capability into the tool itself.

We believe it is still an open question – but a very important question – whether students, and introductory students in particular, can feel the same level of ownership over a model, matched to data, as they can over an image (or conclusions reached from an image) that they took themselves.

But if this is the case with introductory students as well, this opens the door to all sorts of opportunities to use archival resources.  This is particularly the case now, as facilities like Rubin and Argus come online, opening up a plethora of time-domain investigations to student audiences.

\textbf{NITARP}

NITARP partners small groups of largely high school educators with a research astronomer for a year-long authentic astronomy research project (Rebull et al, 2018). We work primarily with educators, and through them, their students. At the educator’s discretion, students may be involved throughout the year. Most of our participants are public high school classroom educators (though we also have had private, middle school, community college, and informal educators participate). The program lasts ~13 months, January to January, and involves three trips, including a week in the summer when the team comes to Caltech in Pasadena, CA to work intensively together on the data analysis.  All NITARP programs use NASA data held at IPAC; fortunately, IPAC houses 3 NASA archives, with petabytes of professional data (as well as tools) waiting to be used by anyone with an internet connection.

Because of the nature of NITARP, all of the data that our participants use are archival data, e.g., none of them requested those data, and a lot of the data are survey data, so no one individual ‘requested’ those exposures. One of the first things that participants do in the program is that the teams have to write a proposal describing what they will do with the data. This proposal has all the same components that a professional archival research proposal does: a review of the literature, a description of the existing data, and a description of what they will do with the data. During this process, the teachers are developing a sense of “ownership.” Many will express “ownership” by the time the proposal is done, so just 3 months into a 13 month program.

By the time the teams begin seriously, intensively working on the data (during their summer visit, typically about 6 months into the program), when they are making decisions such as what broadband colors we will use to select young stars, or how much of an infrared excess is significant, or how will we create a spectral energy distribution (SED), or how will we do photometry to fill this gap in the SED, ownership is firmly established. Note that some of these sample questions have unambiguous answers and some don’t. Making an SED is something where the goal is clear and process reasonably well-established, and this falls closer to the “slightly more sophisticated cookbook inquiry” end of the distribution, but the NITARP participants still need to convert the units of the brightness measures and get everything into the same units and make log-log plots – all things that they haven’t necessarily done before, and are challenging, and yield a sense of accomplishment upon completion. A question that falls closer to the “make significant decisions” part of the distribution is figuring out which photometric colors to use to select young stars. While various approaches have been published in the literature (see, e.g.,  Gutermuth et al. 2008ApJ...673L.151G, 2009ApJS..184...18G or Koenig \& Leisawitz 2015AJ....150..100K), there are many different combinations that can work, with varying degrees of success, and, like any investigator, NITARP participants need to make decisions that have significant impact on what young star candidates they identify. Incorporating all of the data that are pulled together to assess the reliability of the selected young star candidates is yet another step that has real impact on the science that results. During the summer visit, when the teams are actively making these kinds of decisions and working intensively on their projects, at least 8 hours/day with their teams all in the same room at the same time, their sense of ‘ownership’ skyrockets.  One participant told us: “One evening, while working on some homework, I had the realization that THIS WAS REAL. There is no right answer, in fact, no one knows the answer. I can't just go and ask someone the answer. It was like a light bulb went off and I experienced a feeling of excitement and also felt a little bit scared. I thought to myself -- Is this how astronomers feel about their work? It was a great feeling and exciting that I too am part of this now.”

After they go home, the NITARP participants (teachers and students) finish their work remotely, they write up their results, and the whole group takes a poster presentation to the American Astronomical Society (AAS) meeting the following January. By this point, the project is ‘theirs’ forever, because they talked to professional astronomers at the AAS about their work, and there is a publication with their names on it. Additionally, some projects continue beyond the poster stage to become journal articles. For examples of NITARP projects that resulted in refereed astronomy journal articles, see Rebull et al.(2023, 2015, 2012) among others.

An important part of NITARP, we are told by the participants, is that they use professional data, with professional interfaces and professional resources and tools, and consulting the refereed literature. They are not using “watered down” versions of the tools or data; they are using the real thing. Even though it is hard, that’s what they want. Anything less seems “inauthentic.”

One subtlety we note here is that an important part of developing skills as a scientist is a healthy skepticism of one’s own work. Students who are just “working to get a number” might quit when they, well, get a number, but \emph{scientists} take the final step of checking the sensibility of that number, and realizing when they have come up with nonsensical results (calculated density of a Bok globule higher than that of a proton; velocity of a star in the Milky Way three times the speed of light; a planet mass many orders of magnitude larger than its host star or that its diameter is larger than its orbital radius; etc.). We have seen some of our novice NITARP scientists leap to a conclusion based on these incorrect answers of “I have overturned this famous result,” as opposed to “I may have screwed something up in my analysis and I probably should go back and check!” Specifically because of the importance of making real and significant decisions in the process of analysis, an important goal to strive for in the context of ‘ownership’ is inclusion of this final step of a ‘reality check.’

\textbf{Conclusion}

In conclusion, whether dealing with introductory, and often non-STEM, students, or STEM educators, we feel that the key ingredients for fostering a sense of ownership are (a) engagement, (b) use of professional resources (no watering down!), and (c) the ability to make important decisions, decisions that impact the product of the work, whether that is as simple as a target for taking a new image or as substantial as what question to investigate or whether or not certain objects or measurements should be included in the analysis.

\begin{figure}[htbp]
  \centering
  \includegraphics[width=0.8\textwidth]{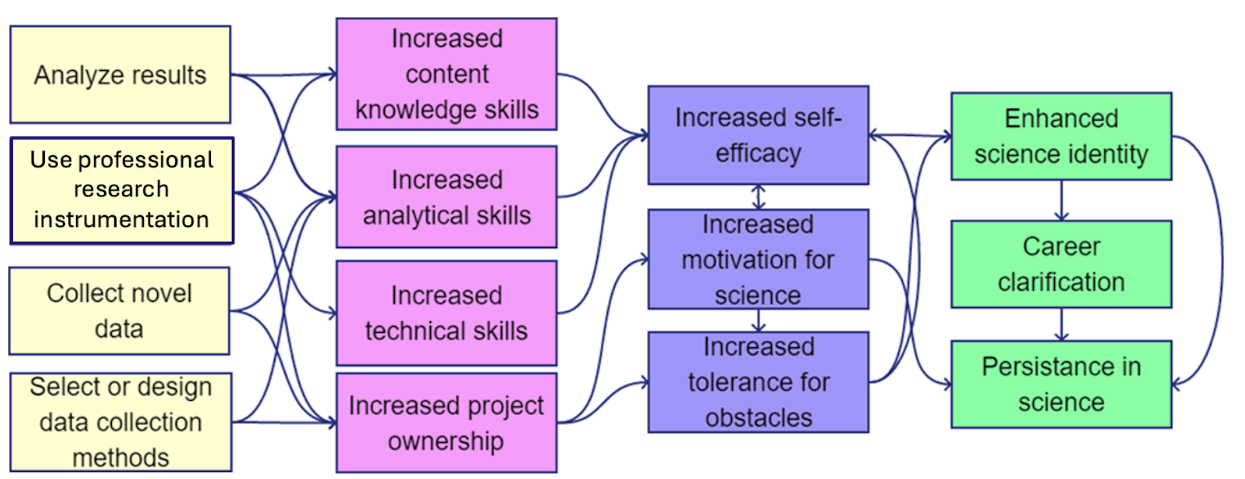}
  \caption{Schematic of the subgroup of the CUREs pathway model that we will focus on. }
  \label{fig:curemodel}
\end{figure}

\textbf{References}

Hanauer, D. I., Frederick, J., Fotinakes, B., \& Strobel, S. A. (2012). Linguistic Analysis of Project Ownership for Undergraduate Research Experiences. CBE Life Sciences Education, 11(4), 378–385.

Rebull, L. M., Anderson, R. L., Hall, G., Kirkpatrick, J. D., Koenig, X., Odden, C. E., ... \& Roosa, N. (2023). Young Stellar Object Candidates in IC 417. \emph{The Astronomical Journal}, \emph{166}(3), 87.

Rebull, L. M., Carlberg, J. K., Gibbs, J. C., Deeb, J. E., Larsen, E., Black, D. V., ... \& Lince, M. (2015). On infrared excesses associated with Li-rich K giants. The Astronomical Journal, 150(4), 123.

Rebull, L. M., Johnson, C. H., Gibbs, J. C., Linahan, M., Sartore, D., Laher, R., ... \& Tilley, C. M. (2012). New young star candidates in brc 27 and brc 34. The Astronomical Journal, 145(1), 15.

Rebull, L. M., French, D. A., Laurence, W., Roberts, T., Fitzgerald, M. T., Gorjian, V., \& Squires, G. K. (2018). Major outcomes of an authentic astronomy research experience professional development program: An analysis of 8 years of data from a teacher research program. Physical Review Physics Education Research, 14(2), 020102.

\section{ Engagement }

\subsection{Can astronomy be used to improve scientific literacy and the STEM workforce?}

\begin{flushright}
  \itshape
    Jack Howard \\
    Shanil Virani \\
    Rhone O’Hara
\end{flushright}

We live in a time where humans have access to more information at their fingertips than ever before. Our cell phones have more computational power now than the computers that were used to put 12 men on the Moon between 1969 and 1972. Despite living in an ever-increasing technology and science-driven culture, scientific literacy appears to be decreasing at an alarming rate (e.g., Thomson et al. 2019). For example, nearly 1 in 2 Americans think there is no evidence for human-induced  climate change or think it is due to natural causes (Pew Research Center, 2016). Approximately 1 in 4 Americans (and 1 out of 3 Europeans) think the Sun revolves around the Earth (Pew Research Center, 2016). Five decades ago we sent 12 men to the Moon. Many Americans now think that the Moon landings are a Hollywood hoax. Conversely, the societal demand for a scientifically literate society, both as a requirement for 21st century jobs as well as among our citizens as we face critical challenges as a society, has never been higher nor more important. General education ASTRO101-level classes are a powerful way to improve science literacy and build STEM capacity in the workforce (Borne et al., 2009).

Astronomy is not only a “gateway” to increasing STEM majors on college campuses, it is widely popular as it is estimated that approximately 10\% of university students now take an ASTRO101 class and the demand is increasing. Fraknoi (2001) estimates that approximately 250,000 students take ASTRO101 in the U.S. each year. However, anecdotally that number is only increasing.  At the University of North Carolina Chapel Hill, for example, there has been explosive growth in enrollment in introductory astronomy classes (see Figure 1, Reichart 2024). Moreover, for many university students, an introductory astronomy class may be the \emph{last} science class in their life. These classes, therefore, represent a novel opportunity to increase science literacy and 	provide a pathway to illustrate \emph{how} science is done, by whom, and at the same time, build capacity in a STEM workforce.  

\begin{figure}
    \centering
    \includegraphics[width=1.0\linewidth]{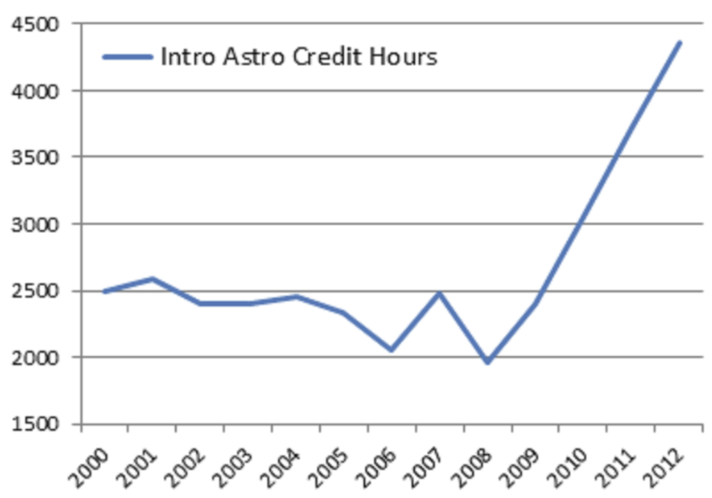}
    \caption{Credit Hours over Time}
    \label{fig:enter-label}
\end{figure}

Not only has the growth in ASTRO101 enrollment grown significantly in the last two decades, the number of astronomy bachelor’s degrees awarded at US institutions has been rising sharply and is now at an all-time high. Nicholson \& Mulvey (2020) report that the 820 astronomy bachelor’s degrees conferred in the class of 2020 represent a 23\% increase over the previous year. They found that over the last two decades, the number of astronomy bachelors conferred in the US has increased by over 300\%. Data from the National Center for Education Statistics (NCES) on the number of degrees earned in astronomy between 2011 to 2021 shows significant growth whether the data is examined by gender or ethnicity (see Figure 2). Data and growth in this discipline demonstrate the fallacy that astronomy is a “limited” discipline with no transferable skills. Indeed, it illustrates that the process and tools of astronomy develop transferable skills that can be applied across STEM fields and can be applied to many careers outside of astronomy. A recent AIP report investigating the employment sectors that recent undergraduateastronomy majors have entered is diverse and not limited to employment only in traditional astronomy sectors (see Figure 3).    

\begin{figure}
    \centering
    \includegraphics[width=1\linewidth]{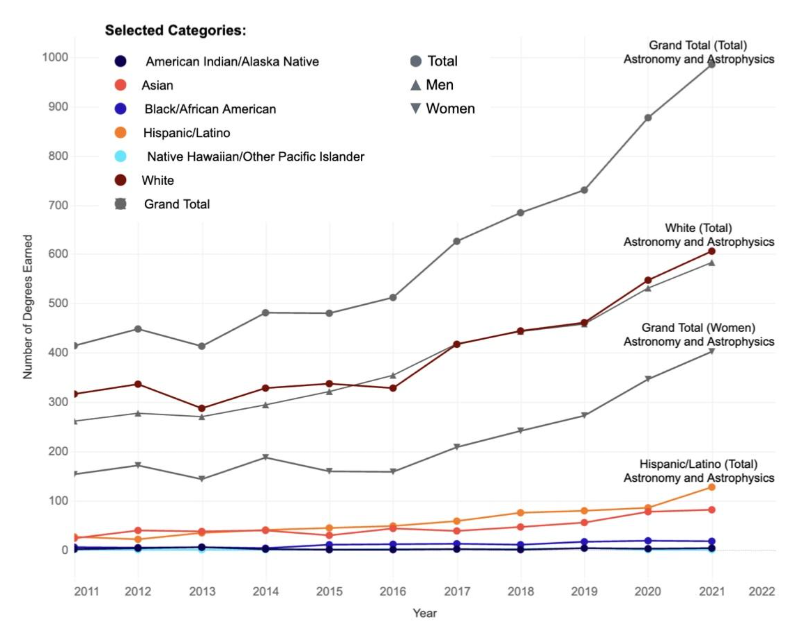}
    \caption{Degrees earned over time}
    \label{fig:enter-label}
\end{figure}

\begin{figure}
    \centering
    \includegraphics[width=1\linewidth]{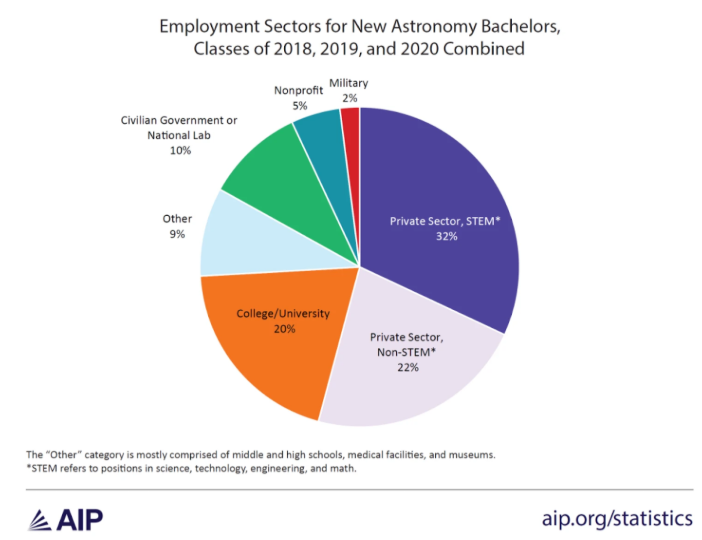}
    \caption{Employment sectors for graduates}
    \label{fig:enter-label}
\end{figure}

Astronomy is a starting point for the development of skills that carry over into many fields. For university students taking their first science class, it is where students can develop their critical thinking skills. For others, it's an opportunity to sharpen skills first learned in high school. In astronomy, as in many sciences, we always have the question before us … \emph{how do you know what you know}? For any conclusion, we ask: what is the supporting evidence; this is the basis of critical thinking. Discussion of the question before us then becomes the foundation for collaborative effort and interpretation.

Astronomy draws on modern technology and the sciences, often physics and engineering but now also biology, underlying that technology. Exploring these aspects of astronomy enables students to get started in useful skills like coding and data analysis, electronics and instrument building, designing experiments, and collaborating with others in project and mission planning. It is also an opportunity to build and foster communication skills that are attendant when working in groups or delivering presentations.

 Data from different sources show, for example, that the proportion of women and people of color studying astronomy has increased in recent years. Data from the American Institute of Physics has even identified the small percentage of astronomy students who identify as nonbinary. Many astronomy projects now include participatory scientists and engage people across different identities and experiences. On a larger level, astronomy now also benefits from large-scale international collaboration. This pan-cultural growth also means that textbooks need to take a broader view when introducing the subject rather than beginning only with the Greeks or during the Renaissance.

\begin{figure}
    \centering
    \includegraphics[width=1\linewidth]{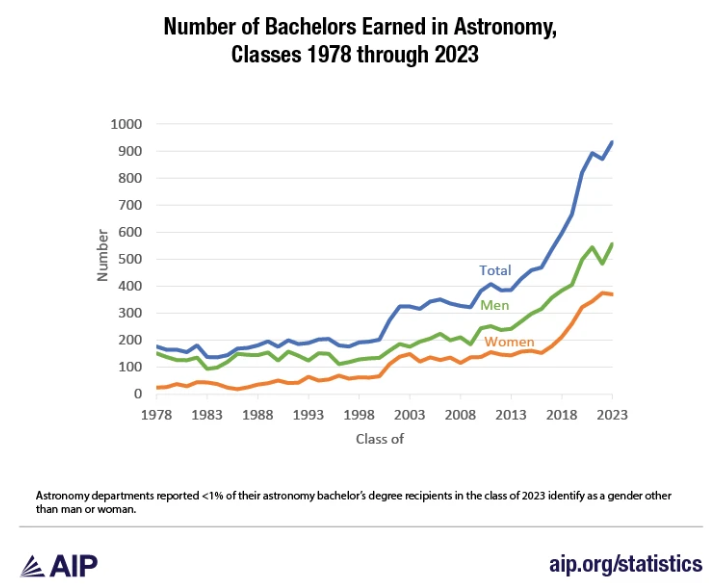}
    \caption{Number of Bachelors Earned in Astronomy 1978 to 2023}
    \label{fig:enter-label}
\end{figure}

Practitioners often state that astronomy lends itself to improving scientific literacy, as a general rule, however, research to support this is lacking.

Taken altogether, astronomy, as a field of study and a scientific practice, clearly lends itself to advancing scientific literacy (Aristeidou \& Herodotou, 2020; Bhathal, 2009; Buxner et al., 2018) and, consequently, the STEM workforce as indicated by recent research from the American Institute of Physics and the National Center for Education Statistics. Between consistent and increasing enrollment trends and promising employment data over the last 50 years, astronomy – and by extension the skills it develops – remains interdisciplinary in nature, technologically forward, and perhaps most importantly, relevant and revolutionary in its application. From coding, data analysis, and project/product management to public policy, aerospace engineering and teaching, astronomy has a unique ability to inspire and shape the forefront of various disciplines and professions. In recent studies it has been shown to improve scientific literacy (Buxner et al., 2018).

However, broadening our understanding of the impact of astronomy must extend beyond scientific literacy and the traditional STEM workforce. Scientific literacy through astronomy can and should also encompass media/digital literacy as well as cultural literacy to best address the world students inhabit today and equip them for their rapidly changing, and unpredictable, future. As such, the same approach should be taken when considering the STEM workforce. If we know that astronomy directly contributes to careers within the government, the military, education and the private sector, can astronomy also help to advance the non-STEM workplace? The short answer is yes. According to the AIP, 22\% of graduates with an astronomy degree are working in non-STEM related fields (see figure 3). As a result, the workforce at large, both STEM and non-STEM, consists of interdisciplinary roles, emphasizing the need for and value of transferable skills in today's diverse and dynamic job market. By embracing the interdisciplinary nature of astronomy and its broader implications and contributions to scholarship, collaboration, and inquiry, we can harness its potential to cultivate a more informed, inclusive, and skilled workforce in the years to come.

By and large, astronomy emphasizes and develops a set of learning, thinking, and collaborating skills that can be transferred across disciplines and careers – even as tools and modalities for work and scholarship change over time. Astronomy, then, serves as a conduit, not only connecting students with relevant content and the history of the content itself, but also to the larger workforce. The field is situated in a pivotal position that underscores and often facilitates opportunities for cross collaboration among research, cultural understanding, and can even diversify ways of working. In order to achieve such ends, it requires ongoing commitment and intentionality on the part of the astronomy community – researchers, educators, students, and professionals – to foster an inclusive and innovative environment where the benefits of astronomy can be felt in the classroom, in our personal lives, and in our professional careers.

Astronomy is not simply a collection of facts and formulas, although some students may see it that way. Astronomy is a mode of thought that allows us to critically engage our world and attempt to understand why things work the way they do. Through initiatives such as those outlined here, students, motivated by curiosity and authentic learning experiences, can see science at work in their everyday lives and careers and develop the confidence that they too can understand how the Universe works and recognize their place within it.

\textbf{References}

Aristeidou, M., \& Herodotou, C. (2020). Online citizen science: A systematic review of effects on learning and scientific literacy. \emph{Citizen Science: Theory and Practice}, \emph{5}(1), 1-12.

Bhathal, R. (2009). Improving the scientific literacy of Aboriginal students through astronomy. Proceedings of the International Astronomical Union, 5(S260), 679-684.

Borne, K. D., Jacoby, S., Carney, K., Connolly, A., Eastman, T., Raddick, M. J., ... \& Wallin, J. (2009). The revolution in astronomy education: data science for the masses. arXiv preprint arXiv:0909.3895.

Buxner, S. R., Impey, C. D., Romine, J., \& Nieberding, M. (2018). Linking introductory astronomy students’ basic science knowledge, beliefs, attitudes, sources of information, and information literacy. Physical Review Physics Education Research, 14(1), 010142.

Employment Sectors for New Astronomy Bachelors, Classes of 2018, 2019, and 2020 Combined. (2023, July 26). AIP.

Fraknoi, A. (2001), Enrollments in Astronomy 101 Courses: An Update, Astronomy Education Review, Vol. 1, Issue 1.

Funk, C. \& Kennedy, B., 1. Public views on climate change and climate scientists. Retrieved from Pew Research Center. Website:

Nicholson \& Mulvey (2020), Roster of Astronomy Departments with Enrollment and Degree Data 2020. American Institute of Physics.

Thompson, S., (2019), Australia: PISA Australia—Excellence and Equity?, Improving a Country’s Education (pp.25-47).

Number of Bachelors Earned in Astronomy, Classes 1977 through 2022. (2023, September 13). AIP.

Number of Bachelors Earned in Astronomy, Classes 1977 through 2022. (2023, September 13). AIP.

\subsection{Is it more important to engender curiosity and change attitudes or to teach a certain body of content (or develop scientific literacy)?}

\begin{flushright}
  \itshape
    Michelle Wooten \\
    Ken Brandt \\
    Tiffany Stone Wolbrecht
\end{flushright}

\textbf{Curiosity and why it matters (research perspective and personal narrative)}

During our AstroEdUNC Conference we collectively discussed that learning the content of astronomy alone is not the desired outcome of ASTRO101.  While there are many topic areas that we may cover in our courses, not all the same, we agreed that it is ideal if we also support students with increased positive affect toward and skill in doing science.  Previous studies in  science education research (Bruckermann et al., 2021; Mao, 2021; Zeidan \& Jayosi, 2014) suggest that these three things - knowledge, affect, and skills - evolve in relation to one another rather than independently.  In this section, we focus on one desired outcome of astronomy education: the possibility for inducing a sense of \textbf{curiosity }about the Universe in our students through course instruction.

As an astronomy education community, we related curiosity to cherished experiences of wonder and awe that seem to come naturally when observing the Universe.  In the psychology literature curiosity has been proposed as a construct based on \textbf{exploration }and \textbf{acceptance }(Berlyne 1954), where exploration regards gaining new knowledge and experience, and acceptance, openness to the unknown (Ahmad et al.  2023).  While some have proposed that curiosity is an intrinsic quality that requires no external recognition (Raharaja et al 2018; Shiau and Wu 2013), others have suggested that it is important to pay attention to its socio-cultural underpinnings.  For example, Nyet Moi \& Ahmad (2023) suggest that learning occurs in contexts where students find a need to apply conceptual knowledge to their experiences relating with others, especially when a moral question is involved.  In this case, curiosity arises from wanting to learn from the experiences and perspectives of others.  For example, when provided structured means to collaboratively organize their (differing) ideas about how to address social issues related to the scientific topic of study, they demonstrated greater increases in curiosity to topics, than if they were provided no structure, or no issue to discuss related to the topic.

Rigorous psychological or educational study of curiosity is still in development, and is mostly focused on younger ages of learning, especially in STEM contexts.  Some examples include: 

\begin{itemize}
  \item W.L.Ostroff, \textit{Cultivating Curiosity in K-12 Classrooms} (Virginia: ASCD, 2016)
  \item I.Inan, L.Watson, D.Whitcomb, and S.Yigit, \textit{The Moral Psychology of Curiosity} (New York: Rowman \& Littlefield International, 2018)
  \item Ahmad, Jamilah \& Siew, Nyet Moi. (2023). “The Effects of Socioscientific Issues with Wheel Thinking Map on Curiosity towards STEM among Year Five Students.” \textit{International Journal of Education, Psychology and Counseling}, 8, 20–35.
  \item D.E. Berlyne, “A Theory of Human Curiosity,” \textit{British Journal of Psychology. General Section}, vol. 45, no. 3, 1954, pp. 180–191, DOI: 10.1111/j.2044-8295.1954.tb01243.x
\end{itemize}

\textbf{How to Affect Change in ASTRO101 Students (From a Research Perspective)}

From a research perspective, affective states of a student include attitudes, self-efficacy, STEM interest, and career intentions. The path of a student’s journey through school and into the STEM workforce is often referred to as the “STEM pipeline” (Blickenstaff, 2005). Understanding the impacts of STEM on students at different points along this pipeline allow education researchers to identify and explore when students lose interest in STEM, which can inform practitioners in the development of education intervention strategies. There is little evidence of changes in attitudes towards STEM or STEM interest at the undergraduate level, but some education strategies have shown self-efficacy and content knowledge gains in ASTRO101 courses.

\textbf{Attitudes Towards STEM: }Research shows that attitudes towards STEM most often sees positive gains among elementary and middle school students (Zhou et al., 2019) and during field experiences (Scinski, 2014). Once students enter college, attitudes are more difficult to change (Bartlett et al., 2018). While attitudes towards STEM and, more specifically astronomy, are positively correlated with items such as attitudes towards the astronomy course and course grade expectations (Trotter et al. 2019), systematic research consistently suggests that schools and universities more often than not actually extinguish students’ interest in science and lead to a decline in positive attitude (Osborne, Simon \& Collins, 2003). This may be influenced by the well-documented discrepancy between student interest in “science” versus “school science” all the way from elementary (Hasni \& Potvin, 2015) to undergraduate institutions (Rennie et al., 2001; Smyth et al., 2016),\textbf{ }and more specifically in astronomy classes (Freed et al., 2022).

\textbf{STEM Identity: }In an astronomy identity framework developed by Colantonio et al. (2021), interest in astronomy in middle school students has a greater effect on identity for girls than for boys and that the effect on boys was mediated through utility value. Additionally, they found that interest in astronomy, perceived utility, and identity decrease significantly as grade level increases. Like attitudes, STEM identity does correlate positively to other indicators such as student career intentions, but changes in identity are most often seen earlier in a student’s education journey.

\textbf{Self-efficacy: }“Self-efficacy is defined as person’s belief that they can succeed in a particular task or activity" (Freed et al., 2023) and is domain or even task-specific (Lent et al., 1997; Bandura, 1997; Pajares, 1996; Bandura, 1977). Research on the ASTRO101 “Our Place in Space!” (OPIS!) curriculum, which allows students to utilize robotic telescopes to collect and analyze their own data as they learn course concepts, show a significant increase in both astronomy personal self-efficacy (APSE), or identifying personally with their own understanding of astronomy and astronomy concepts, and instrumental self-efficacy (ISE), or the individual's belief in their capacity to use astronomical instrumentation to reach specific goals, with particularly high gains among female students (Freed et al., 2024).

\textbf{Scientific Literacy: }Scientific literacy has been defined as the ability to engage with science-related issues, and with the ideas of science, as a reflective citizen, emphasizing competencies such as explaining phenomena scientifically, evaluating and designing scientific inquiry, and interpreting data and evidence (PISA, 2017). Similarly, the U.S. National Research Council describes it as "the knowledge and understanding of scientific concepts and processes required for personal decision making, participation in civic and cultural affairs, and economic productivity" (National Academies of Science, Engineering \& Medicine, 2016). These definitions highlight the importance of scientific literacy in enabling individuals to make informed decisions, participate in public discourse, and navigate an increasingly complex, science-driven world.

\textbf{How to Affect Change in ASTRO101 Students (From a Practitioner Perspective)}

Practitioner should model the expected behaviors in the students being taught. For example: 

\begin{itemize}
  \item Have students bring a pre-written “question I’ve always wanted to know more about in astronomy” to class. Collect responses on arrival, then read and answer some anonymously to set the tone and encourage further questions.
  \item Gather activities that engender curiosity (open-ended, non-cookbook investigations of topics and concepts), especially during the “first impression” phase of the course.
  \item Demonstrate genuine interest in, and encouragement of, student success—acting as an ally rather than a taskmaster—to boost engagement.
  \item Provide feedback that is specific, positive, and formative: grade early submissions in detail, reinforce successes, and suggest concrete ways to improve.
\end{itemize}

\textbf{Practical ways to engender curiosity: model the expected behavior}

\begin{itemize}
  \item Scientific literacy is a value—curiosity can be a tool
  \item AUI: to create a more scientifically literate society (Education Public Engagement Team)
  \item Engineer a better future through excellence in STEM education
  \item One way is to have a more scientifically literate population
  \item David: wonderful if every student mastered knowledge; outcomes alone are hard to assess—knowledge must be constructed, attitudes changed, patience and understanding developed to address misconceptions
  \item Tiffany: false dichotomy in the framing of the question
  \item The eclipse was a good catalyst for curiosity and shaping attitudes
  \item Good vs. bad curiosity: aiming for the right answer versus fostering lifelong learning
  \item Curiosity can lead to personal happiness and achievement
  \item Students need context and the right attitude—whether in the STEM workforce or not
  \item Curiosity: are students asking the right questions that make sense?
  \item Curiosity must be cyclical and inspired
  \item Instilling permanent knowledge can protect against superfluous information
  \item It’s a luxury to be curious; astronomy can be the vehicle to instill scientific literacy
\end{itemize}

I think to some extent it could be helpful to remember that differentiating content from attitudes is something that is distinguished in science education researchers.  In science education research, it is useful to study one construct at a time – through surveys or concept inventories, and then how they correlate with one another.  However, as someone who has used such inventories and surveys in my teaching, sometimes I question whether it is really wise or useful to differentiate content and attitudes.  Even though it makes some immediate sense that objective knowledge is different than one’s attitude toward that knowledge, in a given learning environment, how much does it make sense to parse these two separate things apart when they are always already ongoing, connected, and beholden to the relationships in the specific set of students-instructor in the one class?  In other words, are we creating a false binary when we ask these questions? Maybe it’s not as important to hierarchize these concepts – “which is more important” – and ask how these could be fostered together.

Another thing I think about when I hear this question is why it matters so much to us to engender curiosity.  Perhaps it is because being curious to us feels like one of the most lovely parts of being alive, and/or at least something we long for others.  When I am asked about whether I want curiosity for my students, I feel beholden to ask, Am I a lifelong learner?  When do I feel curious? What exactly fostered my senses of curiosity?  As someone who is modeling science, in what ways do my students see my sense of curiosity unfold, if at all?  Have I made the right conditions in my classroom for our mutual curiosity to unfold?  One of the greatest challenges I currently experience in this effort is our other responsibilities toward instructional alignment between what we “cover” and what we “test.”  If I am wanting to engender curiosity, it seems that there should be some bit of uncertainty in what we come to do or understand in the course.  How then can we reach a prescribed alignment, a fair course, if we also want to leave the possibility for mutual discovery?

\textbf{References}

Bandura, A. (1997). \emph{Self-efficacy: The exercise of control}. Macmillan.

Bandura, A. (1977). Self-efficacy: Toward a unifying theory of behavioral change. \emph{Psychological Review, 84}(2), 191-215.

Bartlett, S., Fitzgerald, M. T., McKinnon, D. H., Danaia, L., \& Lazendic-Galloway, J. (2018). Astronomy And Science Student Attitudes (ASSA): A Short Review And Validation Of A New Instrument. Journal of Astronomy \& Earth Sciences Education (JAESE), 5(1), 1–22. https://doi.org/10.19030/jaese.v5i1.10190

Britner, S. L., \& Pajares, F. (2006). Sources of science self-efficacy beliefs of middle school students. \emph{Journal of Research in Science Teaching, 43}(5), 485-499.

Bruckermann, T., Greving, H., Schumann, A., Stillfried, M., Börner, K., Kimmig, S. E., Hagen, R., Brandt, M., \& Harms, U. (2021). To know about science is to love it? Unraveling cause–effect relationships between knowledge and attitudes toward science in citizen science on urban wildlife ecology. Journal of Research in Science Teaching, 58(8), 1179–1202.

Freed, R., McKinnon, D. H., Fitzgerald, M. T., \& Salimpour, S. (2023). Confirmatory factor analysis of two self-efficacy scales for astronomy understanding and robotic telescope use. Physical Review Physics Education Research, 19(2), 020164.

Lent, R. W., Brown, S. D., \& Gore, P. A., Jr. (1997). Discriminant and predictive validity of academic self-concept, academic self-efficacy, and mathematics-specific self-efficacy. \emph{Journal of Counseling Psychology, 44}(3), 307-315.

Mao, P., Cai, Z., He, J., Chen, X., \& Fan, X. (2021). The Relationship Between Attitude Toward Science and Academic Achievement in Science: A Three-Level Meta-Analysis. Frontiers in Psychology, 12, 784068.

National Academies of Sciences, Engineering, and Medicine. 2016. Science Literacy: Concepts, Contexts, and Consequences. Washington, DC: The National Academies Press. https://doi.org/10.17226/23595.

Pajares, F. (1996). Self-efficacy beliefs in academic settings. \emph{Review of Educational Research, 66}(4), 543-578.

PISA 2015 Assessment and Analytical Framework. (2017, August 30). OECD.

Scinski, L. (2014). \emph{Beyond the classroom: The impact of informal STEM experiences on student attitudes and interest} (Doctoral dissertation, UC San Diego).

Trotter, A. S., Reichart, D. E., LaCluyzé, A. P., \& Freed, R. (2019). Factors contributing to attitudinal gains in introductory astronomy courses. arXiv preprint arXiv:1910.12630.

Zeidan, A. H., \& Jayosi, M. R. (2014). Science Process Skills and Attitudes toward Science among Palestinian Secondary School Students. World Journal of Education, 5(1), p13.

Zhou, S. N., Zeng, H., Xu, S. R., Chen, L. C., \& Xiao, H. (2019). Exploring Changes in Primary Students' Attitudes towards Science, Technology, Engineering and Mathematics (STEM) across Genders and Grade Levels. Journal of Baltic Science Education, 18(3), 466-480.

\subsection{How to improve engagement to create lasting learning in astronomy?}

\begin{flushright}
  \itshape
    Art Borja \\
    Kevin Lee \\
    Digesh Raut
\end{flushright}

\textbf{What is lasting learning?}

Let’s start by posing a question to the reader: What is your most vibrant memory from a classroom experience as a student? You are encouraged to identify the characteristics that made this a lasting learning moment. You may not recall precisely all the details. Still, you remember having fun memories of things that left a lasting impression – things that connected with you and could be incorporated within the framework of your existing knowledge. Educators call this building of knowledge from our experiences constructivism (Larochelle et al., 1998)

We define “lasting learning” as an understanding that students can access in their memory many years after the subject matter is covered in the classroom. We know that lasting learning is NOT obtained by telling students information (Druckman \& Bjork, 1994; Schmidt et al., 2015). Students are far more likely to retain information if it is obtained by “doing” (Kolb, 2015). Students should interact with astronomy content proactively, manipulating and creating on a matter of importance to them somehow. We will summarize some possible factors that could make educational experiences memorable and likely to be retained.

\textbf{What makes astronomy content engaging?}

Research on “lasting learning” is limited as all longitudinal studies are challenging.  Some research has shown that interactive engagement techniques lead to “deep learning” – the retention of content years after instruction (Francis et al.  1998). No matter what the age group, students can be engaged in the study of astronomy if they are given opportunities for inquiry (Plummer, 2012). Students should be “doing” science while “learning” science in other words they should be engaged in genuine scientific activities. Certain characteristics are more likely to make content engaging to today’s students. Let’s explore these characteristics:

\textbf{How will students interact with the content?}

Considerable research illustrates that interactive engagement techniques greatly outperform traditional lecture techniques (Freeman et al., 2012, Hake, 1998, Prather et al. 2009). One way that lasting learning can be strongly facilitated is through active learning. While interactive engagement techniques are necessary, it may not be sufficient to achieve lasting learning. Although the need to keep students engaged has always been with us, today's challenges are more substantial post-covid.

\textbf{Will the materials grab student’s attention?}

Content should be visually vibrant as students have high standards regarding their attention. Colorful graphics, video segments, and simulations are helpful. A two-minute elevator pitch that avoids jargon and excessive details should focus on hooking the students. It should spark curiosity and end with a question that the students will cover through active learning in the rest of the class.

\emph{Example: Showing a short cinematic clip from a science fiction movie that illustrates the science concept. The visuals, theme music, and characters in stressful situations may get students involved. This could lead to discussing realism with students and whether liberties have been taken with the science content. }

\textbf{Students often ask: Why is this important to know? How does it affect them? Why is this important to me? Why is it useful to me?}

Astronomy content can be more appealing to students across the varied spectrum by appealing to what is important to them as individuals. One should also discuss how astronomy research has led to technological advances that have benefited them and the society.

\emph{Example: Stellar evolution and the formation of heavy elements. As Sagan used to say, “We are star stuff.}

\emph{Example: Discussion of cooking with Teflon pans – developed accidentally when looking for a refrigerant for astronaut suits.}

From the perspective of the educator, the educator tends to look at a text or other materials to see if they fulfill a purpose within the material in adequately covering content that the course is intended to cover. Usually, there is no one textbook or resource that fulfills the educator’s requirements, unless, of course, the educator is the author of the text. Some readers may remember when Kinko’s offered a service called “Professor Publishing” where instructors could pick and choose sections from various texts and sources and in essence create a unique text for their class. Unfortunately a lawsuit in the early 90’s put an end to that wonderful tailoring of texts for classes and ensured that the profits for the sales of school and college textbooks continued.

\textbf{Students often struggle to forge connections with content which they can carry beyond the classroom.}

We need to “humanize science” to make it more relatable to issues the students care about, such as race, gender, social justice, human nature, etc. This will create a bond with the content that extends beyond the classroom.

Embedding references and external resources in our lectures extends or supplements the course material to topics that might be of interest to our students. This will also create opportunities for students to reconnect with the content after taking the course. Although scientific events and developments in astronomy have been well documented, the records and stories pertaining to social issues such as social justice, race, and gender often get lost. These are issues that most of our students can relate to. Therefore, we need a coordinated effort from the broader astronomy and scientific community to keep and maintain such records and make them accessible to the wider community.

\emph{Example:  Coverage of the injustice the “Harvard Computers” experienced – Annie Jump Cannon, Henrietta Swan Leavitt, etc. – and how their brilliance was taken advantage of, but not rewarded. }

\emph{Examples: Arthur Bertram Cuthbert Walker was a Black American solar physicist and educator who developed the x-ray and ultraviolet telescopes used in 1987 to capture the first detailed photographs of the Sun’s outermost atmosphere, the corona,[cite }\emph{].}

\textbf{Can the students feel “ownership” over what they create/develop?}

The course and its content should be designed to lead students on a path of discovery. This is essential to enable students to take ownership of their learning experience, thereby allowing them to reconnect with it long after taking the class. Any scientific inquiry starts by asking probing questions. If we plan our lectures focusing on the “questions” rather than the answers, it will help improve ownership.

\emph{Example: A simple }\emph{excercise}\emph{ of asking students to predict outcomes of a physics demonstration forces students to commit to an answer, and they “get some skin in the game.” They then pay rapt attention during the demonstration to determine if their prediction is right.  [see "Classroom demonstrations: Learning tools or entertainment?", C. Crouch, A, Fagen, J. }\emph{P.Callan}\emph{, and E. Mazur, American Journal of Physics 72, 835 (2004)}]

\emph{Example: Let the students invite their friends to observation nights and ask your students to lead these sessions. }

\emph{Example: Require students to research and make presentations as part of the course. Provide choices for them regarding topic and presentation format to choose from. }

\textbf{Can the student meaningfully collaborate with other people to explore astronomy content?}

Many methods of collaboration can be incorporated into traditional lecture settings. These include peer instruction voting/discussion, collaborative worksheets like lecture tutorials, and asking discussion questions. Can we make our lecture content more accessible to enable equitable learning experience to all our students? Teach using different modalities to address the learners’ needs since everybody learns differently. We should present each topic in several different manners.

\emph{Example}\emph{:  Present}\emph{ extrasolar planet orbits as a sonification (where the orbital period of the ESP is proportional to a sound frequency). }

\textbf{Can the content be delivered using technology that the student owns personally?}

\emph{Example: Smartphone Technology}

From our experience, college students today are incredibly attached to their smartphones, and it is hard to find a student who doesn’t have one. They have had one in their pocket every day of their adult lives. This presents a wonderful opportunity to forge connections with our students by delivering interactive engagement materials on their beloved devices – a high-powered computer in their pocket.

Many instructors and education systems in various countries today ban smartphone use in their classrooms because of the potential for distractions (see Bottger \& Zierer, 2024 for a literature review). There is a lot of research on the frequency of non-academic smartphone usage in the classroom and its negative effects on student performance (Elder, 2013; Tinder \& Bohlander, 2012, ). Smartphones also present new challenges related to academic honesty (Best \& Shelley, 2018).

In general, the positive impact of smartphones have been little explored due to the small number of tools available that run on all smartphones. One such tool is smartphone simulations with embedded questions for students to work on. These can be efficiently assigned to students through QR codes and listing the assigned questions.  Examples of such simulations are available at.The motivations, design framework, usage guidelines, and significance of this pedagogical approach are detailed in the AAS-sponsored YouTube videos series at:

Formative Assessment tasks such as ranking and sorting can also be easily delivered on smartphones. Students can quickly complete these tasks, grade their performance, and receive feedback. The WorldWide Telescope also has a web client that functions well on smartphones.

\textbf{References}

Best, L., \& Shelley, D. (2018). Academic Dishonesty: Does Social Media Allow for Increased and More Sophisticated Levels of Student Cheating? International Journal of Information and Communication Technology Education, 14(3), 1–14.

Böttger, T., \& Zierer, K. (2024). To Ban or Not to Ban? A Rapid Review on the Impact of Smartphone Bans in Schools on Social Well-Being and Academic Performance. Education Sciences, 14(8), Article 8.

Druckman, D., \& Bjork, R. A. (Eds.). (1994). Learning, remembering, believing: Enhancing human performance (pp. x, 395). National Academy Press.

Elder, A. D. (2013). College Students’ Cell Phone Use, Beliefs, and Effects on Their Learning. College Student Journal, 47(4), 585–592.

G. Francis, J. Adams and E. Noonan, ”Do They Stay Fixed?”, Phys. Teach. 36, 488-490 (1998)

Hake, R. R. (1998). Interactive-engagement versus traditional methods: A six-thousand-student survey of mechanics test data for introductory physics courses. American Journal of Physics, 66(1), 64–74.

Kolb, D. A. (2015). Experiential Learning: Experience as the Source of Learning and Development. Pearson Education.

Larochelle, M., Bednarz, N., \& Garrison, J. W. (Eds.). (1998). Constructivism and education. Cambridge University Press.

Prather, E. E., Rudolph, A. L., Brissenden, G., \& Schlingman, W. M. (2009). A national study assessing the teaching and learning of introductory astronomy. Part I. The effect of interactive instruction. American Journal of Physics, 77(4), 320–330.

Tindell, D. R., \& Bohlander, R. W. (2012). The Use and Abuse of Cell Phones and Text Messaging in the Classroom: A Survey of College Students. College Teaching, 60(1), 1–9.

\subsection{Are authentic experiences or active engagement (in class, active learning techniques, such as think pair share) more effective?}

\begin{flushright}
  \itshape
    Rachel Freed \\
    Ardis Herrold
\end{flushright}

There is a large body of research on the impact on learning gains through active learning techniques, such as think pair share and lecture tutorials (Prather \& Wallace 2021; Smith et al., 2018). Closing the achievement gap (Burke et al, 2020) is crucial within science, and active learning strategies have been shown to do that (Burke et al, 2020). Furthermore, when instructors become integrated into a community of practice they are more likely to implement student-centric practices leading to more actively engaged students (Tomkin et al., 2019). In one study, active learning approaches led to greater completion success of science courses for Hispanic students while having a negative impact on Asian students, compared to a traditional lecture-based coursework (Burke et al., 2020). Clearly there are cultural implications to consider. In an earlier study moving to active learning closed the gap in learning gains between non-underrepresented minority (non-URM) and URM students, with an increase in science self-efficacy for all students (Ballen et al., 2017). Again, cultural differences are important to consider as in this study social belonging increased only for non-URM students.

John Dewey wrote in the 1880’s about a philosophy of teaching that was called “learn by doing”. This parallels everyday human experiences such as learning to walk, talk, playing a musical instrument, participating in a sport, etc. Involving students in active engagement is certainly not a new or novel idea, yet why is it not embraced by so many educators? From our experience this inertia can partly be attributed to the culture that has persisted for many years. Instructors default to teaching the way that they have been taught, which (especially at the college/university level) is primarily a traditional lecture format.

Traditional teaching requires less time and effort to prepare a lecture delivery than designing a class session that incorporates active learning techniques. Content pressure is another contributing factor. It is far more effective to communicate new knowledge or demonstrate processes in a short amount of time. But beyond the communication, do students really learn when they are not actively involved?

Students must be properly introduced to active learning so that an atmosphere of trust and respect and an expectation of consistent participation is established. Yet, once the routine is in place, active learning techniques take little time to implement.One can argue that the traditional lecture approach to teaching has been effective for many years. But the real question is, how effective? How many more students would have experienced higher achievement if they were involved in active learning techniques?

The Covid pandemic was exemplary in demonstrating that passive learning techniques lead to lesser degrees of engagement, attendance and achievement. In a short time, all students were in effect enrolled in “online classes”, many of them asynchronous in nature. Learning for many during Covid was a combination of watching recorded lectures, videos, completing worksheets and problem sets, and doing solitary writing. The extended period of time when students worked in isolation led to major learning gaps at all levels of education. Additionally, students suffered a setback in social skills which even now is still reflected in altered communications with both instructors and peers.

Why does interaction enhance the learning experience? Cognitive Learning Theory (Nabavi, 2011, Schunk, 2012) considers the factors that are necessary for optimized thinking (reflection) about new knowledge or skills, and how these factors play into the brain’s ability to categorize, relate and store information. The degree of engagement for a task is a key factor in enhancing cognitive processes, because it contributes to how focused the mind is and how long the mind can persist in following or executing a process (attention span). When a task is pleasurable, relevant, and presents the sweet spot of challenge (not trivial or overwhelming), it optimizes the conditions for developing a deeper understanding and an increased retention of new concepts.

While social interaction is commonly thought to be necessary for active learning, it is not essential. The key factor in active learning is what happens in the brain of the learner. Reflection on new information, and contexting this information in light of prior knowledge and experiences, is an important part of the process. Other people can enhance the process of learning by providing feedback, asking questions and interjecting new information into the conversation. This interactive element can also be achieved by a well designed scaffolded activity that provides self-pacing, feedback, and creates opportunities to confront and resolve common learning confusions.

Authentic experiences that use real data more appropriately reflect the nature of science than a passive delivery of knowledge from a lecture, video or textbook. These open-ended, sometimes messy enterprises lend credibility to the task and can create a heightened sense of curiosity. Students do not have the perception of following a cookbook lab or working with simulated “data” for which an answer is already known. Authentic experiences enable students to parse a problem in unique ways. Since the path forward may not be linear, students must engage in critical thinking, construct mini experiments and assess their results, and have permission to make mistakes and reflect on them. The focus is shifted from “did I get this right?” to “what does this mean and how can I investigate more”?. Students immerse themselves more deeply in the activity and develop a sense of ownership and often an increased sense of self-efficacy in their ability to “do science” (Freed et al. 2024).

It is imperative that we as an educational community develop more experiences and teaching techniques that engage students in active learning. At the same time, we need to be mindful of typical classroom contexts and constraints. These may involve required curriculum, limited class time, less- than-robust background knowledge of the instructor, limited preparation and assessment time, financial considerations,  available physical resources, and most importantly the cultural context.

There needs to be a free universal tool developed that can query surveys and retrieve data all in a cloud-based format, in order to eliminate the need for downloads. The graphical interface for the tool should be designed for non-technical (student) users, if it is to see widespread use. A library of online data visualization tools should be developed to interrogate and explore data. These tools should incorporate the simple functions of spreadsheets and plotters, interactive spatial (three dimensional) display tools, and specialized tools, such as fitting light curves, isochrones, etc. These tools should when possible be constructed to address the needs of low vision users. Preliminary work in this area that really takes into consideration the pedagogy has been done in the context of cosmology but has the potential to be expanded to other topics (Salimpour, 2021; Salimpour, Fitzgerald, Tytler, et al. 2021)

It is clear that the teaching of STEM has not had the desired impact on the general population’s scientific literacy  and therefore new approaches must be implemented on a large scale. There is a large body of research on different pedagogical approaches yet there is a gap between research and practice . Authentic experiences and experiential learning have been shown to have many of the benefits that lead to scientific literacy. However, within the classroom, authentic experiences are difficult to implement when a teacher’s own experience as a learner was based on traditional lecture courses. Combined with the lack of a pipeline for information transfer between educational research and the teachers, STEM education doesn’t change quickly enough to have a population equipped to deal with the major scientific issues faced by the planet.

Authentic learning experiences in science lead to increased self-efficacy and science identity (Singer et al, 2020). Authentic experiences provide opportunities for students to make choices within the learning environment, empowering them and providing a sense of ownership. These affective domains are all important in helping students persevere within STEM and choose STEM careers. Authentic experiences and experiential learning provide students with opportunities to engage with the scientific process. Within these contexts students will encounter opportunities to make decisions, engage creatively with their data and learn from mistakes and successes.

\textbf{References}

Ballen, C. J., Wieman, C., Salehi, S., Searle, J. B., \& Zamudio, K. R. (2017). Enhancing Diversity in Undergraduate Science: Self-Efficacy Drives Performance Gains with Active Learning. CBE—Life Sciences Education, 16(4), ar56. https://doi.org/10.1187/cbe.16-12-0344

Burke, C., Luu, R., Lai, A., Hsiao, V., Cheung, E., Tamashiro, D., \& Ashcroft, J. (2020). Making STEM Equitable: An Active Learning Approach to Closing the Achievement Gap. International Journal of Active Learning, 5(2), 71–85.

Kolb, D. (1984). Experiential Learning: Experience As The Source Of Learning And Development. In Journal of Business Ethics (Vol. 1).

Nabavi, R. T. (2012). Bandura’s social learning theory \& social cognitive learning theory. Theory of Developmental Psychology, 1(1), 1-24.

Prather, E. E., \& Wallace, C. S. (2019). Lecture-Tutorials in Introductory Astronomy. In Astronomy Education, Volume 1: Evidence-based instruction for introductory courses. IOP Publishing.

Salimpour, S. (2021). Visualising the Cosmos: Teaching cosmology in high school in the era of big data [Doctoral Thesis]. Deakin University.

Salimpour, S., Fitzgerald, M. T., Tytler, R., \& Eriksson, U. (2021). Educational Design Framework for a Web-Based Interface to Visualise Authentic Cosmological “Big Data” in High School. Journal of Science Education and Technology, 30. https://doi.org/10.1007/s10956-021-09915-2

Schunk, D. H. (2012). Social cognitive theory.

Singer, A., Montgomery, G., \& Schmoll, S. (2020). How to foster the formation of STEM identity: Studying diversity in an authentic learning environment. International Journal of STEM Education, 7(1), 57.

Smith, D. P., McNeil, L. E., Guynn, D. T., Churukian, A. D., Deardorff, D. L., \& Wallace, C. S. (2018). Transforming the content, pedagogy and structure of an introductory physics course for life sciences majors. American Journal of Physics, 86(11), 862–869.

Tomkin, J. H., Beilstein, S. O., Morphew, J. W., \& Herman, G. L. (2019). Evidence that communities of practice are associated with active learning in large STEM lectures. International Journal of STEM Education, 6(1), 1.

\section{ Beyond the Classroom}

\subsection{What can instructors do that provides sufficient scaffolding for students who enter a course with weak prior knowledge and experiences while maintaining their course standards?}

\begin{flushright}
  \itshape
    Mariel Meier \\
    Carl Schmiedekamp \\
    Ann Schmiedekamp \\
    Oliver Fraser \\
    Christine Russell
\end{flushright}

Many ASTRO101 instructors say they do not expect their students to have any prior astronomy knowledge, but there are many related concepts we depend on in our teaching. These include fundamental mathematics and science knowledge, familiarity with technology tools, and executive functioning skills (e.g. organizing and planning). This section lists knowledge and experiences we might assume students have, and ways we can support them in the context of teaching an ASTRO101 course. In creating this, we realized that we’ve created a list of all the things that ASTRO101 instructors might reasonably have to consider beyond astronomy!

\textbf{Expectations }\textbf{of  Students}

\begin{itemize}
  \item Knowledge of class expectations:
    \begin{itemize}
      \item Syllabus, rubrics, announcements
      \item Note differences between high school and college grading culture (e.g., make-up tests, dropped tests, extra credit)
    \end{itemize}
  \item Syllabus should cover basics:
    \begin{itemize}
      \item How to contact instructors and how they’ll contact you
      \item Test policies (make-ups?)
      \item Late-work policies
    \end{itemize}
  \item Syllabus quiz (can also ask for info from the first announcement to ensure students read announcements)
  \item Review the Learning Management System and direct students to course resources
  \item Demonstrate how to use the rubric to complete assignments (recorded video or live); or have students do peer evaluations using the rubric
  \item Knowledge of institutional policies (dropping a class, withdrawals, credit/no-credit grading, etc.)
  \item Maintain an up-to-date gradebook
  \item Refer to key dates in your syllabus
  \item Use advising resources (what do local advisors recommend?)
  \item How to take advantage of academic accommodations for disabled students:
    \begin{itemize}
      \item Syllabus should direct students to local resources and explain how approved accommodations work in your course and on campus
    \end{itemize}
  \item Engagement:
    \begin{itemize}
      \item Attendance, attention to due dates, completing assignments, awareness of course policies
      \item For first-year students, create good habits via a flexible but supportive attendance policy
    \end{itemize}
  \item Messages to students who miss multiple assignments:
    \begin{itemize}
      \item Send email or text (if university system allows)
      \item Require office-hours visit for those who miss multiple assignments or perform poorly on a high-stakes task
    \end{itemize}
  \item Encourage responsibility for identifying knowledge gaps and seeking remediation (office hours, focused study, etc.)
  \item Incentivize office hours (extra credit, exam-related problems, stickers)
  \item Provide tutoring through affinity groups (e.g., Office of Minority Affairs, Student Athlete Academic Services) and “free help sessions” (office hours with prof, TA, or trained peers)
  \item Use ELI5 (Explain It Like I’m 5) summaries to force students to confront their understanding gaps
  \item Provide links to Khan Academy topics, YouTube videos, Wikipedia articles, etc.
  \item Demonstrate effective note-taking
  \item Foster collaboration among students:
    \begin{itemize}
      \item Train for teamwork: explain how to start, how to work in a group, and make goals explicit
      \item For larger groups, assign roles: reporter, notetaker, facilitator (ensures everyone contributes)
      \item For group projects, have students create a group contract with contact info, response-time expectations, communication modes, and consequences for unmet expectations
    \end{itemize}
\end{itemize}

\textbf{Mathematical Knowledge}

\begin{itemize}
  \item For major-specific or math-intensive courses, consider a pre-test or pre-semester math modules to review necessary topics.
  \item Number sense: powers of ten, scientific notation.
  \item Show calculations step-by-step or have students work in groups and share their solutions.
  \item Focus on concepts rather than heavy math in out-of-class assignments.
  \item Algebraic manipulation:
    \begin{itemize}
      \item Give low-stakes assignments for algebra practice.
      \item Develop practice sets to be done with a TA or LA (e.g., for nominal extra credit).
    \end{itemize}
  \item Reading graphs:
    \begin{itemize}
      \item Demonstrate (live or video): explain axes, interpret one data point’s x/y values, then discuss the overall trend (macro–micro–macro).
    \end{itemize}
  \item Plotting data:
    \begin{itemize}
      \item Provide a rubric outlining expectations for plots.
      \item Practice in small groups and have students peer-evaluate using the rubric.
    \end{itemize}
  \item Math vocabulary:
    \begin{itemize}
      \item Create low-stakes writing tasks requiring use of mathematical and content-specific terms.
    \end{itemize}
  \item Geometry basics: radius, diameter, circumference.
  \item Graphing circumference vs.\ diameter:
    \begin{itemize}
      \item Ask for predictions, measure circular objects, plot results, then discuss findings and uncertainties.
    \end{itemize}
  \item Translating 2D diagrams to 3D structures:
    \begin{itemize}
      \item Use 3D-printed models (e.g., segments of the celestial sphere) for tactile learning.
      \item Leverage web or app-based 3D models (e.g., Exoplanet App with 3D Milky Way).
    \end{itemize}
  \item Population distributions:
    \begin{itemize}
      \item Relate cluster CMDs as snapshots of evolving populations.
      \item For students without AP Statistics, use familiar examples (e.g., height vs.\ weight plots) to introduce the concept.
    \end{itemize}
\end{itemize}

\textbf{Computer Literacy}

\begin{itemize}
  \item Engagement activities can require computer simulations, online tools, and software.
  \item Provide links to online tools that don’t require downloads or installations.
  \item Include multi-modal access instructions—written guides, videos, and live walkthroughs in class.
  \item Utilize former students as learning assistants to provide tech support.
  \item File management and software installation:
    \begin{itemize}
      \item Provide activities with small files that are quick to download and easy to save/compare.
      \item Host files on your LMS to avoid firewall or login issues.
      \item Clearly state technology requirements before semester start (laptop vs.\ tablet) and campus resource availability.
    \end{itemize}
  \item Spreadsheet proficiency:
    \begin{itemize}
      \item Have students create simple spreadsheets to calculate and graph data.
      \item Require proper graph labeling and axis ranges to highlight key properties.
      \item Begin with pre-loaded templates; gradually remove scaffolding so students add calculations themselves.
    \end{itemize}
  \item Calculator skills (scientific notation, parentheses, powers, roots):
    \begin{itemize}
      \item Assign problems involving algebraic expressions with powers of ten, using contextually relevant values.
    \end{itemize}
\end{itemize}

\textbf{Physics Knowledge}

\begin{itemize}
  \item Units (metric system: km, kg, etc.)
  \item Discuss units when a new quantity is introduced
  \item Require units on all answers
  \item Show derived units and the base units that compose them
  \item Vocabulary: volume, density
  \item Relate to everyday notions of “heavy” and “big”
  \item Demonstrations:
    \begin{itemize}
      \item Density: identical containers with heavy \& light contents
      \item Volume: large and small containers of the same weight on a balance
    \end{itemize}
  \item Relationship between distance, velocity, and time (constant velocity model of motion)
  \item Multiple ways of representation:
    \begin{itemize}
      \item Describe in words how one quantity depends on the others
      \item Show as a graph
      \item Find a movie clip that demonstrates motion
      \item Live demo (pace the room to a metronome)
    \end{itemize}
\end{itemize}

\textbf{Know}\textbf{ Other Student Resources}

\begin{itemize}
  \item Healthcare resources:
    \begin{itemize}
      \item Physical Health: nurse or clinic on campus?
      \item Mental Health: counseling availability; how to contact the counseling office; student-of-concern check-in forms for at-risk students
      \item Reproductive care: specific clinic available?
    \end{itemize}
  \item Public Safety:
    \begin{itemize}
      \item Title IX (sexual harassment, discrimination, etc.)
        \begin{itemize}
          \item What is Title IX?
          \item Who to report incidents to; know your campus representatives
        \end{itemize}
      \item Reporting theft: where to report stolen items; location of the safety office
    \end{itemize}
  \item Conduct Codes:
    \begin{itemize}
      \item Academic misconduct/plagiarism policies
      \item Student conduct and honor codes
    \end{itemize}
  \item Alternative Tutoring Resources:
    \begin{itemize}
      \item Writing Center (sometimes combined with a reading center)
      \item Math Help Center
      \item Tutoring Center: live tutors, online tutoring, etc.
      \item Incentives: extra credit or mandatory visits for assignments; verification notes from tutors
    \end{itemize}
  \item IT Office: where to get computer help beyond the classroom
\end{itemize}

\subsubsection{Conclusion}

This section highlights general expectations instructors have of their students when they enter the introductory astronomy course, as well as the expectations with respect to mathematical knowledge, computer literacy, and physics knowledge. For each of these, suggestions are provided for addressing student challenges in specific areas.

\subsection{What are the hidden expectations in Astronomy 101 that represent barriers to access to diverse student groups?}

\begin{flushright}
  \itshape
    Sandy Liss \\
    Elise Weaver \\
    Ulrike Lahaise \\
    Amy L Glazier \\
    Marta Dark-McNeese
\end{flushright}

Introductory Astronomy students. Typically, these students are assumed to be: 

\begin{itemize}
  \item Of traditional age (18–22)
  \item In possession of little to no science background
  \item Taking these classes to fulfill general education requirements
  \item Middle- to upper-middle class
  \item Urban or suburban
  \item Unmarried, with no children or pets
  \item Residing on or near campus
  \item Unemployed or working fewer than a few hours per week
\end{itemize}

These students tend to be well served by traditional college experiences and expectations. We want to consider ways to engage with students who are not well served by traditional college experiences. To do so, we must first examine the barriers to access. We consider these issues in three broad groups: Accessibility Barriers in ASTRO101, Hidden Expectations and Potential Solutions.

\textbf{Accessibility Barriers in ASTRO101:} Economic disparities, learning differences, work-school balance, caregiving responsibilities, health issues, visual impairments,hearing impairments, and lack of knowledge about utilizing resources can hinder diverse student groups from accessing astronomy education effectively.

\textbf{Potential Barriers}

\begin{itemize}
  \item Astronomy has a classism problem, which can be a root of accessibility concerns in the field
  \item Insecurity (food, housing, financial)
  \item Imposter syndrome (imposter environments)
  \item Learning differences
  \item Work interfering with schoolwork
  \item Caregiving responsibilities
  \item Parenting responsibilities
  \item Illness or injury
  \item Undiagnosed or untreated mental health issues
  \item Concern about the deficit-model focus
  \item How can we integrate and accommodate differences in the classroom, especially with unsupportive administrations?
  \item Students are scared—fear inhibits the learning process
  \item Students fear failure and admitting it
  \item Faculty and staff face administrative and policy pressures that can indirectly create barriers for students
  \item Visual impairments pose a barrier since astronomy is highly visual
  \item First-generation college students may not know how to leverage resources (help, office hours, tutoring, etc.) and navigate “college culture”
\end{itemize}

\textbf{Hidden Expectations:} Students are expected to have strong mathematics and study skills, familiarity with technology, prior knowledge and interest in astronomy, positive past experiences with science, physical abilities, time management skills, reading comprehension skills, and an understanding of college culture/resources and Western-centric views of astronomy.

\begin{itemize}
  \item Mathematics skills and knowledge of calculating tools.
  \item Ability to easily interpret 2D representations (diagrams) of 3D situations.
  \item Note-taking ability and study skills.
  \item Solar System background and interest in the subject (is it a general ed requirement? Is this something we assume of our students?).
  \item Students have had positive prior mathematics and science classes.
  \item Students will have technology access and knowledge of how to use it.
  \item Students possess time-management skills and have chunks of available quiet time for coursework (much harder if, e.g., they are parents or don’t live on campus).
  \item Students have the ability to see, hear, and manipulate objects.
  \item Students are “blank slates” without preconceptions coming in.
  \item Assumption that all students know how college works before they get there, and that they share the same broad cultural background.
  \item Assumption that the Western history of astronomy is the only history of astronomy taught.
  \item Students knowing names of, e.g., Copernicus and Galileo and basic astronomy facts.
  \item Reading/learning style and preference matches the instructor’s.
\end{itemize}

\textbf{Potential Solutions:} Implement flexible course loads, engage faculty in promoting accessibility, use collaborative tools, offer flexible deadlines, use frequent low-stakes assessments, foster a sense of belonging, use clear language in materials, and provide advocates to help students navigate institutional challenges.

\begin{itemize}
  \item Facilitate part-time course loads without loss of financial aid or other benefits
  \item Focus on changing faculty policies and practices
  \item Use beginning-of-semester student information sheets (pronouns, names, prior math/science comfort, other commitments)
  \item Employ course tools that foster collaboration and conversation (e.g., Perusall)
  \item Emphasize approachability of due dates rather than mere flexibility: communicate late policies but encourage students to discuss extensions with the professor
  \item Implement frequent low-stakes formative assessments
  \item Foster regular substantive interaction (teacher–student, student–student)
  \item Promote a sense of belonging
  \item Use plain, simple, positive language in syllabi, instructions, and feedback
  \item Provide ombudspersons or advocates to help students communicate their needs to faculty and administration
  \item Apply universal design principles
\end{itemize}

Bowman, N., Logel, C., Lacosse, J., Canning, E. A., Emerson, K. T., \& Murphy, M. C. (2023). The role of minoritized student representation in promoting achievement and equity within college STEM courses. \emph{AERA open}, \emph{9}, 23328584231209957.

\emph{Graham, M., Frederick, J., Byars-Winstone, A., Hunter, A-B., Handelsman, J., Increasing Persistence of College Students in STEM, Science, 2013, 341, 6153}

\subsection{What are the best practices to attract and retain students from underrepresented communities, groups, and identities?}

\begin{flushright}
  \itshape
    Lancelot Kao \\
    Madeline Shepley \\
    Anna DeJong \\
    Katherine Hunt \\
    Kate Meredith
\end{flushright}

\textbf{Introduction}

In evaluating this question, we discovered that the process of attracting underrepresented students and retaining them are two concepts that are unique enough in practice to be tackled independently. While there is some cross pollination between the two, the execution of each is distinct enough to warrant a more individualized evaluation.

It was also acknowledged that these two practices most likely will need to be handled differently at different levels of execution: educators and administrators. To elaborate, educators are often restricted in the meaningful changes they can make in their own practice by the decisions and resources of their administrators and localized communities. Therefore, we are also dividing the practice of attraction and retention into micro/macro recommendations. Micro recommendations pertain to the smaller scale changes within an educator’s control. Macro recommendations are for our leadership entities who may be seeking guidance on how to support meaningful changes on a larger scale.

Ultimately, all of the micro recommendations appeared in the work related to: What are the hidden expectations in ASTRO101 that represent barriers to access to diverse student groups? For a deeper dive into the micro level of this question, please refer to this section.

\textbf{Recruitment Recommendations}

Attracting students from an underrepresented group starts with going to the source: children. Children are the future of humanity, and the things they encounter as they grow up inspire their wonder and shape their future path. Therefore, giving children the opportunity to encounter scientific work gives them the opportunity to explore the field and begin to contemplate and see themselves as a future professional (Aduriz-Bravo, 2020). In this way, it is believed that recruitment really begins much earlier than the last year of high school (Anderhag et al., 2016). This means that institutions should invest resources into K-12 outreach. This can include a variety of visible community programs such as:  open house events targeting diverse interests and needs, planetarium/star/sun parties on campus, relevant public lectures, and funding field trips if there are field trip appropriate features to one’s campus. Additionally, it is important in these recruitment activities to represent the diversity that is desired on campus by highlighting student role models. For these role models, some experience or training should be a prerequisite so that they may be as effective and intentional in their work as possible.

Guardians, often an underserved group in the recruitment process, should also be targeted in these efforts as familial support influences how likely a student will enter a STEM major or go to college. In recruitment efforts, it is important to frame college and STEM careers in particular, as a positive path for their child. This can be done in a variety of ways including alumni and undergraduate research highlights and concrete career examples. Guardians should also be encouraged to participate in these recruitment efforts as meaningful partners in their child’s education. This can be done in a variety of ways that include programs during hours guardians can access outside of work and in meal accommodations that can serve them and their whole family during the program. Naturally, community partners will be instrumental to institutions to help identify and access these communities in the recruitment process.

Lastly, the very programs on campus that can help retain students, can also be leveraged as a recruitment tool. Prospective students will crave a feeling of security as much as current students need to be secure, therefore, investing in robust student services will be instrumental in both the recruitment and retention of a diverse and healthy student body.

Below are some of the services that would be attractive to students with diverse needs: 

\begin{itemize}
  \item Academic Services
  \item Academic Counseling
  \item Tutorial Services
  \item Learning-assistance Program
  \item Infrastructure Support
  \item Childcare Solutions
  \item Housing and Food Solutions
  \item Medical Services
  \item On-Campus Medical Treatment Centers
  \item On-Campus Mental Health Support
  \item Group Insurance Policies Students Can Access
  \item Disability Services
  \item Institutional Support for Accommodation Assessments
\end{itemize}

It is also not enough to simply have these options available, but to make sure all faculty that interact with students are adequately aware of them. Every educator on a campus should not only feel informed but empowered to communicate what resources are available when they see needs in their students.

In addition to the above structures that can continue recruitment and retention, removing economic barriers to entry will go a long way. At the enrollment stage, having free or reduced hardware loans and rethinking textbooks and how they are accessed are just a couple of economic supports that are relatively straightforward to implement but have a major impact on a student's perceived ability to attend. Later in their academic journey, being offered paid internship opportunities and stipends for career development, this will require both institutional and community buy-in to achieve.

\textbf{Supporting Educators and Students to Self-organize}

The educational journey can have ups and downs, and it is imperative to not feel alone in the process. Being in community allows you to bond with fellow human beings in times of both joy and suffering. Fortunately, there are a multitude of ways to do that in a college environment.

From the class context, your faculty, as content experts, can encourage study groups among your students. In this way, they can get your students to practice their ability to collaborate. As individual human beings, we are not good at everything, so students can patch their various different skill sets to help each other succeed. They can use these skills to coach each other through the difficulties they encounter on homework, projects, and studying for exams. This also has the added benefit of allowing them to strategize to make the most of their resources, including using office hours efficiently. There’s usually many more students than the professor(s), which makes it hard for professors to give each student one-on-one help. Students helping each other gives them the power to help themselves and coordinate their questions when asking the professor for help. They can do this by figuring out common areas of struggle and getting the professor to address and fix the confusion a minimal amount of times, which creates more time for answering other questions. Additionally, since it can be hard to schedule office hours that are at a free time for \emph{all} students, organized study groups allow students to go to office hours to seek help on behalf of group members who can’t make the office hours and then report back for their compatriots. This mirrors the adage that “if you can teach it, you can understand” as the students become like “extra professors” for each other.

It is also important for students to have role models through which they can view the life of someone in the profession they aspire to so that they can see that this is a profession that they can see themselves in. This can be done in a variety of ways, whether via connections with older students or even professors themselves. In the professors, students can see what the life of a seasoned professional is like and learn from the wisdom their professors have gained in their years working in their job. This wisdom can guide them on what skills they should learn that will make themselves competitive and what pitfalls to avoid. Older students can also provide this example, but given their frequent closer proximity age-wise, they have the benefit of having been more recently in the shoes of the their younger compatriots and can give targeted advice (e.g. which professors teach a subject best if applicable, common student challenges in specific classes, skills that will be helpful for suggest) for navigating challenges as a student of your institution from a very similar point of view.

In the role model thread, it is also important to strive to be a good model of work-life balance. As human beings, we are not designed to work 24/7. Sleep, nutrition, hydration, and even spending time with friends and connecting with life-giving activities rejuvenates us and prepares us to do good work in our fields. Each of these things has a proper order, and neglecting even one of the areas can lead to deterioration of health that can bring a complete halt to the work and eventually burn someone out. If we model unbalanced work-life situations, students will think that they have to prioritize their work at the expense of their health. Thus, by showing students how to set boundaries, they can see that it’s okay to say no to things to make sure they don’t ruin their health over their work.

In addition, we can support students by encouraging student-led clubs, service learning opportunities, mentoring programs, and electronic forums. Research has shown that the creation of learning communities based upon identity is a powerful tool for retaining students from underrepresented groups in STEM.  These learning communities can be subject matter specific or STEM in general. Student-led clubs and mentoring programs will allow students to build community for themselves by giving students an opportunity to connect with students with similar interests and connect younger students with older students so that younger students may glean wisdom from their older peers. Service-learning will allow students to learn important non-astronomy specific skills (i.e. public speaking) and practice them in a way that helps out the local community. Finally, electronic forums will allow students to form communities and solve problems more easily by allowing them to connect even when not in person.

\textbf{References}

Adúriz-Bravo, A., \& Pujalte, A. P. (2020). Social Images of Science and of Scientists, and the Imperative of Science Education for All. In H. A. Yacoubian \& L. Hansson (Eds.), Nature of Science for Social Justice (pp. 201–224). Springer International Publishing.

Anderhag, P., Wickman, P.-O., Bergqvist, K., Jakobson, B., Hamza, K. M., \& Säljö, R. (2016). Why Do Secondary School Students Lose Their Interest in Science? Or Does it Never Emerge? A Possible and Overlooked Explanation. Science Education, 100(5), 791–813.

\section{Astronomy Education Research}

\subsection{Is content knowledge and cognitive processes really that important to understand compared to attitudes, self-efficacy and career intention?}

\begin{flushright}
  \itshape
    Jackie Milingo \\
    Abbas Mokhtarzadeh \\
    Sean Moroney \\
\end{flushright}

In this article, we are interpreting this question to be what is more important to assess - content knowledge and cognitive processes OR attitudes, self-efficacy, and career intentions.  Our discussions questioned whether they are inextricably linked\textbf{;} at some point one of the groups pointed out they lead to each other in an interrelated way.

Curiosity → engagement → literacy → attitude → engagement → literacy → and so on …. How are they linked?

Formative and summative assessments would be useful in responding to this question.  Low-stakes and high-stakes assignments/exams, surveys, interviews, etc. whatever would be useful for education researchers.

Does this question depend on whom we are assessing?  Majors or non-majors?  At what level?  Somewhere in K-12?  If only at the intro level, unsure of whether AST 101 is a vehicle for prospective majors, then it would be interesting to understand and assess all of these things - content knowledge (this requires thought considering\textbf{ }that AST 101 can encompass a variety of appropriate topics), cognitive processes (what and how?), attitudes (about the subject, identity, belonging?), self-efficacy, and career intention.  The intersection and relationship of all these things would be useful to analyze.

The five concepts associated with this question need to be carefully defined before we can assess them.  The following are our starting points.

\textbf{Content Knowledge}

This refers to the actual subject matter or information that students learn. It includes facts, concepts, theories, and skills related to a specific field (e.g., mathematics, science, or engineering).  It includes all the facts as known so far, but should also include projections of that knowledge outward from those facts into the unknown intellectual spaces around that knowledge.

\textbf{Cognitive Processes}

These are mental activities involved in learning, problem-solving, and critical thinking. Examples include analyzing, evaluating, and applying knowledge.  They could include memory, perception, decision-making,  cognition, metacognition, interacting, working memory, attention, language, critical thinking, learning, executive functioning,  problem solving, dreaming, creativity, reading,information processing, planning, and reasoning\textbf{, etc}.

\textbf{Attitudes:}

These are internal postures that are present at the beginning of the class.  They determine receptivity and can be shaped positively or negatively by the educational experience. Students’ attitudes toward a subject influence their engagement, motivation, and persistence. Attitudes may be positive or negative; positive attitudes can enhance learning experiences.

The \textbf{student’s }internal posture \textbf{is} adopted \textbf{here} toward an external situation, in this case, a learning situation.   We may also want to consider how attitudes are related to \textbf{the student’s} science identity, \textbf{sense of} belonging, etc.  What exactly is assessed when it comes to “attitude”?

\textbf{Self-Efficacy: }

This refers to a student’s belief in their ability to succeed in a specific task or domain. Does high self-efficacy lead to greater effort and persistence, improved content knowledge and cognitive processing\textbf{?}  Can self-efficacy change through the educational process \textbf{with} intentional learning and interventions?

\textbf{Career Intention:}

Career Awareness and Intention.  Students’ awareness of STEM careers \textbf{and their level of interest in pursuing} them impact\textbf{s} their educational choices. Positive career intentions can drive motivation and focus.

Relevant comments from group discussion follow, along with how we think they connect to the question at hand.

\emph{We've used Saundra McGuire's Teach Students How }\emph{To}\emph{ Learn and the student version, Teach Yourself How }\emph{To}\emph{ Learn to introduce this to our students.  I typically dedicate one class per semester to addressing it directly in class (the one after I return their first graded in-class assessment). I also have them do quiz/exam reflections throughout the semester.}

Perhaps this comment is more relevant to what we can intentionally do in the classroom to connect attitudes to pedagogy and other interventions.  This could be assessed (formative).

\emph{Student's}\emph{ don't know what content knowledge is in their best }\emph{interest  (}\emph{usually, that would be what is necessary to pass the next exam).  We [need more] knowledge of how knowledge is acquired and of how culture and prior knowledge impacts learning and cognitive processes.  Students need to understand how they learn - metacognition.}

Again, this comment is more relevant to what we can intentionally do in the classroom to connect attitudes to pedagogy and other interventions.  This could be assessed (formative).

\emph{We should all be researching our own teaching and our own students all the time --> action research (which is cyclical). }

\emph{For Q2. There is something more important in the classroom, and we need to resist just partitioning the subject of study into just }\emph{studens}\emph{, just instructors or just populations. AER has a dark matter problem and that is adequately capturing and analyzing the dynamic web of interactions present in the classroom that are modifying the learning experience and enhancing or damping the effects of pedagogical intervention, but that the instructors and the students individually might not be aware.}

This is a comment in response to the question about which populations to study - K-12 students, college students, instructors/teachers.

\emph{Cognitive processes are important for increasing self-efficacy; how do cultural components tie in? More research }\emph{needed}\emph{ here.  }

This addresses the question directly.

In summary, a holistic approach to future AER research should consider content knowledge, cognitive processes, and factors such as attitudes, self-efficacy, and career intention, as all of these are essential for fostering successful STEM learning and career pathways.  We need to capture other factors such as science identity, belonging, and other cultural components to know what affects attitudes and self-efficacy.  We need to understand how content knowledge, cognitive processes, attitudes, self-efficacy, and career intention are related and affect each other.  And finally we need to consider which populations we’re studying to know if content knowledge and cognitive processes are as important as understanding attitudes, self-efficacy, and career intention.  

\subsubsection{Should we be focussing our research more on students, more on undergraduate instructors or more on school-level teachers?}

\begin{flushright}
  \itshape
    Jim Buchholz \\
    John B Taylor \\
    Nicole Gugliucci \\
    Debbie French
\end{flushright}

One of the key aspects of research and practice is that each informs the other. With this in mind, this section synthesizes the discussions focussed on identifying where research efforts need to be focussed on with regards to target groups (students or educators). To achieve this we identify several stages of astronomy education and provide recommendations with regards to important outstanding research questions.

\textbf{Elementary Students (Age 5-10) \& Teachers (K-5 grade)}

The research shows students decide to pursue STEM vs non-STEM careers as early as grade 4 (Murphy, 2011; Ricks, 2012). Thus, exposing students to these topics in earlier grades is essential. While scant science instruction occurs in elementary school (approximately two hours per week on average), it is more likely to occur in districts with more resources (Blank, 2013; Banilower, 2018). As STEM careers typically pay more (Edwards, et al., 2021; ) than non-STEM careers, teaching elementary STEM is an act of social justice (Early Childhood STEM Working Group, 2017; National Academies of Science, Engineering, \& Medicine, 2021). Offering science in earlier grades may also help increase science literacy overall (National Academies of Science, Engineering, \& Medicine, 2021). No Child Left Behind has given precedence to mathematics and English Language arts. Because of these points, we advocate for more research to be done in the following areas:

\begin{itemize}
  \item How can astronomy be integrated with mathematics and English Language Arts curriculum in the elementary grades?
  \item What additional training or professional development do in-service teachers need?
  \item What high-quality, research-based resources need to be developed to support astronomy integration in elementary school?
  \item Is astronomy integration an effective tool for increasing students’ engagement in science?
  \item Can astronomy be used as a vehicle for teaching STEM in elementary school as an act of social justice by exposing students to future careers in astronomy and related STEM fields?
\end{itemize}

\textbf{Middle School(}\textbf{ish}\textbf{) (4 - 8 grade, age 9-13)}

We make the claim that while astronomy (and science education) need improvement at all levels (K-16) to help engage and encourage students to enter STEM fields for careers, grades 4 - 8 may be the ideal time to heavily engage and infuse science into a student's studies. Middle school teachers are often overlooked as precedence is given to elementary and high school teachers. This is problematic as Earth science instruction (where astronomy is typically included as the NGSS lists Earth \& Space Sciences as a category for the Disciplinary Core Ideas) begins and ends in middle school. Only about 5-15\% of high school students take Earth science in high school (Gonzales \& Keane, 2011; Worth, 2021). It has often been said that astronomy is in the unique position of being a “Gateway” science to STEM (Salimpour, Bartlett, Fitzgerald et al., 2021). With this postulate, we believe that the standard ASTRO101 classes taught in college are far too late for introducing students to STEM careers. Some of this reasoning comes from many college students taking ASTRO101 in their junior or senior year as their lab science general education requirement. Also, even collegiate freshmen and sophomores taking ASTRO101 who find an interest in STEM via astronomy, may discover they are lacking the mathematical preparation necessary to continue. The most productive way to attract the largest increase in student interest in STEM is to engage the students at a much earlier time in their studies. High school may be semi-late as a structured mathematical and science program needs to be set out at the beginning of one’s high school experience. Therefore, the middle school years may be the most advantageous time to engage students.

With the assumption that astronomy can be used as a gateway to STEM, here are some possible paths to studying and changing how students learn and engage with astronomy/science education:

\begin{itemize}
  \item Pre-surveys on knowledge and attitudes of students and teachers at the beginning of the semester/year
  \item Post-surveys on knowledge and attitudes of students and teachers at the end of the semester/year
  \item Analyze surveys (with college professors and select 4th–8th team leaders) to evaluate weaknesses and needs in 4th–8th grade classrooms
  \item Develop workshops (with college professors and select 4th–8th team leaders) on new astronomy education methods and pedagogies (e.g., inquiry-based, project-based learning, research methodology)
  \item Assess students’ observation and questioning skills
  \item Conduct teacher in-service/continuing education/professional development in conjunction with college professors
  \item Repeat pre-surveys for the next class
  \item Implement activities learned in continuing education with students
  \item Conduct post-surveys
  \item Analyze new survey results
  \item Share survey results and analysis with teachers and counselors
  \item Coordinate with district Curriculum \& Instruction and/or Assistant Superintendents to align K–12 science across grades and support the 4–8 transition
\end{itemize}

 Thinking research (Liljedahl 2016) has revealed how many instructional processes develop an atmosphere that promotes non-thinking students within the traditional classroom's mathematics instruction.  His research comes from a multiyear study and focuses on the importance of moving classrooms to “thinking” classrooms.  But, his ideas easily translate to the sciences, with potential to compound gains when utilizing fields such as astronomy.  Beyond classroom practices, he emphasizes the needs of utilizing Professional Learning Communities (PLC’s) which have been incorporated as a common component of many schools nationally.  Research should consider utilizing these existing networks to help facilitate positive change in schools.

He also stresses the importance of vertical teaming between grade levels.  To successfully impact grades 4-8 (ages 9-12), the teachers in the grades directly above and below need to be connected for successful impact. This capstone could prove important to the ongoing success of students, as they could better prepare students heading into 4th grade, meeting teachers with realistic and data based information, and then provide a smooth transition into high school.

Two other symbiotic points he makes are the utilization of external partners and allowing students to have real-world experiences to support the growth of thinking students.  Although he refers to many community groups, this may be where higher academia science, such as astronomy, may make sense to provide a sensible support.  Mathematics supporting mathematics, ironically lacks the inquiry, and investigative experiences that are a component of astronomy education.  Despite his focus on improving mathematics instruction, astronomy not mathematics, offers the fun hands-on inquiry investigations to do mathematics.

\textbf{High School Teachers \& Students (Grades 9-12)}

The NGSS includes astronomy in the Disciplinary Core Ideas under Earth \& space science section. Thus, the NGSS gives equal importance to Earth \& space sciences as physical and life sciences (National Research Council, 2013). Yet, few students (less than one fifth) take Earth \& space science in high school (Gonzales \& Keane, 2011; Worth, 2021).

Suggestions for future research for high school students:

\begin{itemize}
  \item As solar-system and galactic astronomy is typically taught in high schools, further research needs to be conducted on students’ perceptions about and misconceptions when learning about extragalactic astronomy. (Salimpour, Fitzgerald, Hollow, 2024)
  \item How are students impacted by learning from teachers who have participated in a research experience for teachers? (Rebull et al., 2018)
  \item Assessing students’ changes in attitudes during an astronomy course. (Bartlett, 2018)
  \item Assessing students’ perceptions of STEM and their understanding of the nature of science as a result of engaging in student-centered lessons using authentic/archival research.
\end{itemize}

Suggestions for future research for high school teachers:

\begin{itemize}
  \item As the NGSS advocates for more authentic science instruction through three-dimensional learning, what additional \emph{training} do high school teachers need in order to effectively teach astronomy either as a separate course or embedded in other courses such as physics or Earth science? (Framework, 2012)
  \item What additional training do teachers need in order to incorporate authentic/archival data into their curriculum?
  \item As the NGSS advocates for more authentic science instruction through three-dimensional learning, what additional \emph{resources} do high school teachers need in order to effectively teach astronomy either as a separate course or embedded in other courses such as physics or Earth science? (Framework, 2012)
  \item What type(s) of professional development programs best support teachers in increasing their self-efficacy with teaching authentic scientific inquiry and being comfortable with the iterative process of science? (Rebull, 2024)
  \item How can we support teachers in incorporating science research into their courses? (National Research Council, 2013; Rebull, 2024)
\end{itemize}

We also advocate for Astronomy Education Research to be included in the Handbook of Research in Science Education, Volume III (or similar work). We also advocate for the status of Earth Science courses (and, thus, astronomy) to be elevated (Orion \& Libarkin, 2023; Shaffer, 2012).

\textbf{Undergraduate Instruction, Research, Mentoring, and Advising}

\textbf{Recommendation: }\emph{Investigate the efficacy of implementing evidence-based astronomy education techniques in different environments. Reframe investigations to center the assets and skills of instructors. }

Much astronomy education literature focuses on techniques for teaching material in “ASTRO101,” as is the case across the various STEM discipline-based education research areas (). As seen from physics education research, efficacy of evidence-based tools and techniques depend on the implementation in the classroom (e.g. ), and, though more and more instructors in physics are aware of active learning techniques, a substantial amount of class-time still uses lecture (). Astronomy instructors may follow similar patterns, so an accurate accounting of this should be pursued. In fact, a stronger connection between astronomy education researchers and astronomy education practitioners is needed at every level as discussed in the sub-section on “Feedback Mechanisms” below to address such questions of implementation.

The “teaching-method-centered paradigm” is only one way of looking at discipline-based education research, proposing an alternative “asset-based agentic paradigm” which focuses on individual faculty teaching practices and agency. Research with such a lens is an important step in bringing feedback back to the astronomy education researchers, thus informing a cycle of refining pedagogical strategies in their natural context in different classroom environments with different instructors. Even a single instructor can see vastly different results in different sections of the same course indicating that a wide range of environmental factors need to be considered.

\textbf{Recommendation: }\emph{Expand astronomy education pedagogy research to include upper-division coursework. }

Research into the implementation of pedagogical techniques should explore beyond “ASTRO101” to ensure the health of the discipline. The typical “ASTRO101” course may not be taken by potential astronomy and astrophysics majors and minors. As noted above, this is often “too late” for recruiting new astronomy/astrophysics majors. For students who have made it to the point of declaring (or declaring interest in) the major, equitable and effective teaching strategies are needed to ensure retention of students and appropriate preparation for studies and careers in astronomy.

This lack of literature on upper-division astronomy has been noted by  in describing the opportunities within an upper level quantitative question where active learning strategies are likely to improve student understanding and skills. This may be complicated by the fact that astronomy does not have a “standard” curriculum, either formally, such as with chemistry programs that are certified by the American Chemical Society or informally through recommendations such as in undergraduate physics (e.g. ). This complication has been a recent topic of study and discussion by the Education Committee of the American Astronomical Society (AAS) ().

\textbf{Recommendation: }\emph{Investigate best practices for mentoring and advising outside the classroom for undergraduate astronomy/astrophysics majors/minors.}

The number of bachelor's degrees awarded in astronomy has approximately doubled in the last ten years. This counting does not include students who have a primary major in a field such as physics with an astronomy or astrophysics minor. However, doctoral programs in astronomy have not increased commensurately, though this is still often the career path most familiar to undergraduate instructors and advisors. This presents two issues; first, students interested in pursuing an academic career path face very difficult competition for a coveted few spots in graduate programs. How has this increase in selectivity advantaged or disadvantaged certain populations of students? What are undergraduate programs doing to prepare today’s students for that reality? Secondly, this underscores the need for well-informed career mentoring outside of the academic track at the undergraduate level, as 53\% of astronomy bachelors are not enrolled in graduate school one year after their degree ().

All of these questions should be studied in light of various reports and recommendations which aim to increase participation of historically underrepresented groups in the field (e.g. ) and in consultation with the following AAS committees, CSMA, SGMA, CSWA.

After all, though recruitment in middle school grades is crucial to engendering interest in STEM, appropriate steps should be taken at every educational stage to ensure equity of access. This includes academic advising, career advising, involvement in research, and other aspects of education that are likely to happen outside the classroom.

\textbf{All levels: Feedback Mechanisms Between Researchers and Practitioners}

At all levels, we call for a stronger relationship between astronomy education researchers and practitioners. Too often, research on excellent instructional strategies is unknown to many astronomy instructors who would be eager to implement recommendations in the classroom. On the other hand, the research would greatly benefit from the real-world knowledge of instructors, and instructors can better communicate the kinds of classroom constraints that complicate adoption of new techniques.  found that inadequate time, resources, and models were barriers to implementing inquiry-based pedagogies for high school teachers who were positively inclined to use such strategies. This research should be expanded to all levels of education.

We envision a collaborative structure using a common platform that would be accessible to both researchers and practitioners. This would likely include support and engagement of our common professional societies, such as AAS and the American Association of Physics Teachers (AAPT). This could facilitate learning communities, drawing on experiences in faculty development such as with the Next Generation Physical Science and Everyday Thinking (NGPET) Faculty Online Learning Communities (). Researchers and practitioners could use this platform to start and nurture collaborations. Although we cannot necessarily solve the issues of time constraints for instructors, this could remove the barriers of access to resources and access to other members of the community as long as it is properly supported and advertised by professional societies.

\textbf{References}

AIP TEAM-UP Task Force, 2020. \emph{The Time Is Now: systemic Changes to Increase African}

\emph{Americans with }\emph{Bachelor’s Degrees in Physics}\emph{ and Astronomy}.

Andrews, T.C., Speer, N.M. and Shultz, G.V., 2022. Building bridges: A review and synthesis of research on teaching knowledge for undergraduate instruction in science, engineering, and mathematics. \emph{International Journal of STEM Education}, \emph{9}(1), p.66.

Dancy, M., Henderson, C., Apkarian, N., Johnson, E., Stains, M., Raker, J.R. and Lau, A., 2024. Physics instructors’ knowledge and use of active learning has increased over the last decade but most still lecture too much. \emph{Physical Review Physics Education Research}, 20(1).

Dancy, M., Henderson, C. and Turpen, C., 2016. How faculty learn about and implement research-based instructional strategies: The case of peer instruction. \emph{Physical Review Physics Education Research}, \emph{12}(1).

Fitzgerald, M., Danaia, L. \& McKinnon, D.H. Barriers Inhibiting Inquiry-Based Science Teaching and Potential Solutions: Perceptions of Positively Inclined Early Adopters, 2019 \emph{Res Sci Educ} 49, 543–566.

Joint Task Force on Undergraduate Physics Programs, 2016. \emph{Phys}\emph{21:Preparing}\emph{ Physics Students}

\emph{for 21st-Century Careers}.  

Liljedahl, P. (2016). Building thinking classrooms: Conditions for problem-solving. Posing and solving mathematical problems: Advances and new perspectives, 361-386.

More, T., Goldberg, F., Basir, M., Maier, S., and Price, E., 2024. From implementation to reflection: exploring faculty experiences in a curriculum-focused FOLC through multi-case analysis. \emph{Discip}\emph{ }\emph{Interdscip}\emph{ Sci Educ Res} 6, 7.

Mulvey, P. and Pold, J., 2023. New Astronomy Bachelors and Masters: What Comes Next. \emph{AIP Statistics Report}.

Nicholson, S. and Mulvey, P., 2023. Roster of Astronomy Departments with Enrollment and Degree Data, 2022. \emph{AIP Statistics Report}.

Strubbe, L.E., Madsen, A.M., McKagan, S.B. and Sayre, E.C., 2020. Beyond teaching methods: Highlighting physics faculty’s strengths and agency. \emph{Physical Review Physics Education Research}, 16(2).  

Wallace, C.S.; Developing peer instruction questions for quantitative problems for an upper-division astronomy course. \emph{Am. J. Phys.} 1 March 2020; 88 (3): 214–221.

Wallace, C., Coble, K., and Bailey, J., 2021. What Do We Expect Undergraduate Astronomy Majors to Learn? \emph{AAS Education Committee Blog}.

\subsection{Should we be broadening our horizons beyond ASTRO101 as the focus? Is ASTRO101 "too late"? Should we push to year levels beyond astro101?}

\begin{flushright}
  \itshape
  Richard Datwyler\\
  Enrique Gomez\\
  David McKinnon
\end{flushright}

We interpret the term “broadening” to mean undertaking research activities before, during and after the semester when students “experience ASTRO101”. That is to say, we expand the domain of research into that which can happen both before and after the events occurring in ASTRO101 as well as including what might happen \emph{within }the subject.

Over the past thirty years, professional societies and government organizations began to recognize the impact of informal STEM education in supporting formal education. As a result of this, they have invested in this informal component in the form of “science ambassadors” programs and education resources for “informal” or “outreach” educators (Fraknoi et al., Bennet \& Morrow, 1994). The effectiveness of these programs in the preparation of informal educators, and the quality of the education they offer is not known. To what extent do they disrupt or inadvertently perpetuate alternative conceptions (ACs) that undermine the work of formal educators? As an aside, unless we deal with these in our ASTRO101 courses, the products of these will carry these ACs into the future.

In our early discussions we defined both “Formal” and “Informal” ecosystems of astronomy education. In reality, there is a second dimension that has to be considered. We have termed this other dimension “veracity” and which could be considered to also exist on a continuum from “currently accepted science belief” to “pure hokum”.

We define a “Formal System" as one in which students engage, or not, within a formal system of instruction, e.g., schools, universities, Scouts etc. These Formal systems are ones that typically follow a “curriculum”. The “veracity” of the information being presented through these curricula is highly dependent on the instructor delivering the curriculum and the wider social system within which the organization exists.

Using this two-dimensional approach, a “Formal-Dubious-Veracity” system may be one which is informed not by science, but by a “belief system” such as a fundamentalist faith organization. A “Formal-High Veracity” source may be one that reflects the current scientific understanding of the state of our knowledge such as a school, university or the Scouts (where the curriculum is vetted by experts in the field) and where the instructors know the content to a deep level. However, even within such Formal-High Veracity systems, the instructor’s content knowledge and belief systems can interact in unanticipated ways that lead to alternative conceptions (ACs) being promulgated as “scientific truth”.

Educational researchers have highlighted this problem and that it is common in elementary, middle and high schools where astronomy is concerned. This is especially the case where teachers have to teach outside their subject speciality. The literature has many reports of non-astronomy teachers teaching that the phases of the Moon are caused by the Earth’s shadow blocking light. The potential disparate nature of the enacted curriculum content in schools merits more research to investigate what teachers “actually” know and teach. We assume that ASTRO101 instructors will teach “high veracity” formal curricula given that the vast majority have a degree in a field of astronomy.

Informal systems are many and diverse, and range in the veracity of the content they deliver. They include, but are not necessarily limited to: school clubs and camps; astronomy camps; visits to planetariums; museums; talks by Mums and Dads; Youtube influencers; Bill Nye the Science Guy; Mythbusters; “unvetted” amateur content creators on YouTube; and, social media influencers on platforms such as Facebook and “X”. That said, many of these informal sources can engender curiosity that leads to individual learning and further study.

Of these informal systems, nationally funded organizations exist where materials have been vetted by professional astronomical committees: NOVA, NASA Kids club and other similar programs. For example, NASA has supported the work of amateur astronomy clubs in providing informal astronomy education through public outreach through its Night Sky Network. Another example would be the National Science Foundation’s Advancing Informal STEM Learning (AISL) Program that funds research and implementation of a broad range of informal STEM learning experiences. These organizations may be considered to deliver “high veracity” content.

However, the veracity of the information provided by other Informal Systems can vary greatly from the “correct” currently understood best explanation of astronomical and cosmological phenomena to “conspiracy theories” such as the Moon Hoax landings, or the occultation of Mars by the Moon that surfaces regularly, or the alignment of the planets, or the Mayan “end of ages” as well as what will be seen or experienced at these events and commonly promulgated by social media influencers. There is also a potential source of ACs generated by science script consultants who lack scientific knowledge and who work on popular films and/or TV series. Here large distances within the galaxy have to be covered within an episode and can lead to ACs about space travel times and distances in space (e.g., Dahsah et al., 2012; Miller \& Brewer, 2010). Thus, many informal sources can generate widely held “alternative conceptions”.

We use the term “Alternative Conception (AC)” in education rather than the term “Misconception”, which tends to be used by scientists. The latter term implies that there is a right/wrong aspect to a conception. In fact,  students’ ACs can provide a very useful starting point for educators (Salimpour et al., 2020; Salimpour, 2021; Salimpour et al, 2023). For example, students can be put in pairs with varying ACs in class invited to share these with each other and then to generate “experimental tests” designed to disprove the other. These “experiments” can be conducted in the formal setting of the classroom or laboratory where the scientific approach of proving the \emph{other }model/explanation is \emph{wrong }or that a \emph{simpler }explanation is available, a key process of science. See for example:   (Slater et al., 2018; Trumper, 2003).

However good the veracity of the content, both formal and informal systems can spur curiosity in students. The content presented can play the role of the “icebreaker or introductory educational experience” at the beginning of the educational journey where students can become “Engaged” by what is being presented (Bybee, 1997). Following Engagement, students Explore, try to Explain before the teacher helps them Elaborate and then Evaluate what they have learned. If done well, the information and the 5E learning approach can lead to correct ideas and practices that can be further elaborated in schools leading up to college.  But, if incorrect ideas and information is presented as “truth", more work in the ASTRO101 classroom will be required to dismantle the ACs students possess before they learn better scientific explanations using high quality materials and good instruction.

Identifying students’ ACs before instruction can be challenging. If the instructor holds the view that a “misconception” is just plain “wrong” and tries simply to replace it with the “correct one” during the delivery of ASTRO101, h/she runs the risk that various different “mixes” of ideas will be generated in the students’ minds where the scientific explanation being presented is incorporated to varying degrees within the “original AC” (White \& Gunstone, 1992). At one end of this continuum is complete rejection of the instructor’s new (and correct) explanation while at the other end is complete acceptance of the scientific explanation (although the learner may be smart enough to deliver answers that they think the teacher wants). In between, there are mixed cognitive models generated by the student where aspects of the correct explanation are incorporated into the original AC and the student is satisfied with this new \emph{hybrid }conception. For example, in dealing with the AC of distance from the Sun being responsible for seasons, one hybrid conception is that we are still \emph{closer }to the Sun in summer and \emph{further away} in winter “because of the Earth’s tilted axis”. They have thus incorporated one aspect of the correct scientific model but not dealt with the more general ideas associated with the flux of radiation hitting each square meter of the Earth’s surface caused by the tilted axis at different times over a year. They will thus exit the ASTRO101 with this new hybrid, but still alternative, conception believing it to be correct.

Thus, when students arrive at university and perhaps undertake ASTRO101, they will have had a variety of exposure to aspects of Astronomy in both formal and informal settings and likely hold many ACs that contain only certain aspects of the correct scientific explanation. Thus, a pre-test on entry to ASTRO101 can illuminate many of the ACs that students hold as a result of their previous experiences with Astronomy and of their attempts to make sense of the world. A post-test can be employed to investigate how successful we have been in redressing their ACs.

More broadly, can we, as the ASTRO101 community, “investigate” these sources of information, both good and bad, in any sort of meaningful way? Our conclusion is that the short answer is both “yes and no”. Some influencers could be investigated by looking at what they post as content and the number of followers they have. The courses offered by other formal groups, such as the Scouts, could be investigated in a realistic fashion using appropriate methods.

From the perspective of the ASTRO101 instructor, and in contrast to these broader investigative questions about the formal and informal ecosystems, which would require funding, it is the opinion of our group that the first task of the ASTRO101 instructor should be to find out what conceptions the students have about the various phenomena s/he intends to include in her/his course as they arrive into ASTRO101 for the first time. If one adopts this approach, it indicates that a \emph{Participatory Action Research} methodology (MacDonald, 2012) could be employed. The instructor is a participant in improving their own practice and can employ a mixed-method approach to becoming better at teaching. Documenting the investigation can lead to publication in Q1 Journals as well as a case being made for promotion and/or teaching awards.

How can this information collection be done easily and with minimal or zero funding? In order to elicit the current “scientific beliefs” of the students on arrival, various instruments exist in the published astronomy education literature that can be used with minimal time constraints on the course. One such instrument is the Astronomy Diagnostic Test comprising 25 multiple choice items covering topics such as phases of the Moon, Day and Night, Seasons, celestial movements ( ). Another example is the TOAST (Test Of Astronomy STandards) instrument ( ). Various other and more highly focussed “Concept Inventories" exist. Reliable and valid tests for Physics are designed to determine whether a student has an accurate working knowledge of a specific set of concepts in a particular content knowledge domain (e.g., forces, EM phenomena, and electron transitions in atoms). Yet another example of a validated instrument is the Light and Spectra Concept Inventory (LSCI, Bardar et al.,  2007).

Any introductory astronomy college course such as ASTRO101, however, plainly does not exist in an educational vacuum. Rather,  it is a part of both the formal and informal STEM education ecosystem, and which also contains sources of knowledge of varying veracity. ASTRO101 courses thus accept students who have had their concepts of astronomy already shaped by that ecosystem. These ASTRO101 courses also shape the ecosystem by populating it with graduate students who have taken the course. This means that the effectiveness of the ASTRO101 course is dependent on the overall quality of the prior ecosystem including the disruptive effects of ACs that persist within it. Thus, astronomy educators and AER researchers could, if they so desire, inform the dynamics of this ecosystem both in terms of the content that is being delivered, as well as generating a taxonomy of organizations that support and populate it.

There is, to some extent, some pressure on educators on what to cover in their ASTRO101 course. This pressure stems both from internal institutional standards and requirements of course transferability across institutions in the USA. Some release of this pressure on instructors may come if they have trust that the informal STEM education ecosystem can fill in the gaps in coverage of Astronomy content thus realizing the promise of life-long astronomy learning. Dissemination of an updated overview of the STEM ecosystem would help educators in making choices about what richer learning experiences to implement such as robotic telescope labs. However, the instructor’s perceptions of the pressure and the content that has to be covered can generate resistance towards the adoption of pedagogical innovations such as student-centered active learning or the implementation of robotic telescope laboratories. An investigation of these perceptions of pressure would be worth investigating.

Too often in higher education, we can become narrow in our view of the effects that our courses have on the astronomy community. We must consider the formal and informal ecosystems with their varied degrees of content veracity to which our students have been exposed \emph{before }they enter our classes and perhaps even \emph{before }we plan and prepare the curriculum for the class. Additionally, exit surveys at the end of class where students identify their future goals and changed perspectives, along with post-test concept inventories measuring normalized gains could provide the necessary information of the success of our attempts.  Longer term studies are also possible, given appropriate funding, to track students five, ten years and beyond into the future, recording employment and activity in the astronomy community.

\textbf{Conclusion}

In summary, ASTRO101 instructors can undertake research into: 

\begin{itemize}
  \item The informal and formal ecosystems to which their students have been exposed
  \item The veracity of the content these ecosystems have delivered
  \item Their own practice in ASTRO101
  \item Longer-term studies of their graduates post–ASTRO101
\end{itemize}

All of these investigations should be formalized and submitted for publication in the Astronomy Education literature in an attempt to disseminate the findings and avoid the replication of research questions that have already been investigated.

\textbf{References}

The AAS Astronomy Ambassadors Program | American Astronomical Society. (n.d.). Retrieved February 27, 2025

Ayas, A., Özmen, H., \& Çalik, M. (2010). Students’ conceptions of the particulate nature of matter at secondary and tertiary level. International Journal of Science and Mathematics Education, 8, 165-184.

Bardar, E. M., Prather, E. E., Brecher, K., \& Slater, T. F. (2007). Development and validation of the light and spectroscopy concept inventory. Astronomy Education Review, 5(2), 103-113.

Bennett, J. O., \& Morrow, C. A. (1994). NASA’s Initiative to Develop Education through Astronomy (IDEA). Astrophysics and Space Science, 214(1), 237–252.

Bybee, R. W. (2002). Scientific inquiry, student learning, and the science curriculum. Learning science and the science of learning, 3, 25-35.

Dahsah, C., Phonphok, N., Pruekpramool, C., Sangpradit, T., \& Sukonthachat, J. (2012). Students’ conception on sizes and distances of the earth-moon-sun models. European Journal of Social Sciences, 32(4), 583-597.

Fraknoi, A., Fienberg, R. T., Gurton, S., Schmitt, A. H., Schatz, D., \& Prather, E. E. (2014). Training Young Astronomers in EPO: An Update on the AAS Astronomy Ambassadors Program. ASP Conference Series. V 483.

MacDonald, C. (2012). Understanding participatory action research: A qualitative research methodology option. The Canadian Journal of Action Research, 13(2), 34-50.

Miller, B. W., \& Brewer, W. F. (2010). Misconceptions of astronomical distances. International Journal of Science Education, 32(12), 1549-1560.

Salimpour, S. (2021). Visualising the Cosmos: Teaching cosmology in high school in the era of big data [Doctoral Thesis]. Deakin University.

Salimpour, S., Tytler, R., \& Fitzgerald, M. T. (2020). Exploring the cosmos: The challenge of identifying patterns and conceptual progressions from student survey responses in Cosmology. In R. Tytler, P. White, J. Ferguson, \& J. C. Clark (Eds.), Methodological Approaches to STEM Education Research (Vol. 1, pp. 203–228). Cambridge University Press.

Salimpour, S., Tytler, R., Fitzgerald, M. T., \& Eriksson, U. (2023). Is the Universe Infinite? Characterising a Hierarchy of Reasoning in Student Conceptions of Cosmology Concepts Using Open-Ended Surveys. Journal for STEM Education Research.

Slater, E. V., Morris, J. E., \& McKinnon, D., (2018) Astronomy alternative conceptions in preadolescent students in Western Australia, International Journal of Science Education, 40 (17) 2158-2180.

Trumper, R., (2001) A cross-age study of junior high school students conceptions of basic astronomy concepts, International Journal of Science Education, 23 (11) 1111- 1123.

White, R.T., \& Gunstone, R.F., (1992). Probing understanding: London: Falmer Press.

\end{document}